\documentclass[preprint2]{aastex}

\slugcomment{}

\shorttitle{The {\it Spitzer} view of FR--I radio galaxies}
\shortauthors{Leipski et al.}

\begin{document}

\title{The {\it Spitzer} view of FR--I radio galaxies: on the origin of the
  nuclear mid--infrared continuum.}

\author{C. Leipski}
\affil{Department of Physics, University of California, Santa Barbara, CA 93106}
\email{leipski@physics.ucsb.edu}

\author{R. Antonucci}
\affil{Department of Physics, University of California, Santa Barbara, CA 93106}

\author{P. Ogle}
\affil{Spitzer Science Center, California Institute of Technology, Mail Code 220-6, Pasadena, CA 91125}

\and

\author{D. Whysong}
\affil{National Radio Astronomy Observatory, P.O. Box O, Socorro, NM 87801}

\begin{abstract}

We present {\it Spitzer} MIR spectra of 25 FR--I radio galaxies  and
investigate the nature of their MIR continuum emission.   MIR spectra
of star--forming galaxies and quiescent elliptical galaxies are used
to identify host galaxy contributions while radio/optical core data
are used to isolate the nuclear non--thermal emission.  Out of the 15
sources with detected optical compact cores, four sources are
dominated by emission related to the host galaxy.  Another four
sources show signs of warm, nuclear dust emission:  3C15, 3C84, 3C270,
and NGC\,6251. It is likley that these warm dust  sources result from
hidden AGN of optical spectral type 1.  The MIR spectra of seven
sources are dominated by synchrotron emission, with no significant
component of nuclear dust emission.  In parabolic SED fits of the
non--thermal cores FR--Is tend to have lower peak frequencies and
stronger curvature than blazars. This is roughly consistent with the
common picture in which the core emission in FR--Is is less strongly
beamed than in blazars.

\end{abstract}


\section{Introduction}

Large double radio sources come in two radio--morphological types. The
more powerful type is the ``classical double'' or FR--II source, whose
prototype is Cygnus A. This group has strong terminal ``hotspots'',
and generally one--sided arcsecond (kpc) scale. The other type, FR--Is 
like Centaurus A and M\,87, is
lower in radio luminosity, double--jetted, and edge--darkened
\citep{fan74}. These sources  mostly have symmetric kpc jets.

It has been known since the 1960s that some of these giant
($\gtrsim$\,100\,kpc) radio sources show powerful optical/UV continuum
emission (called ``Big Blue Bump'', or BBB) from a central point
source. This optical/UV emission is generally identified as thermal
radiation from dense matter accreting onto supermassive black
holes. The gravitational potential energy of this infalling matter is
thought to power much or all of the observed activity. Objects showing
this strong optical/UV continuum and accompanying broad emission lines
are classified as quasars and broad line radio galaxies (hereafter,
``quasars''). On the other hand, double radio sources without these
components are called (Narrow Line) Radio Galaxies (NLRGs).

Many FR--II NLRGs have hidden quasars detected by spectropolarimetry
\citep[e.g.][]{ant84,ogl97,coh99} and recently also by their strong
hot nuclear dust emission in the mid infrared
\citep[e.g.][]{mei01,haas04,shi05,ogl06,cle07,tad07}.   Thus, many
powerful radio galaxies contain quasars which are hidden from direct
view by dusty equatorial tori -- i.e. as expected in the ``Unified
Model'' \citep{bar89,ant93,urr95}.

The total light optical spectra of many FR--II NLRGs already  indicate
the presence of a ``classical'' AGN (those thought to accrete
radiatively efficiently) by their  high ionization emission line
ratios (high ionization galaxies -- HIG\footnote{Although such objects
are usually referred to as high excitation galaxies (HEGs) the term
HIGs seems more appropriate because the distinction is primarily one
of  ionization and not excitation, as expected for photoionized
gases.}).  Some FR--II RGs and most FR--Is, however, show only weak
line emission of low ionization and are therefore classified as low
ionization galaxies (LIGs).  Here a luminous AGN (if present) has to
be discovered at other wavelengths.  Alternatively, these radio
galaxies may simply lack a hidden quasar. A lack of a BBB means,
according to current wisdom, a lack of a copious and  radiatively
efficient accretion flow. Historically such a situation was thought to
require energy extraction from black hole rotation, and various
mechanisms have been suggested for this \citep[e.g.][]{beg84}.
Alternatively in principle accretion power could make the jet, yet not
radiate significantly. In such a case the BBB is present, but of low
power compared to a (lobe--matched) quasar BBB.

In an optical imaging study of 3CR FR--I sources  using {\it Hubble
Space Telescope}, \citet{chia99} report unresolved central compact
cores (CCCs)  in the centers of most observed galaxies. The
correlation of optical (at $\sim$\,7000\,\AA) compact cores and radio
(at 6\,cm) cores in flux and luminosity is taken as evidence for their
common origin as synchrotron emission from the base of a jet.  
Assuming that the base of the jet is found on scales {\it smaller}
than any  obscuring torus and considering the high detection rate of
CCCs (85\,\%) they argue that  we see directly into the nucleus and
thus no standard, geometrically thick torus can be present in most
low--luminosity (FR--I) radio galaxies. Since no broad lines or strong
BBB emission is observed there could also be no
hidden, radiatively efficient AGN in those cases. However,
e.g. \citet{cao04} note that if the emission from the jet occurs on
scales larger than that of the torus the absence (or presence) of a
torus cannot be directly inferred from the observed nuclear jet
components.  In addition, high--resolution CO maps have revealed
massive nuclear tori like those invoked in the standard unified model
in the center of at least two FR--I radio galaxies: 3C31 \citep{oku05}
and NGC\,3801 \citep{das05}.  These tori have high molecular masses
($>$$10^8$\,M$_{\sun}$), with high molecular column densities and
probably geometrically thick shapes.

From an UV imaging study \citet{zir03} argue that the properties of
FR--Is are consistent with the classic unification models and the
existence of an obscuring torus: While BL Lac objects are well aligned
sources, FR--I sources with nuclear UV components are at a critical
angle for which a torus hides the nucleus but not the jet (or the
latter greatly outshines the former). FR--Is without nuclear UV
components are thought to have a larger viewing angle where a torus
hides the central regions completely.

Recently, some of the compact optical cores discovered by
\citet{chia99} have been studied with imaging polarimetry revealing
fairly large polarization ($<$\,11\%) which further indicates their
non--thermal origin, though such high  polarizations can also result
from scattering \citep{cap07}.

In the X--rays, FR--I sources are dominated by weakly absorbed
non--thermal jet components in most cases
\citep[e.g.][]{bal06b,eva06}. They lack the powerful absorbed,
accretion--related component of a strong AGN (associated with a
luminous accretion disk and circumnuclear obscuring structure). In
many  FR--IIs the latter is found {\it in addition} to the
non--thermal jet component \citep{eva06}.  Thus, in FR--Is any
non--jet accretion--related X--ray component  carries much lower
intrinsic AGN luminosities (10$^{39}$--10$^{41}$\,erg\,s$^{-1}$) or
else their detection is hampered  by column densities of well above
$10^{24}\,{\rm cm}^{-2}$. Also \citet{wu07} argue that the (soft)
X--rays in their FR--I  sample can be accounted for by jet emission
only. We must note however that FR--Is with broad emission lines  in
the optical do exist (e.g. 3C120, \citealt{tad93}; see also
\citealt{ant02a}  for an anecdotal listing of such objects). Such
broad--line FR-Is can extend up to quasar  luminosities
\citep{gow84,blun01,hey07}.

We here present a study of the mid--infrared properties of a sample of
FR--I radio galaxies. The motivation is to possibly detect a hidden
AGN by the reprocessed warm thermal dust emission as found in many
FR--II radio galaxies \citep[e.g.][]{ogl06}. And even in the case of
exceptionally high extinction which attenuates also the MIR
luminosity, spectral features that give clues on the thermal or
non-thermal origin of the emission might still be detectable. On the
other hand, the nuclear part of the MIR spectra bridges the gap in
wavelength between the  optical compact cores  and the radio cores
observed by \citet{chia99}.  If there is in fact no hidden AGN and no
obscuring structure to be found in FR--Is, the MIR spectra should
agree with synchrotron core estimates.

\begin{table*}[t!]
\begin{center}
\caption{The Sample.\label{tab1}}
\begin{tabular}{llccccccccc}
\tableline\tableline
Object      & alt.            &      z   & F$_{178\,{\rm MHz}}$       & total                & F$_{5\,{\rm GHz}}$ & core                 & F$_{\rm CCC}$\tablenotemark{b} & $\lambda_{\rm CCC}$    & map                  & spec \\ 
            & name            &          & Jy                         & ref\tablenotemark{a} & mJy                & ref\tablenotemark{a} & $\mu$Jy                        & filter                     & ref\tablenotemark{a} & ref\tablenotemark{a} \\  
\tableline
3C15        &                 & 0.073000 & 17.2                         &      42  &   32                         &       25 & 38        & F160W     &       22 &      48 \\	
3C29        &                 & 0.045031 & 16.5                         &      42  &   93                         &       24 & 10        & F702W     &       24 &      48 \\	
3C31        &  NGC\,383       & 0.017005 & 18.3                         &      43  &   92                         &       12 & 29        & F702W     &       19 &      48 \\
3C66B       &                 & 0.021258 & 26.8                         &      43  &  182                         &       12 & 97        & F702W     &       14 &      48 \\
3C76.1      &                 & 0.032489 & 13.3                         &      43  &   10                         &       37 & \nodata   & \nodata   &       21 &      48 \\
3C83.1      &  NGC\,1265      & 0.025137 & 29.0                         &      43  &   25                         &       27 & 3         & F702W     &       27 &      48 \\
3C129       &                 & 0.020800 & 51.1                         &      44  &   34                         &       38 & \nodata   & \nodata   &       36 &      48 \\
3C189       &  NGC\,2484      & 0.040828 &  7.4                         &      45  &  195                         &       13 & 76        & F814W     &        7 &       1 \\
3C264       &  NGC\,3862      & 0.021718 & 28.3                         &      43  &  200                         &       12 & 248       & F791W     &       41 &      28 \\
3C270       &  NGC\,4261      & 0.007465 & 56.5                         &      42  &  308                         &       24 & 11        & F791W     &        9 &      33 \\
3C272.1     &  M\,84          & 0.003536 & 21.1                         &      43  &  180                         &       12 & 139       & F814W     &       18 &      48 \\
IC\,4296    &                 & 0.012465 & 16.8                         &      46  &  297                         &       24 & 524       & F160W     &       24 &      40 \\
3C293       &                 & 0.045034 & 13.8                         &      43  &  100                         &       12 & \nodata   & \nodata   &        5 &      34 \\
3C317       &                 & 0.034457 & 53.4                         &      42  &  391                         &       24 & 23        & F814W     &        6 &      48 \\
NGC\,6251   &                 & 0.024710 & 11.6                         &      43  &  350                         &       16 & 191       & F812W     &       30 &      23 \\
3C386       &                 & 0.016885 & 26.1                         &      43  &   14                         &       32 & 2301      & F702W     &       21 &      48 \\
3C403.1     &                 & 0.055400 & 14.7                         &      42  & \nodata                      & \nodata  & \nodata   & \nodata   & \nodata	 & \nodata \\
3C424       &                 & 0.126988 & 15.9                         &      42  &   18                         &        8 & \nodata   & \nodata   &        3 &      48 \\ 
3C465       &  NGC\,7720      & 0.030221 & 41.2                         &      43  &  270                         &       12 & 36        & F702W     &       10 &      11 \\\hline
3C84        & NGC\,1275       & 0.017559 & 66.8                         &      43  & 3100                         &       26 & 3489      & F702W     &       29 &      48 \\ 
3C120       &                 & 0.033010 & 7.4                          &      45  & 1970                         &       39 & \nodata   & \nodata   &       31 &      33 \\ 
3C218       & Hyd\,A          & 0.054878 & 225.7                        &      46  &  217                         &       24 & \nodata   & \nodata   &       35 &      40 \\ 
3C274       & M\,87           & 0.004360 & 1144.5                       &      43  & 4000                         &       12 & 894       & F814W     &       15 &      48 \\ 
BL\,Lac     &                 & 0.068600 & $\sim$\,2.5\tablenotemark{c} &      47  & $\sim$\,4.0\tablenotemark{c} & \nodata  & \nodata   & \nodata   &        2 &      20 \\
E1821+643   &                 & 0.297000 & $\sim$\,0.5                  &      47  &    8.0                       &        4 & \nodata   & \nodata   &        4 &      17 \\
\tableline
\end{tabular}
\vspace*{-5mm}
\tablenotetext{a}{References for total radio fluxes, radio core fluxes, radio maps, and optical spectra, respectively.}
\tablenotetext{b}{Data taken from \citet{chia99} except for 3C189 \citep{cap02}, IC4296 \citep{bal06a}, and NGC\,6251 \citep{chia03}. Data on 3C15 provided by R. Baldi (private communication). }
\tablenotetext{c}{The flux of {\it only} the extended emission is 29\,mJy at 20\,cm \citep{ant86} which correspond to $\sim$\,$151$\,mJy at 178\,MHz 
                  (using $\alpha = -0.8$; $S\propto\nu^{\alpha}$). For the variable 5\,GHz core flux we here give the average of all measurements available 
		  from NED.}
\end{center}
\tablerefs{
  (1) SDSS;
  (2) \citealt{ant86};
  (3) \citealt{bla92};
  (4) \citealt{blun01};	 
  (5) \citealt{bri81};	 
  (6) \citealt{burns90}; 
  (7) \citealt{cap93};	 
  (8) \citealt{chia99};	 
  (9) \citealt{con88};	 
 (10) \citealt{con91};	 
 (11) \citealt{dero90};	 
 (12) \citealt{gio88};	 
 (13) \citealt{gio05};	 
 (14) \citealt{har96};	 
 (15) \citealt{hin89};	 
 (16) \citealt{jon86};	 
 (17) \citealt{kol06};	 
 (18) \citealt{lai87};
 (19) \citealt{lai08};  
 (20) \citealt{law96};
 (21) \citealt{lea91};  
 (22) \citealt{lea97};	 
 (23) \citealt{mil79};	 
 (24) \citealt{mor93};	 
 (25) \citealt{mor99};	 
 (26) \citealt{tay06};	 
 (27) \citealt{odea86};	 
 (28) \citealt{owen96};	 
 (29) \citealt{ped90};	 
 (30) \citealt{per84};	 
 (31) \citealt{simp96};	 
 (32) \citealt{str78};	 
 (33) \citealt{tad93};
 (34) \citealt{tad05};	  
 (35) \citealt{tay90};	 
 (36) \citealt{tay01};
 (37) \citealt{val82};	   
 (38) \citealt{vbreu82}; 
 (39) \citealt{wal87};	 
 (40) \citealt{wil04};
 (41) see http://www.jb.man.ac.uk/atlas/; 
 (42) \citet{kel69}, on \citet{baa77} scale using \citet{lai80};
 (43) taken from http://3crr.extragalactic.info/;
 (44) estimated using GMRT measurements \citep{lal04}; 
 (45) \citealt{kue81};
 (46) estimated using the 160\,MHz flux \citep{kue81} and $\alpha = -0.8$;
 (47) estimated from low frequency measurements in NED; 
 (48) \citealt{but09}}

\end{table*}

\section{Sample selection and observations}

Our sample of FR--I radio galaxies was selected from the 3CR
\citep{spi85} catalog (about half of the sources are also included in
the 3CRR\footnote{http://3crr.extragalactic.info/}).  We selected a
tractable subset of lobe dominated sources with a flux limit of
$\sim$\,15\,Jy at 178\,MHz. To increase the sample size we added some
very similar sources which were close to, but technically below, the
flux cut off. We also observed  two additional sources because they
were of special interest:  3C189\footnote{3C189 is only part of the 3C
catalog \citep{edge59} but not of the 3CR or the 3CRR catalogues. See
e.g. \citet{lai83} for details on these samples.} as well as
IC\,4296. See Tab.\,\ref{tab1} for some basic data of the  selected
sources. We note that our sample is not complete but can be considered
representative for lobe dominated FR--I sources.  No sources were lost
by imposing scientifically relevant selections.

The morphological classification of the sources according to the
Fanaroff--Riley scheme \citep{fan74} was checked using radio maps from
the literature (see Tab.\,1 for references).

The objects were observed in low--resolution mode with IRS
\citep{hou04} on board  {\it Spitzer Space Telescope} \citep{wer04},
mostly under pid--20525. Each source under this pid was observed in
staring mode for a total of 960 sec, 1440 sec, 720 sec, and 480 sec in
SL2, SL1, LL2,  and LL1, respectively. After averaging all the cycles
at one nod position we subtracted the off--order frames to remove the
background. The resulting images were cleaned of residual rogue pixels
and cosmic rays using {\sc Irsclean}. The {\sc Spice} software  was
used to calibrate and extract one dimensional spectra. The extracted
spectra from the two nod positions were then averaged and data from
the different modules were  combined into a single spectrum. Due to
the larger slit width in LL (10.7\arcsec~in LL versus 3.7\arcsec~in
SL),  multiplicative factors ($<$\,1.5) were applied to the SL modules
where necessary in order to  match the flux levels.

Spectra for additional FR--I sources not observed under pid--20525
(see Tab.\,\ref{tab1}, lower part) were also retrieved from the archive and were
subject to the same procedures.

\section{The Spitzer spectra}
In Figures\,\ref{obs_spectra} and \ref{add_spec} we present the rest
frame MIR spectra for our FR--I core sample and the supplemental
objects. Because for 3C15 and 3C29 the SL modules of IRS suffer from 
saturated peak--up areas we present their spectra separately in
Fig.\,\ref{add_corr_spectra}.  
Statements on the spectral behavior of the individual sources refer to
the figures where their observed spectra are shown
(Figs.\,\ref{obs_spectra} and \ref{add_spec}) and are thus considered
in terms of F$_{\nu}$. All sources in our sample have been detected 
except for 3C403.1 which was not detected in either SL or LL.

\begin{figure*}[t!]
\includegraphics[angle=0,scale=.44]{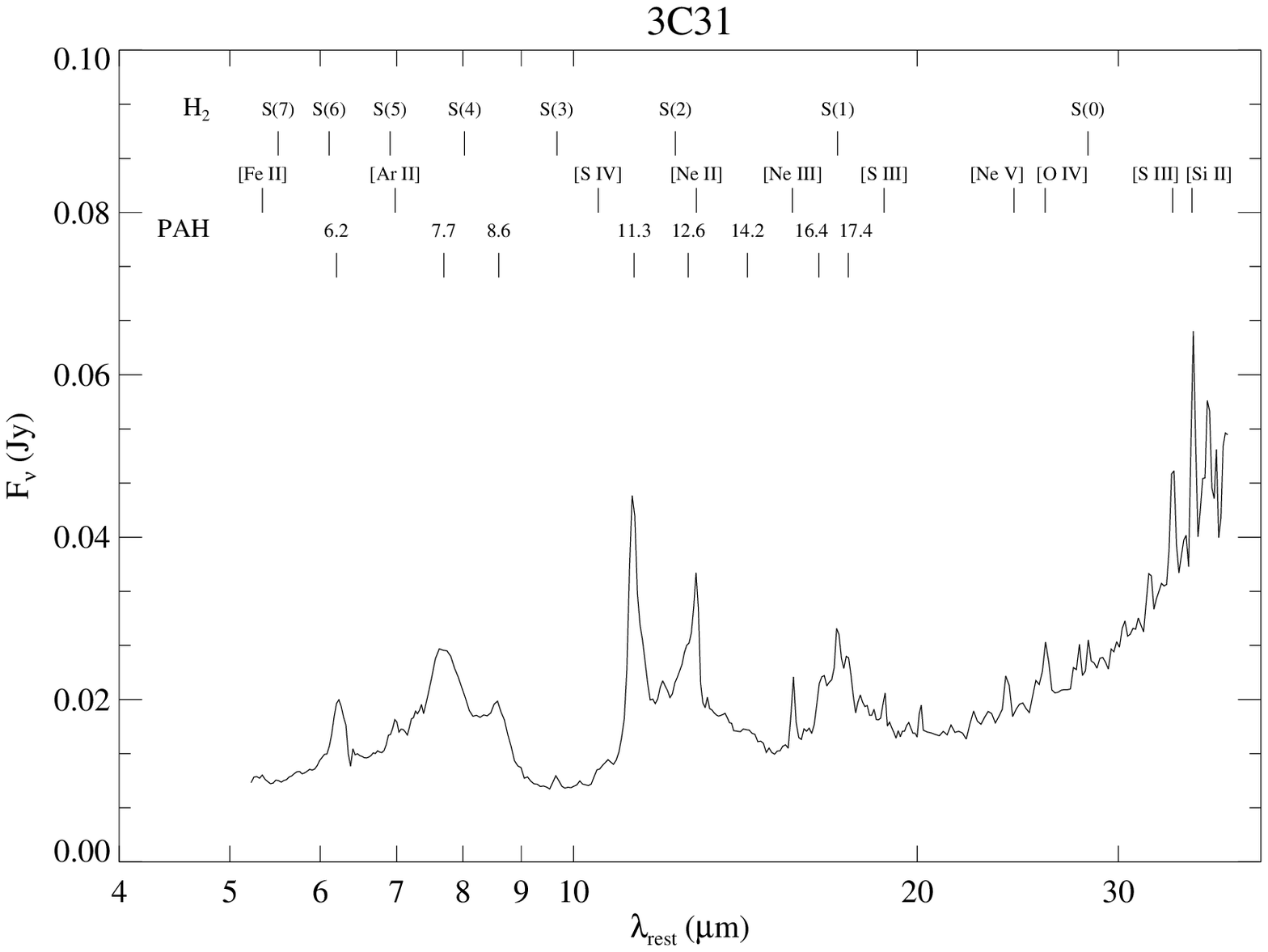}
\includegraphics[angle=0,scale=.44]{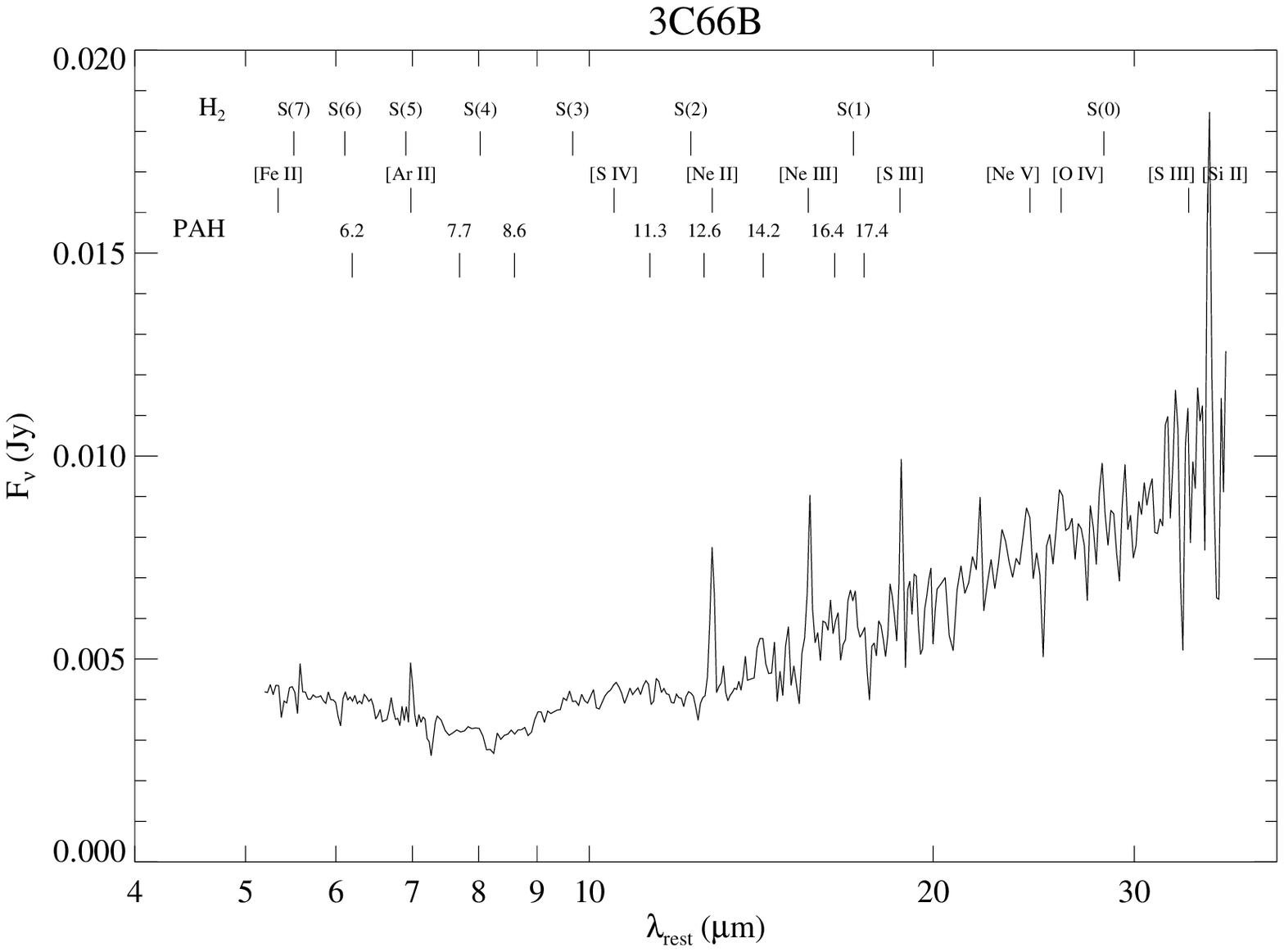}\\
\includegraphics[angle=0,scale=.44]{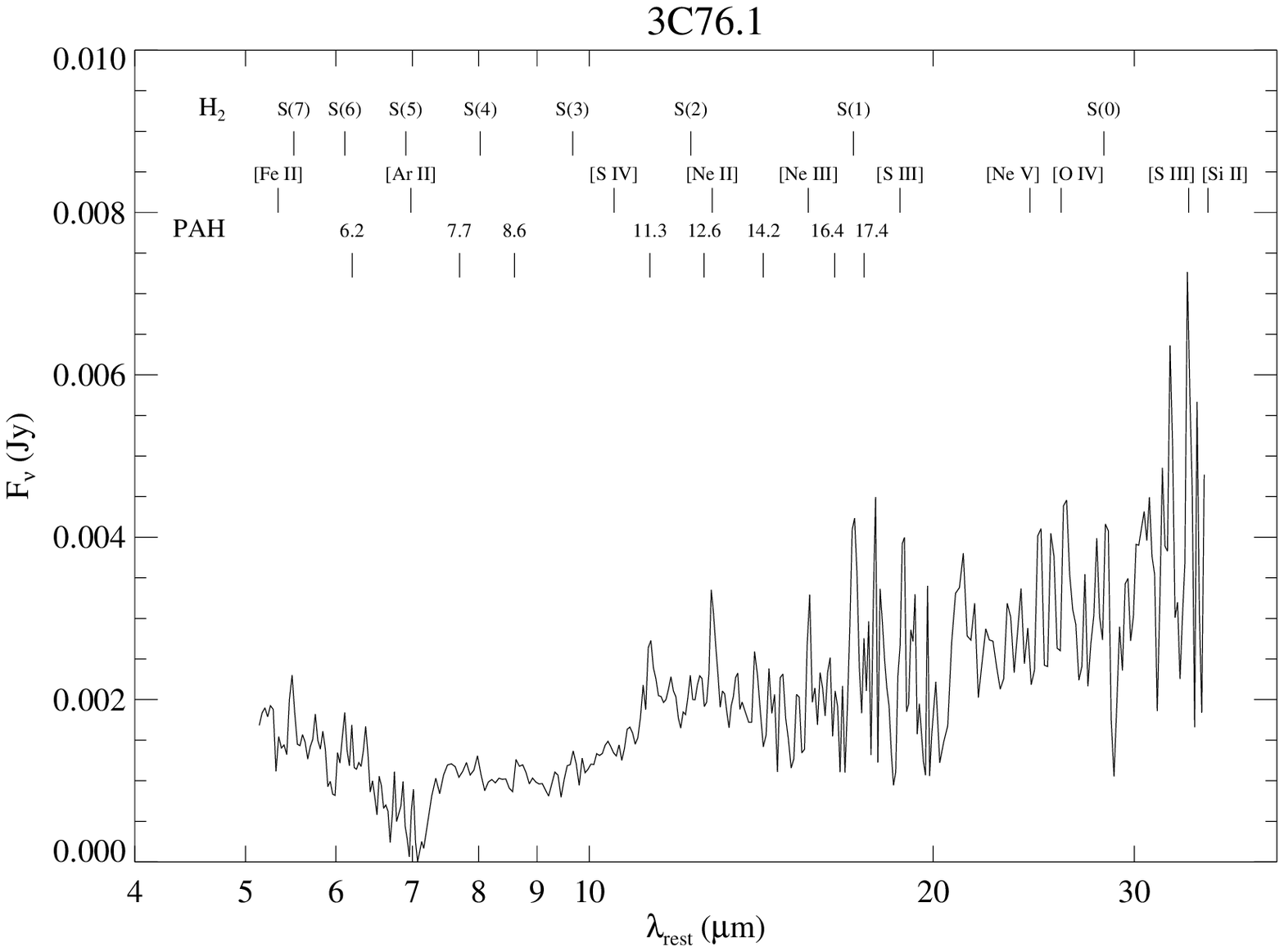}
\includegraphics[angle=0,scale=.44]{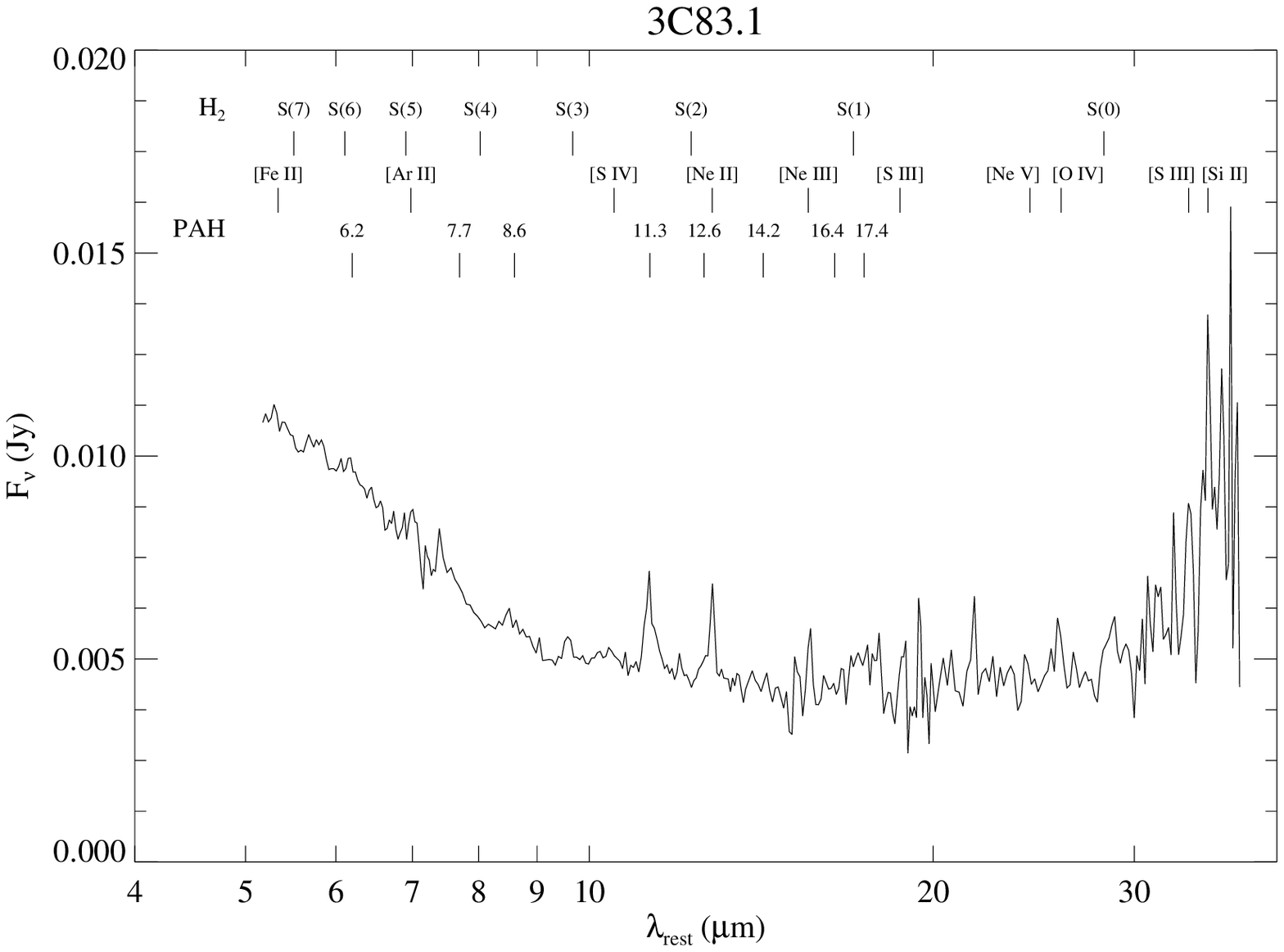}\\
\includegraphics[angle=0,scale=.44]{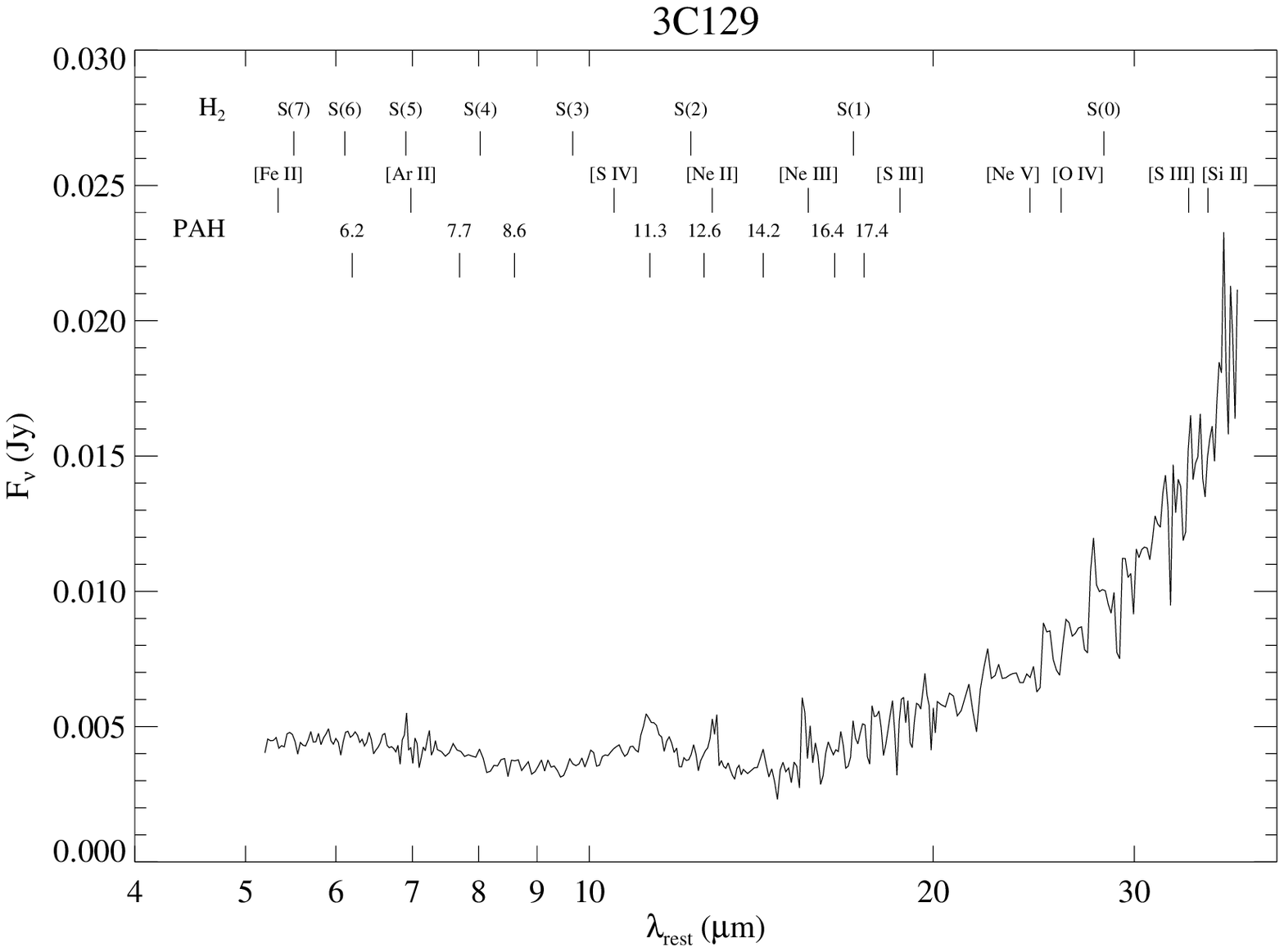}
\includegraphics[angle=0,scale=.44]{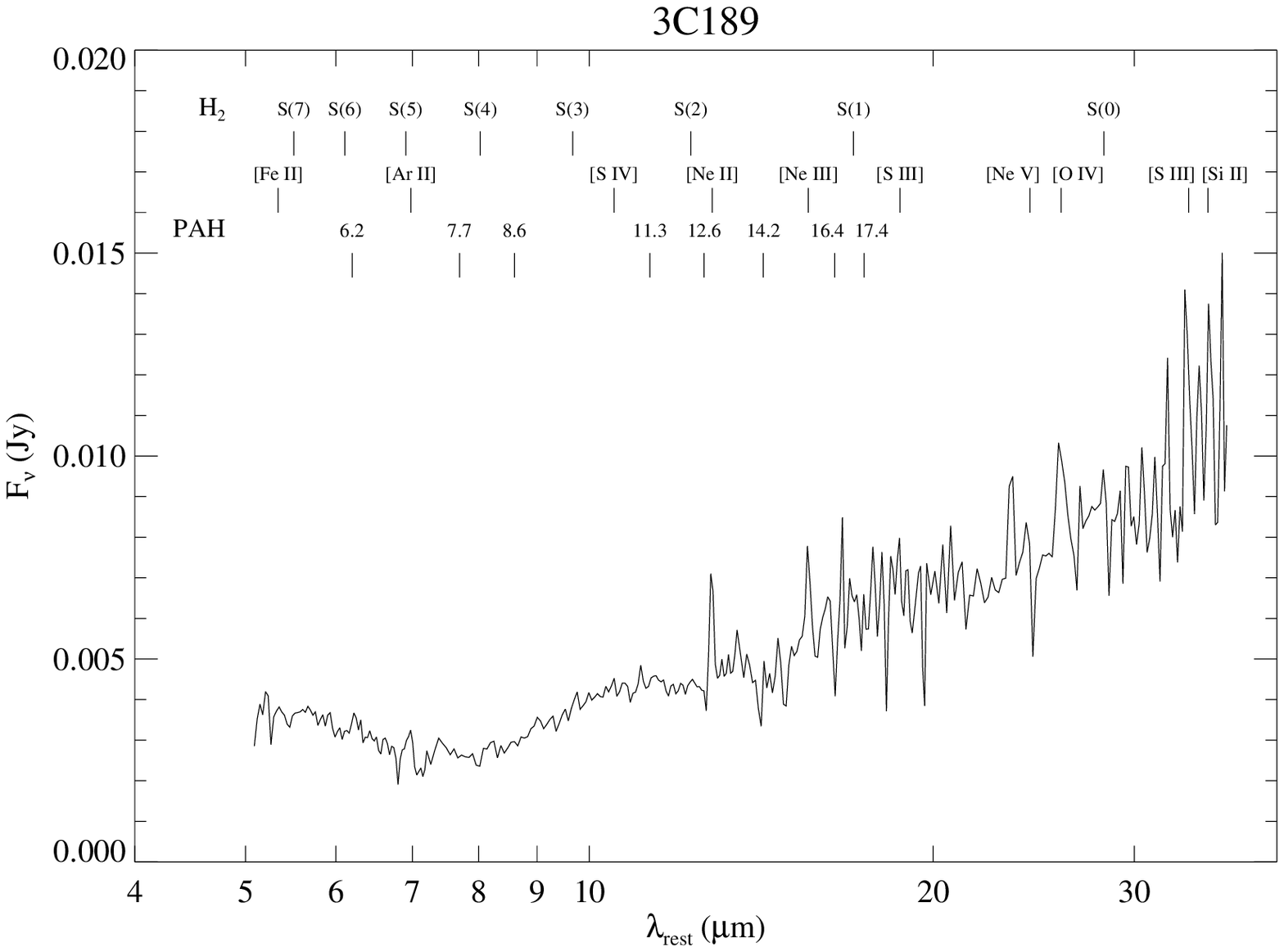}
\caption{Observed MIR spectra of the FR--I sources. Flux in Jy is plotted over
  the rest frame wavelength in $\mu$m.\label{obs_spectra} }
\end{figure*}
\addtocounter{figure}{-1}
\begin{figure*}[t!]
\includegraphics[angle=0,scale=.44]{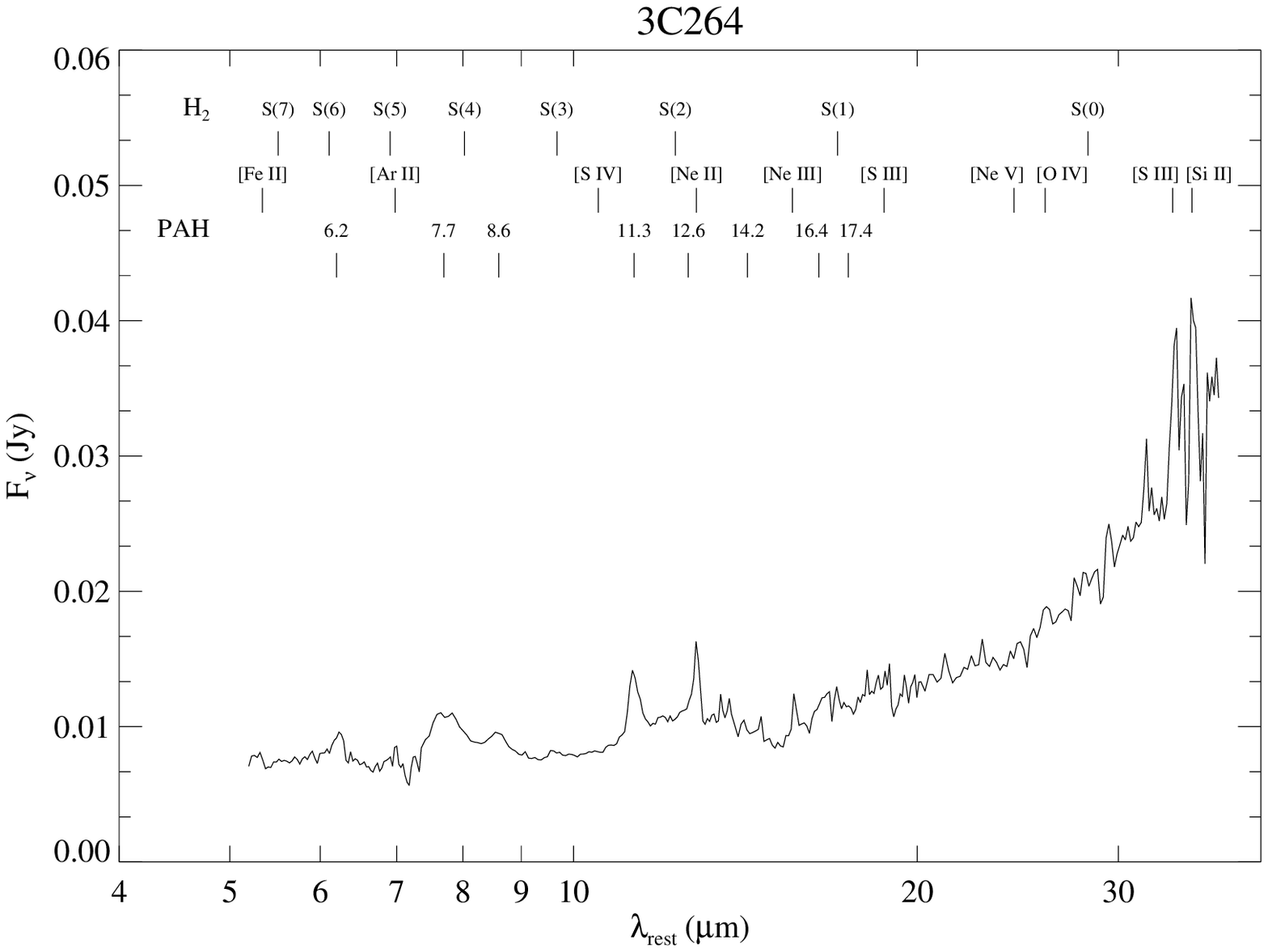}
\includegraphics[angle=0,scale=.44]{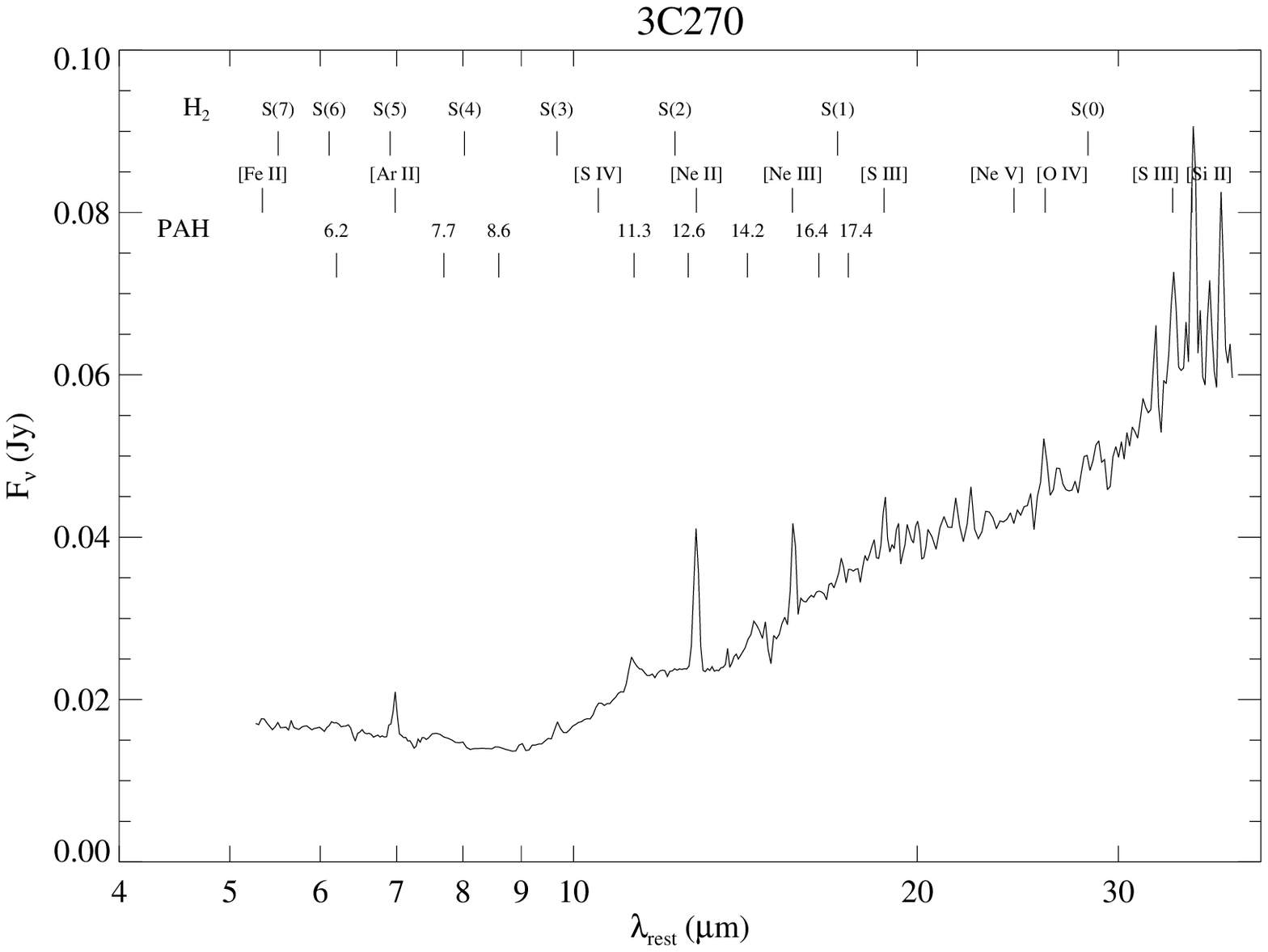}\\
\includegraphics[angle=0,scale=.44]{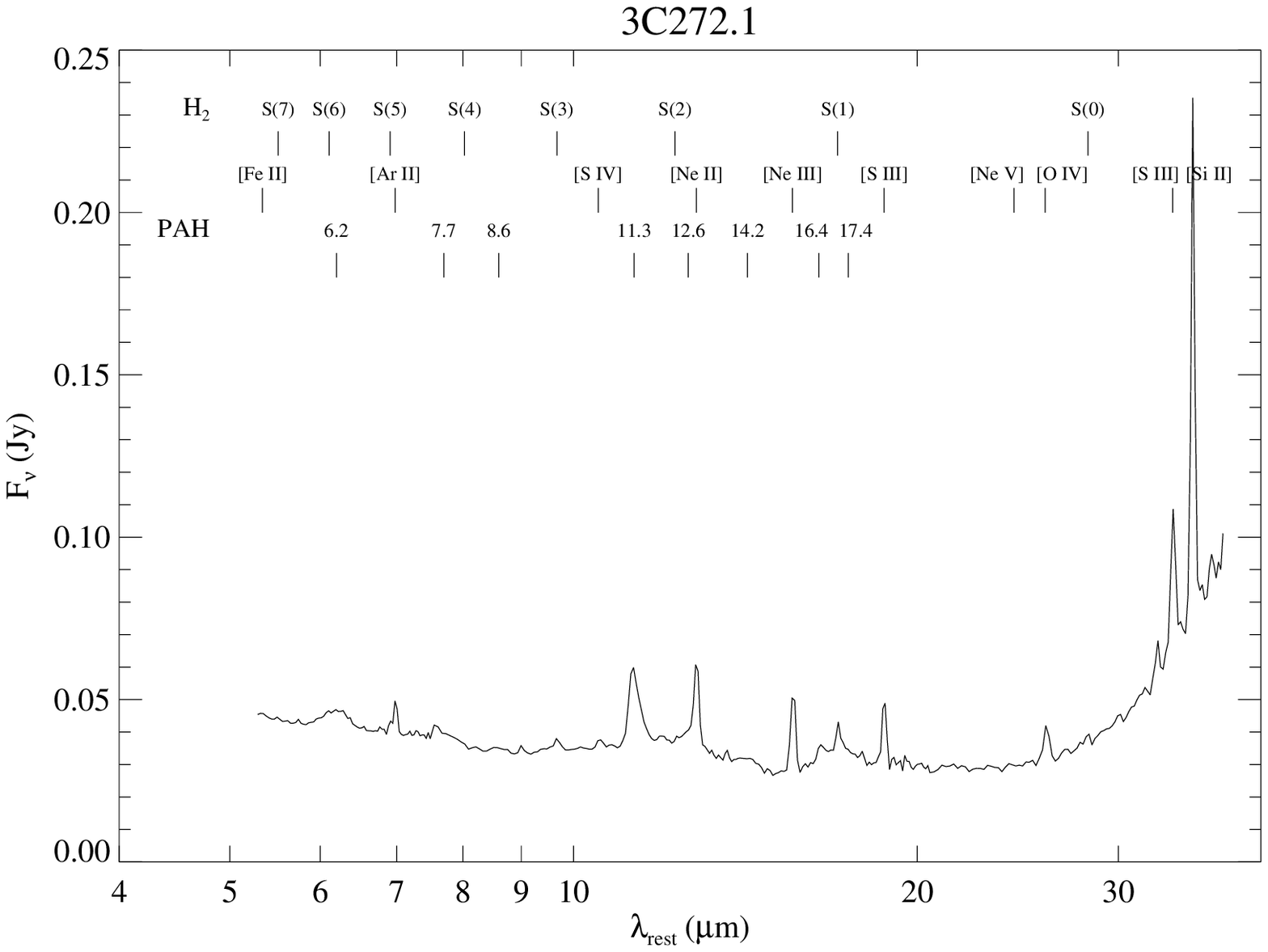}
\includegraphics[angle=0,scale=.44]{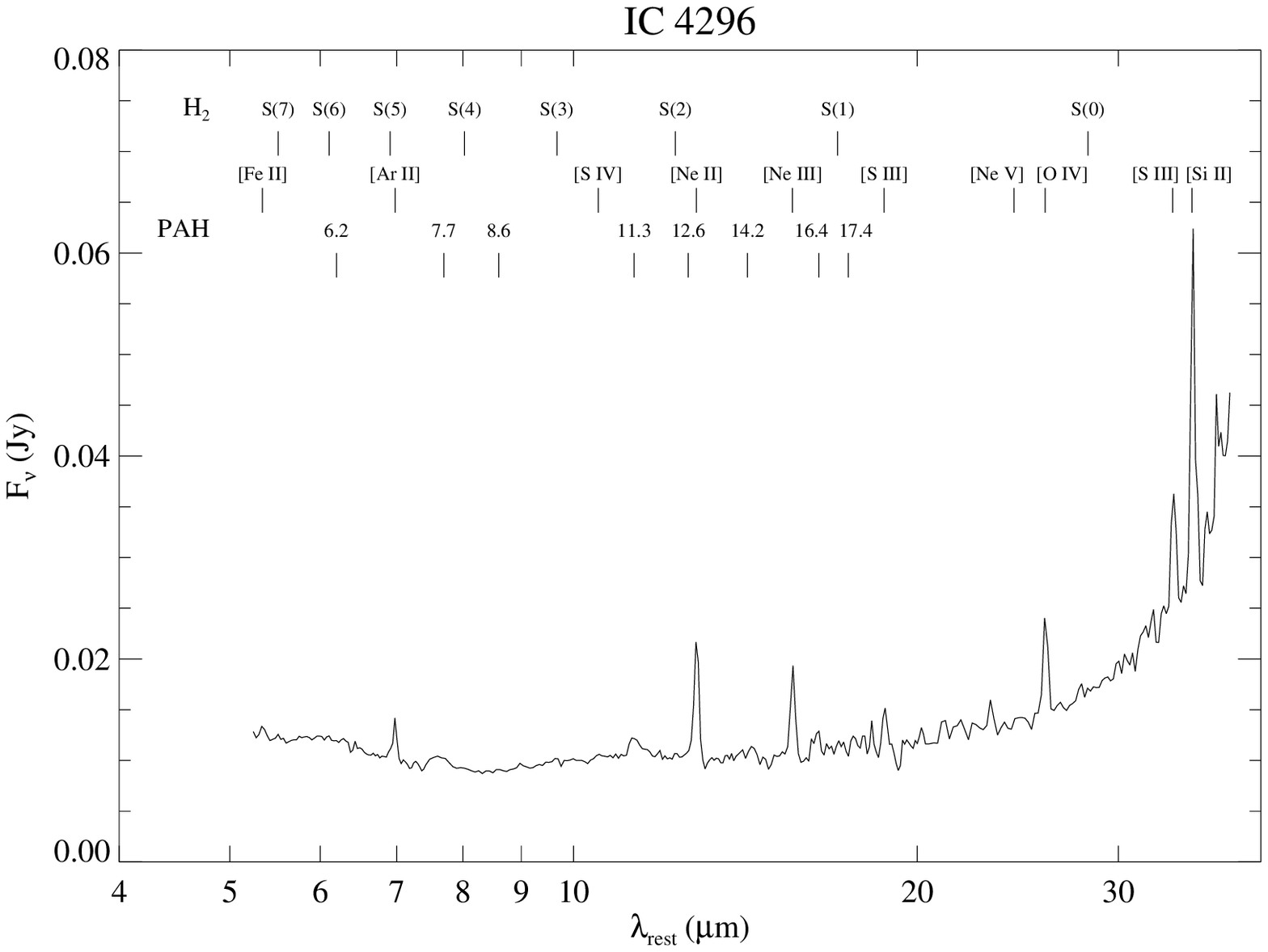}\\
\includegraphics[angle=0,scale=.44]{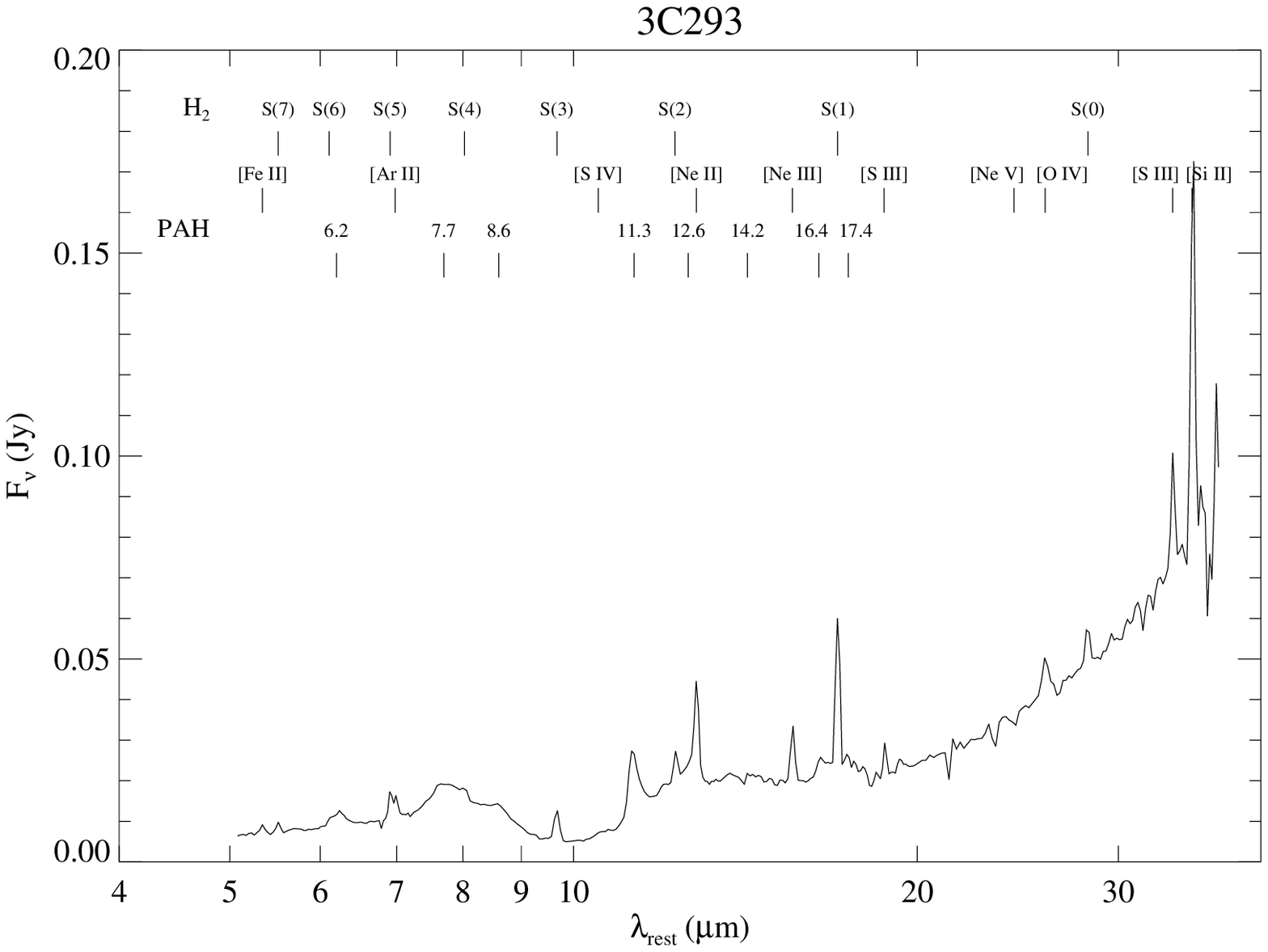}
\includegraphics[angle=0,scale=.44]{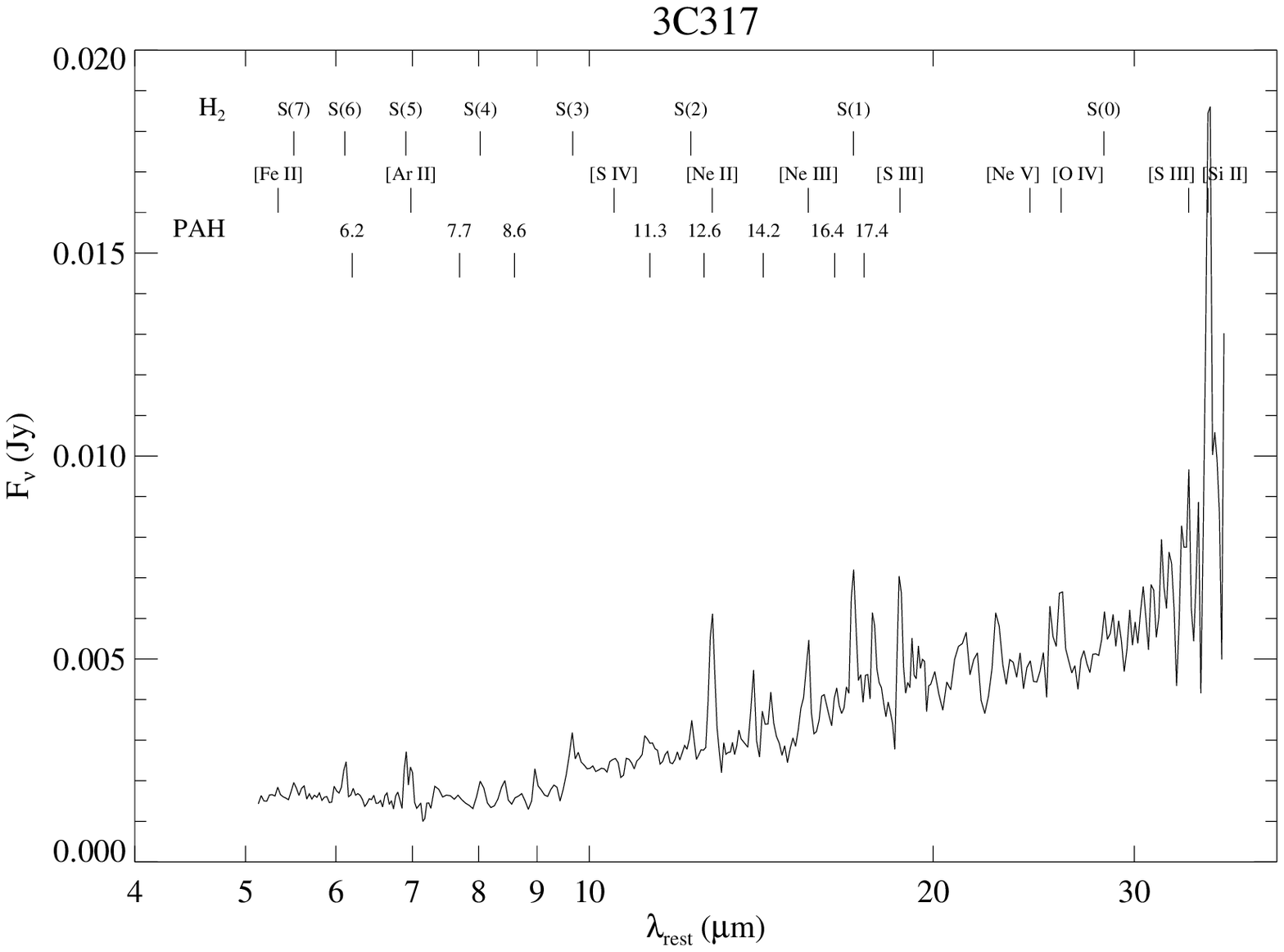}
\caption{{\it continued}}
\end{figure*}
\addtocounter{figure}{-1}
\begin{figure*}[t!]
\includegraphics[angle=0,scale=.44]{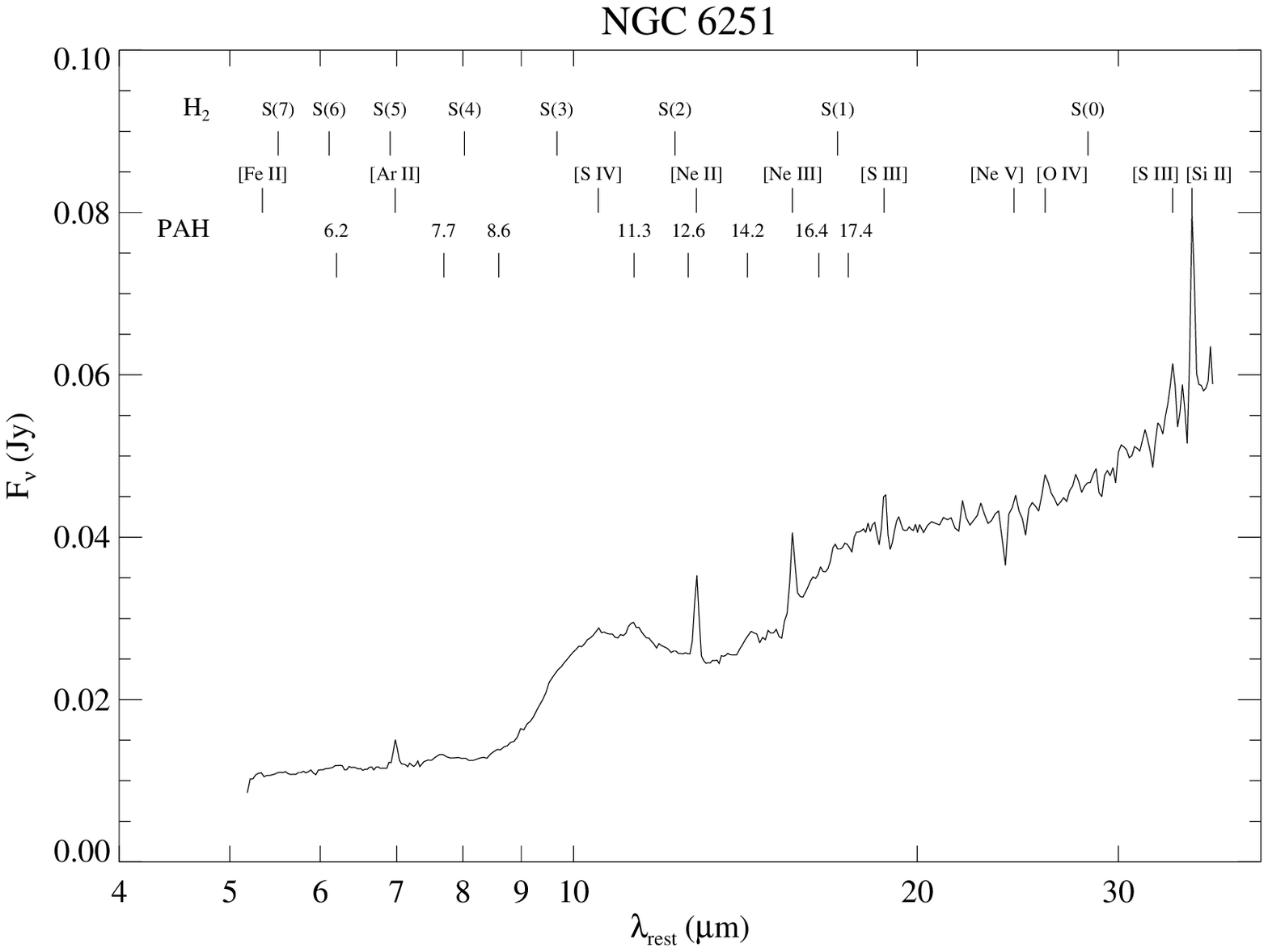}
\includegraphics[angle=0,scale=.44]{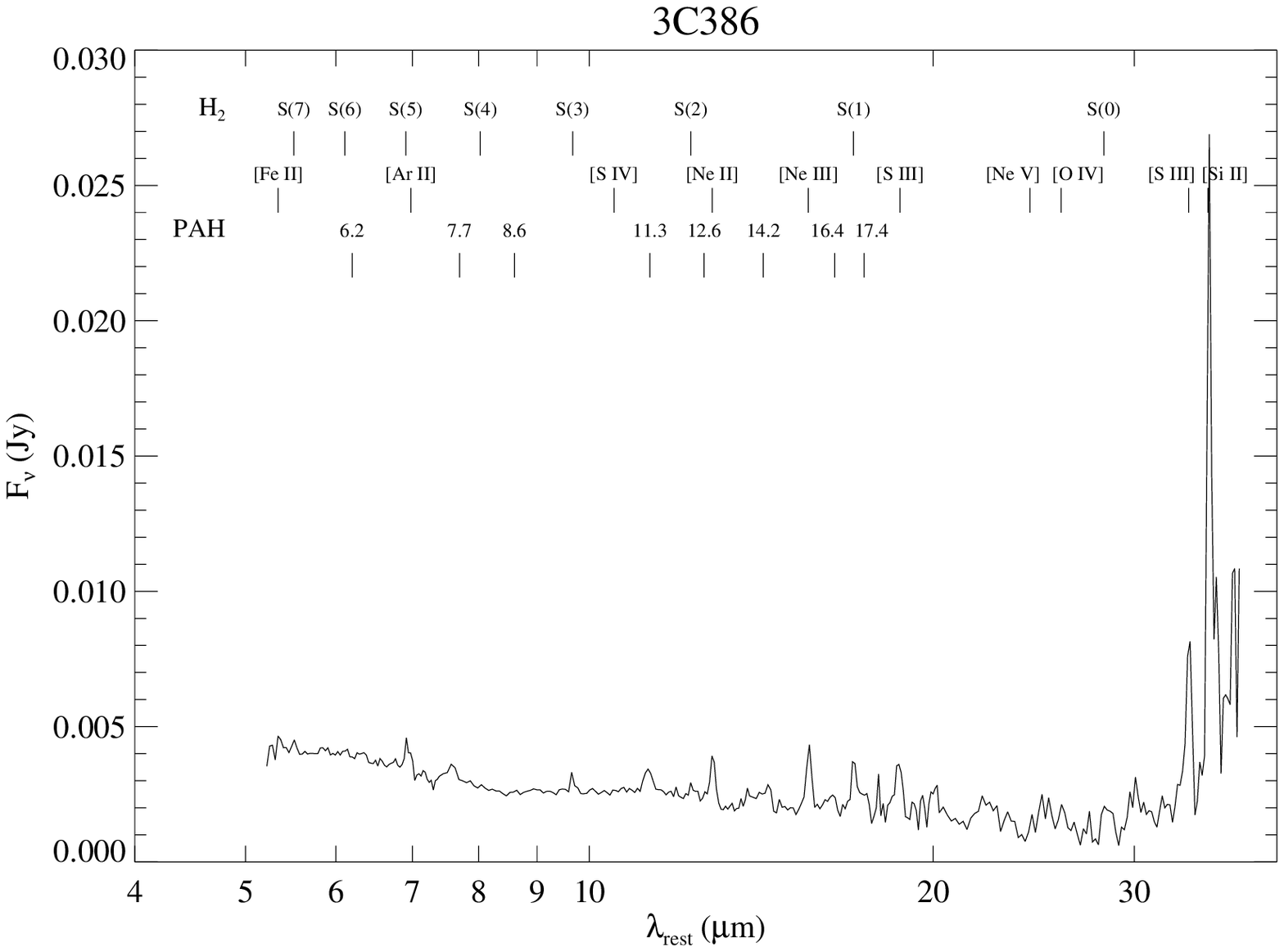}\\
\includegraphics[angle=0,scale=.44]{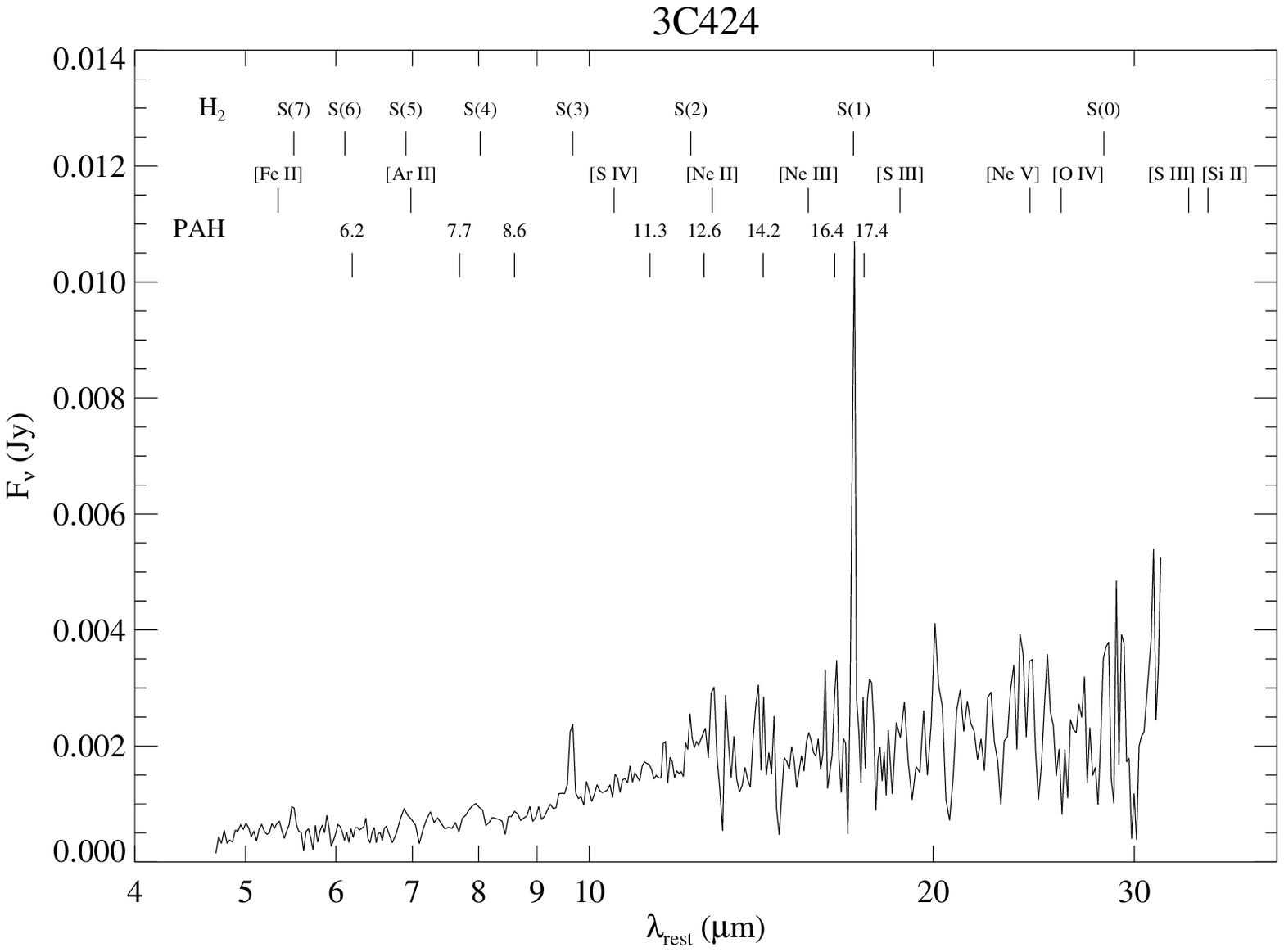}
\includegraphics[angle=0,scale=.44]{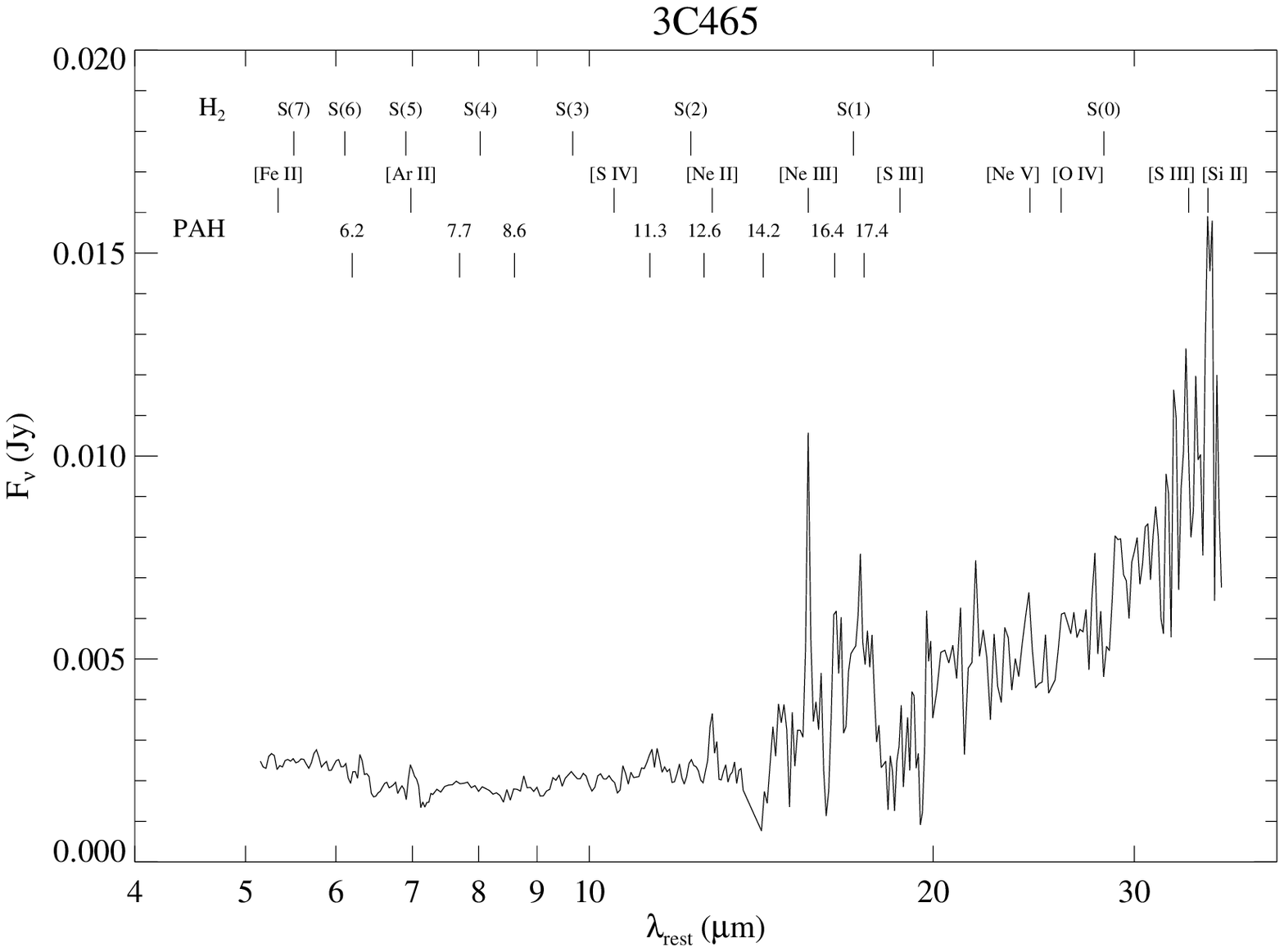}
\caption{{\it continued}}
\end{figure*}

\begin{figure*}[t!]
\includegraphics[angle=0,scale=.44]{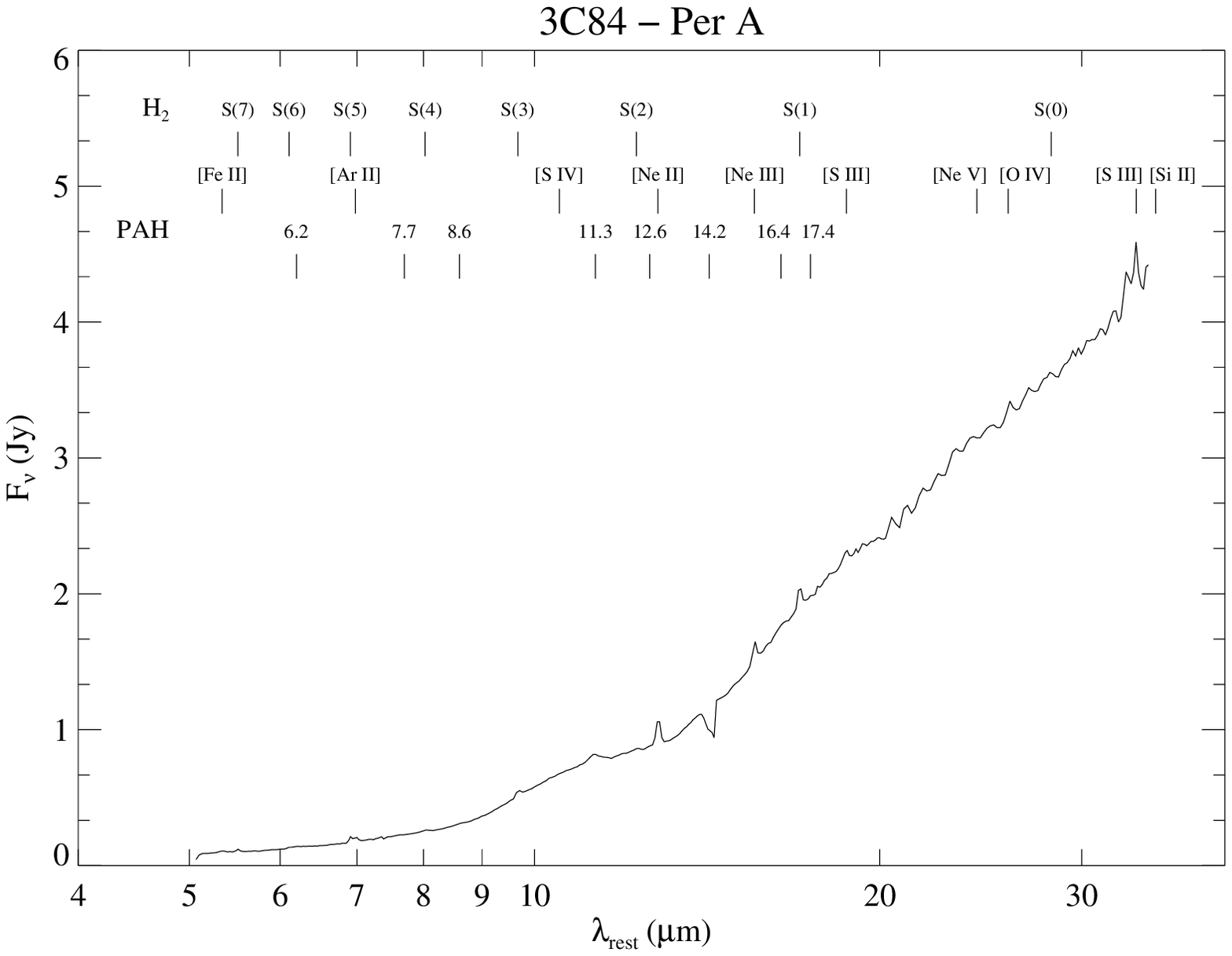}
\includegraphics[angle=0,scale=.44]{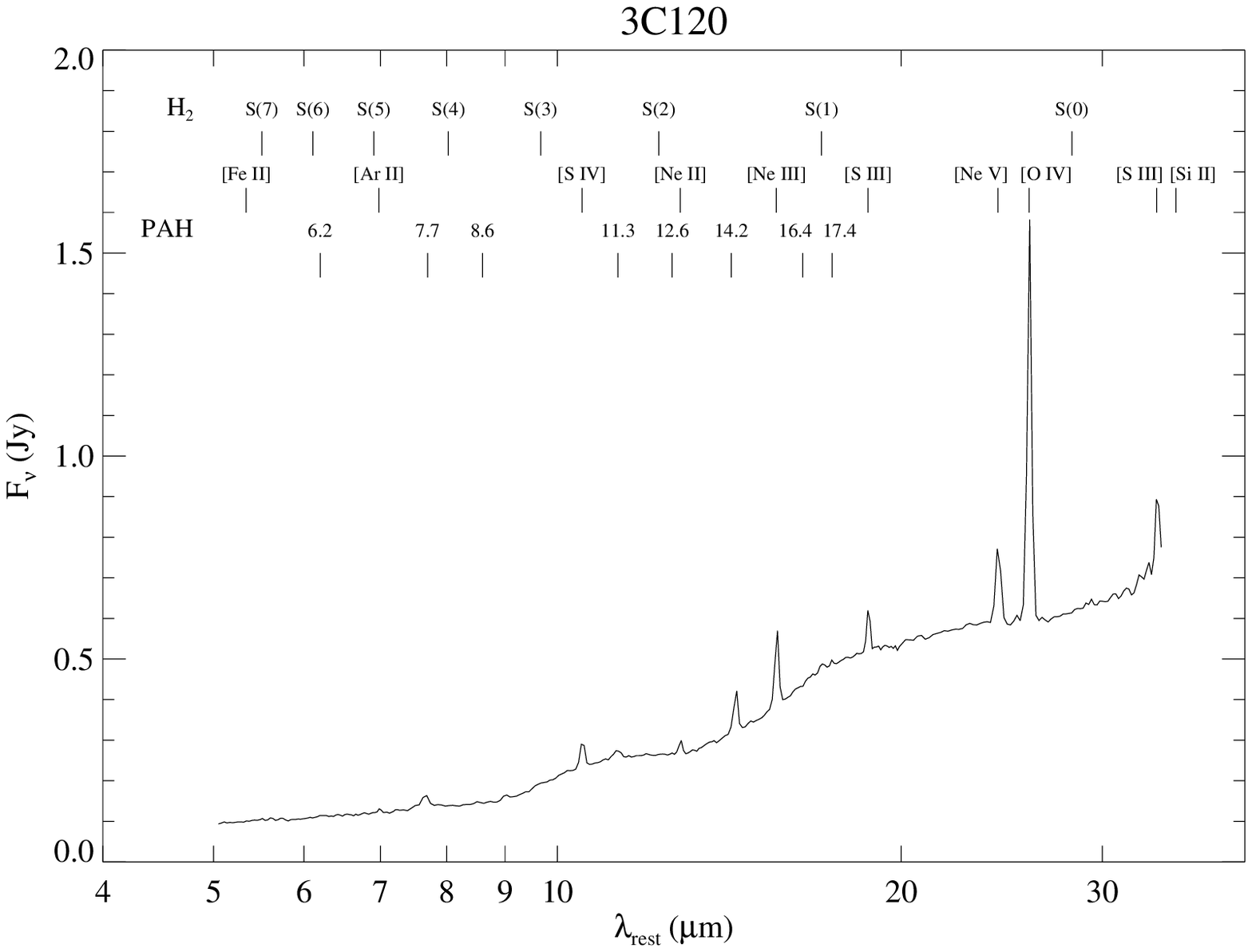}\\
\includegraphics[angle=0,scale=.44]{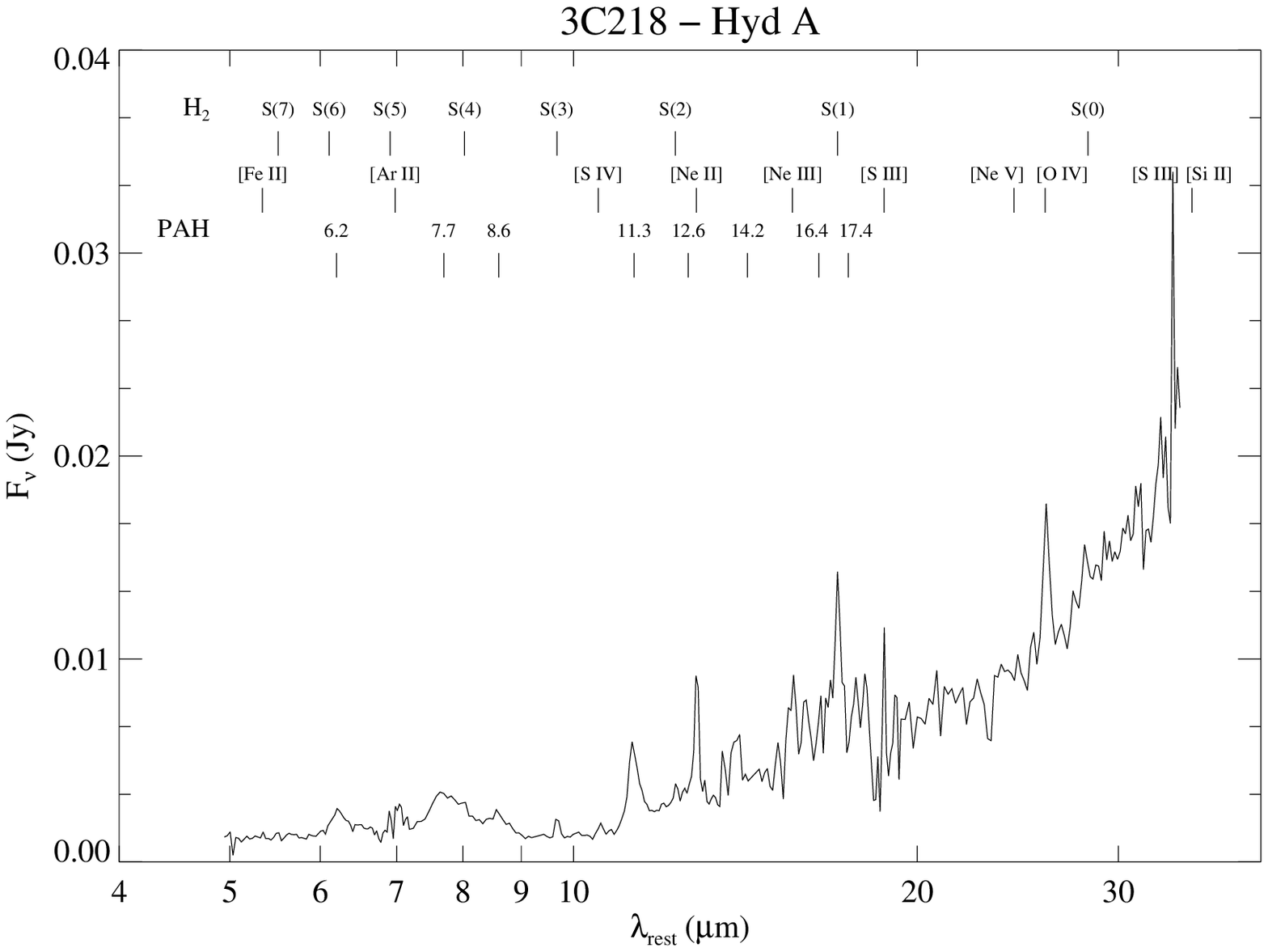}
\includegraphics[angle=0,scale=.44]{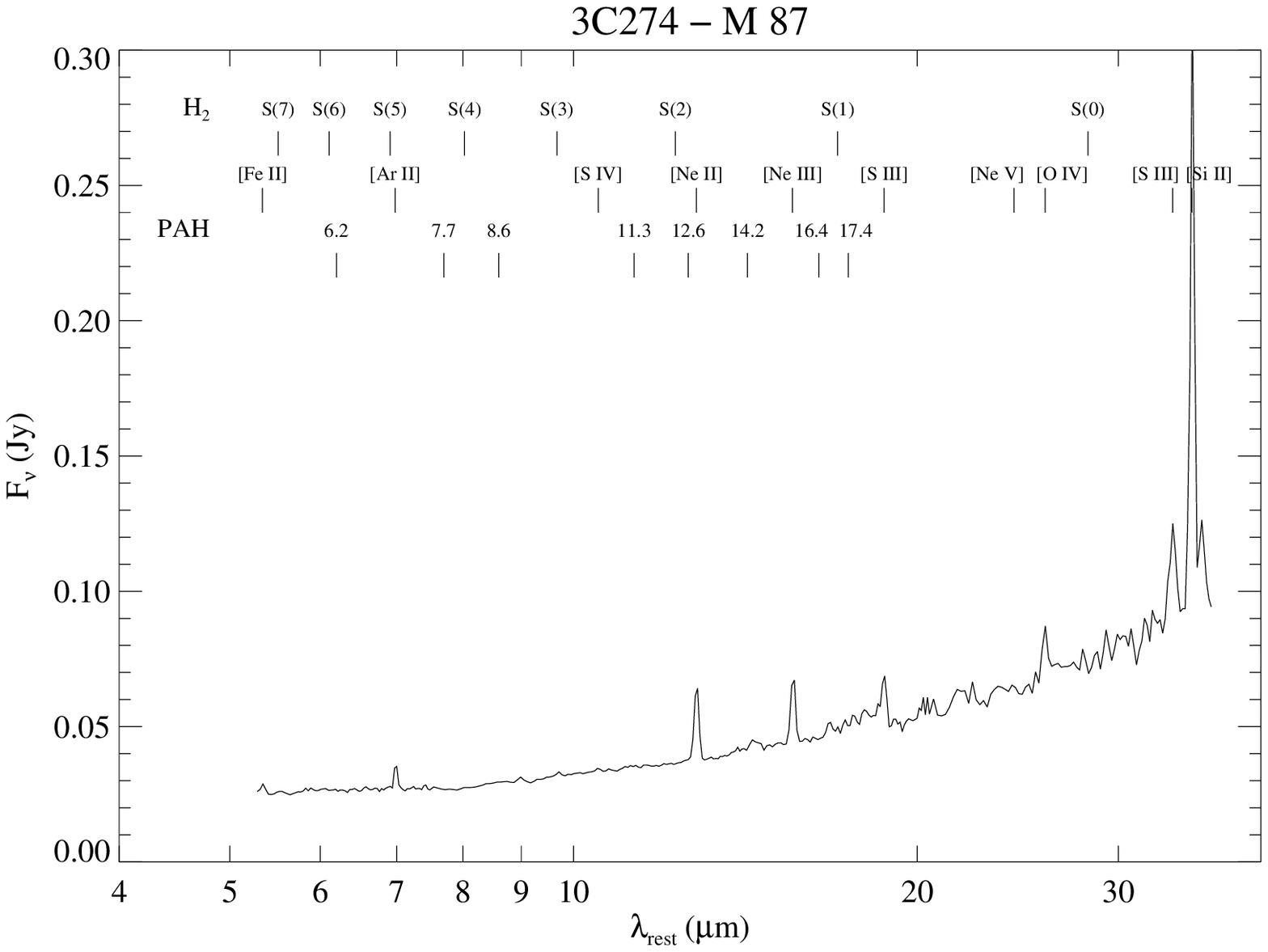}\\
\includegraphics[angle=0,scale=.44]{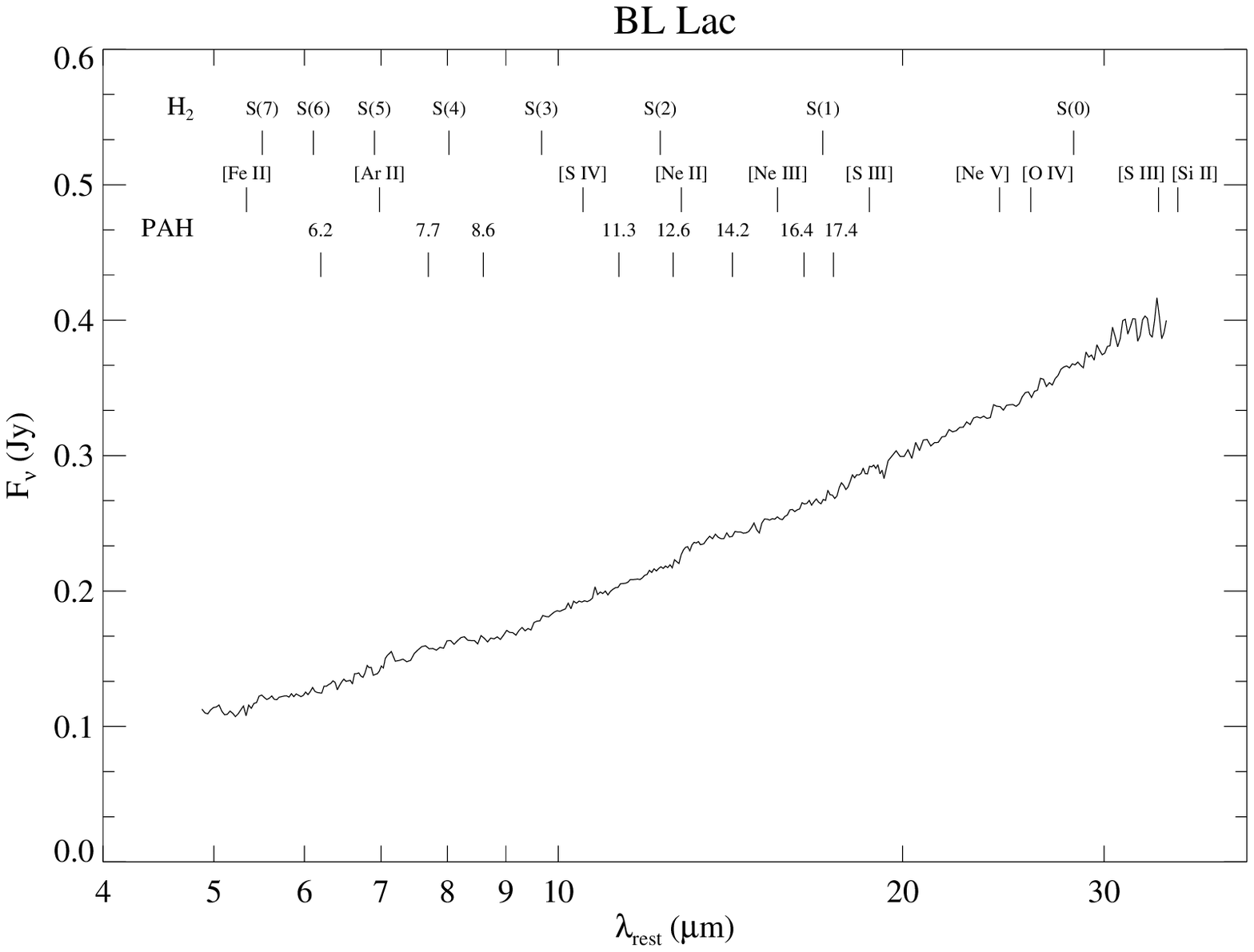}
\includegraphics[angle=0,scale=.44]{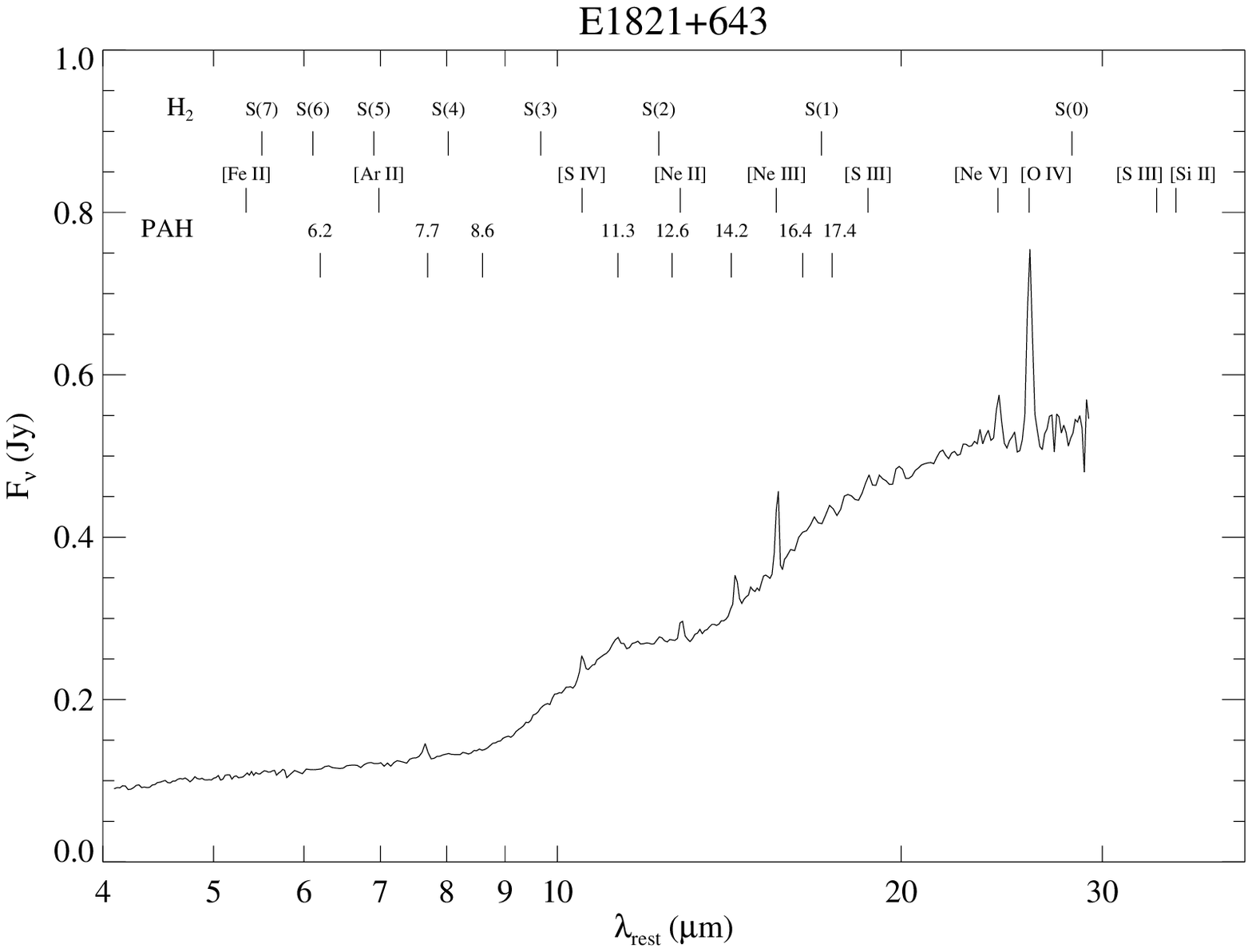}
\caption{Observed MIR spectra of the supplemental objects. Again we plot the flux
  in Jy over the rest frame wavelength in $\mu$m.\label{add_spec}}
\end{figure*}

\begin{figure*}[t]
\centering
\includegraphics[angle=0,scale=.44]{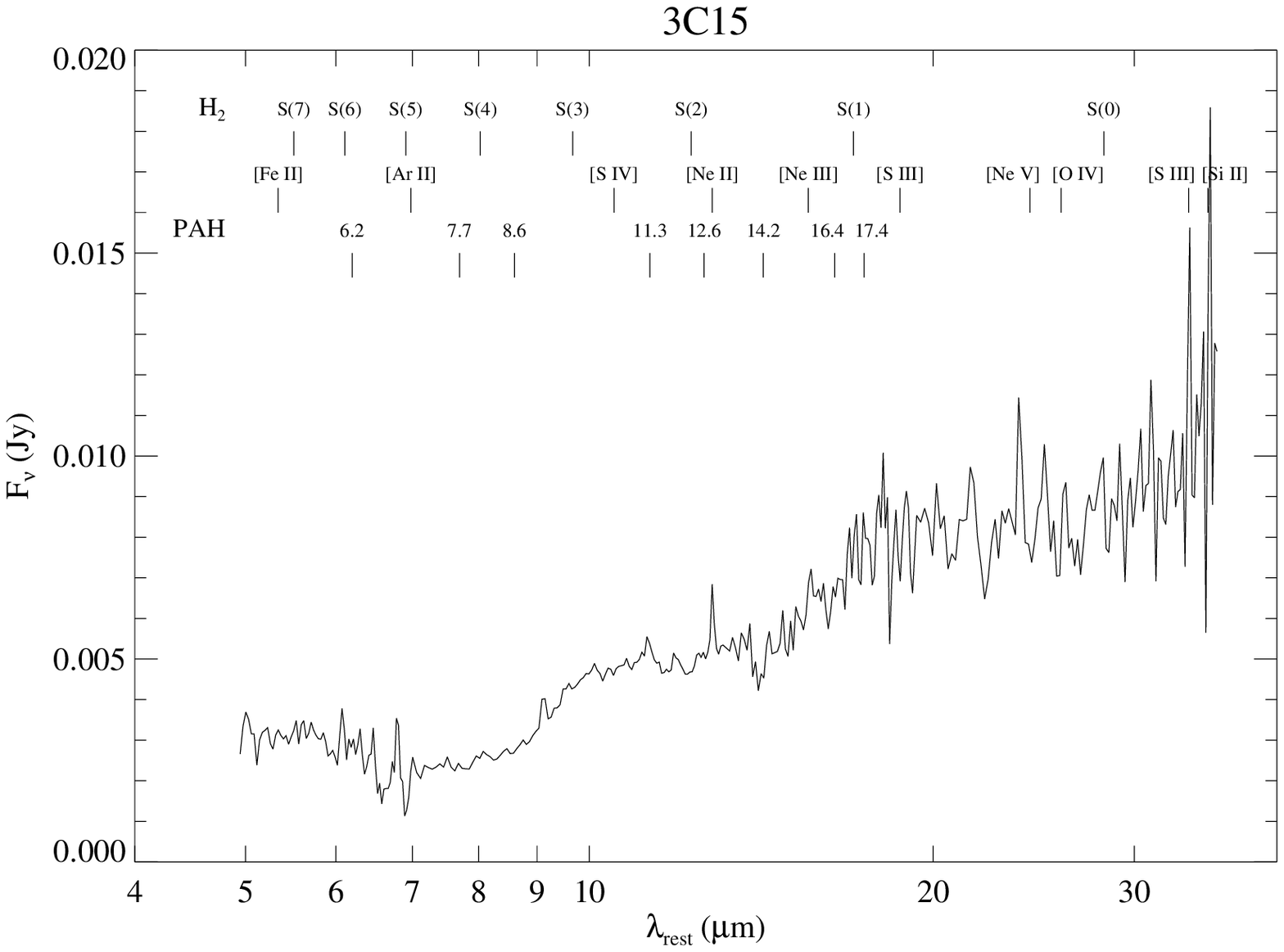}
\includegraphics[angle=0,scale=.44]{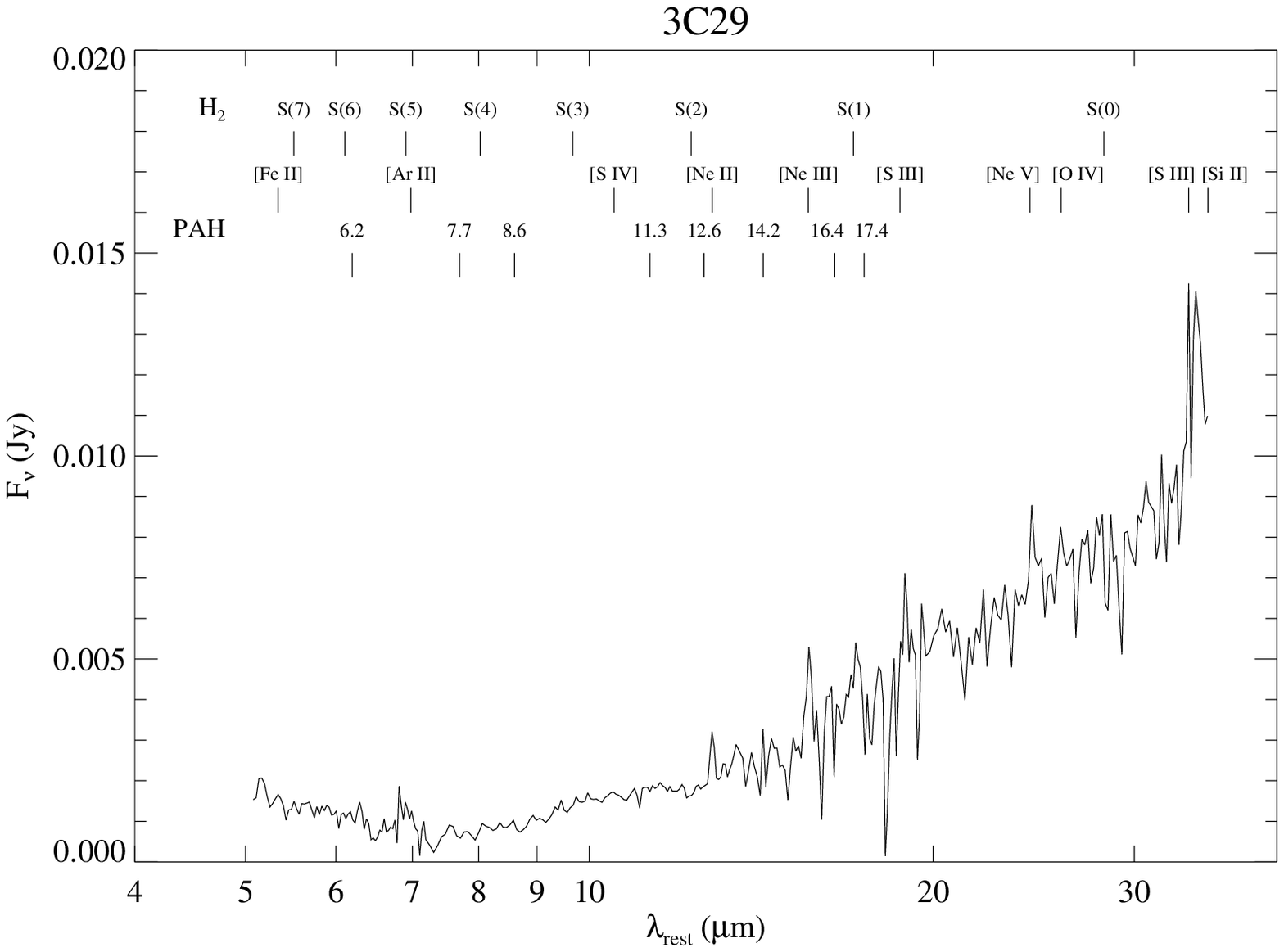}
\caption{The MIR spectra of 3C15 (left) and 3C29 (right). \label{add_corr_spectra} }
\end{figure*}

\subsection{The main sample}

{\bf 3C31 (NGC\,383)\quad}This source shows very strong PAH emission
and the MIR spectrum is likely to be dominated by emission related to
star formation.

A massive molecular gas disk has been detected in CO  in this object
\citep{oku05}.

{\bf 3C66B\quad}This source shows increasing flux towards the shortest
wavelengths  observed (starting at around 8\,$\mu$m) indicating a
dominant contribution from the stellar component of the host galaxy at
these wavelengths.  A weak  silicate feature in emission can be seen
around 9.7\,$\mu$m as well as significant red continuum at longer
wavelengths. Atomic line emission from low ionization species is
detected but no PAH features.

{\bf 3C76.1\quad}While this source is clearly detected (as seen from
the 2--D frames) it has  low flux and shows a noisy spectrum.  No
spectral features can be securely identified. The spectrum shows slowly
increasing flux towards longer wavelengths,  starting at around
15\,$\mu$m.

{\bf 3C83.1 (NGC\,1265)\quad}Emission from the host galaxy stellar
population dominates this source up  to $\sim$\,15\,$\mu$m where the
spectrum starts to flatten in flux. PAH emission at
11.3\,$\mu$m and atomic emission from [\ion{Ne}{2}] at 12.8\,$\mu$m
are clearly detected. The ratio between the PAH features at
7.7\,$\mu$m  and  11.3\,$\mu$m appears very small compared to e.g. 3C31 
(see \S\,4.1).

{\bf 3C129\quad}Here the stellar emission again dominates below
$\sim$\,10\,$\mu$m and for $\lambda>15$\,$\mu$m strong MIR
continuum emission is observed. We also detect weak silicate emission
and PAHs at 11.3\,$\mu$m.

{\bf 3C189 (NGC\,2484)\quad}This source appears very similar to 3C66B
in the MIR spectrum with a  distinct stellar component at short
wavelengths, weak silicate emission around 9.7\,$\mu$m, red continuum
emission for $\lambda>10\,\mu$m, and no PAHs.

{\bf 3C264 (NGC\,3862)\quad}Strong PAH features are detected in this
source which argues for a  considerable, if not dominant, contribution
from emission related to star formation. In fact, the steep continuum
at $\lambda>20\,\mu$m might well be associated with star
formation. At the shortest  wavelengths some minor contributions
from the host galaxy stars can be identified.

{\bf 3C270 (NGC\,4261)\quad}A weak silicate feature and weak
PAH emission is detected.  While the blue spectral slope
towards shorter wavelengths indicates contributions by stellar
emission, a strong red continuum is observed for
$\lambda>10\,\mu$m. In combination with the relatively weak PAH
emission this continuum emission is not likely to originate from
star--forming activity only.

{\bf 3C272.1 (M\,84 -- NGC\,4374)\quad}A very unusual spectrum which
has a blue slope up to  $\sim$\,20\,$\mu$m and shows a very steep red
slope at $\lambda>30$\,$\mu$m. Strong  PAH emission is observed, but
as in the case of 3C83.1 the  7.7\,$\mu$m to 11.3\,$\mu$m ratio
appears very small. Several atomic emission  lines are detected, among
them [\ion{O}{4}], a relatively high ionization line  which can,
however, also be found in star--forming galaxies \citep[e.g.][]{smi07}.

{\bf IC\,4296\quad}Somewhat similar in appearance to 3C272.1 the
continuum shows a red slope for $\lambda>20$\,$\mu$m which
steepens for $\lambda>30$\,$\mu$m.  Weak PAH emission and strong
atomic line emission is detected. The  [\ion{O}{4}] emission line has
much larger equivalent width than in 3C272.1 although the
relative contributions from star formation to the spectrum, as traced by the
PAHs, is apparently lower. 

{\bf 3C293\quad}The MIR spectrum of this object is clearly dominated
by star  formation. The strong PAH features and the shape of the red
continuum are very  similar to what is seen in local star--forming
galaxies \citep[e.g.][]{smi07}.  Notably, we also detect several
rotational transitions from molecular hydrogen  in emission (Ogle et
al., in preparation).

{\bf 3C317\quad}This source shows a blue slope at the lowest
wavelengths  suggestive of stellar emission. Starting at around
10\,$\mu$m the spectrum  turns to a red, but not very steep slope. We
do not detect significant PAH features  and
[\ion{Ne}{2}]\,$\lambda$11.3\,$\mu$m is the strongest atomic emission
line.  We do, however, detect faint emission from molecular hydrogen
from S(1) up to at  least S(6) (Ogle et al., in preparation).

{\bf NGC\,6251\quad}Strong silicate emission at 9.7\,$\mu$m and
18\,$\mu$m is detected  as previously noted in \citet{ogl07}. Strong
atomic and weak PAH emission can also be seen.

{\bf 3C386\quad}This source shows a very starlight--dominated, blue
continuum throughout the whole covered  wavelength range. Some PAH
features (e.g. 7.7\,$\mu$m and 11.3\,$\mu$m) and atomic  lines
(e.g. [\ion{Ne}{2}]\,$\lambda$12.8\,$\mu$m and
[\ion{Ne}{3}]\,$\lambda$15.6\,$\mu$m)  are present as well as weak
molecular hydrogen emission. \citet{simp96} tentatively detected a
broad H$\alpha$ line in the optical and \citet{mad06} find that a
foreground star  falls on top of the nucleus for this galaxy
(previously also noted by \citealt{lyn71}).

{\bf 3C424\quad}The continuum emission is rather weak in this source
but we detect  it in all orders. The most notable feature in this
spectrum is the exceptionally  bright S(1) emission line from
molecular hydrogen. Several other H$_2$  emission lines can be
identified as well.

{\bf 3C465 (NGC\,7720)\quad}While the presence of stellar emission can
be identified at short wavelengths, the spectrum shows a prominent red
slope for $\lambda>20\,\mu$m.  Some atomic emission lines are detected
on top of the rather noisy continuum and no significant PAH features
(besides a possible detection of a weak 11.3\,$\mu$m feature) are
found. Inspection of the 2--D spectral frames  confirms that the
source is detected in all orders, but especially in LL2
($\sim$\,$15-20$\,$\mu$m) it is of very low S/N.

\subsection{The supplemental sample}

In Fig.\,\ref{add_spec} we show the spectra of the FR--I sources that
do not belong to our initial sample. We use them here to demonstrate
the diversity of MIR spectra seen for FR--I  sources.

\begin{figure}[t!]
\centering
\includegraphics[angle=0,scale=.44]{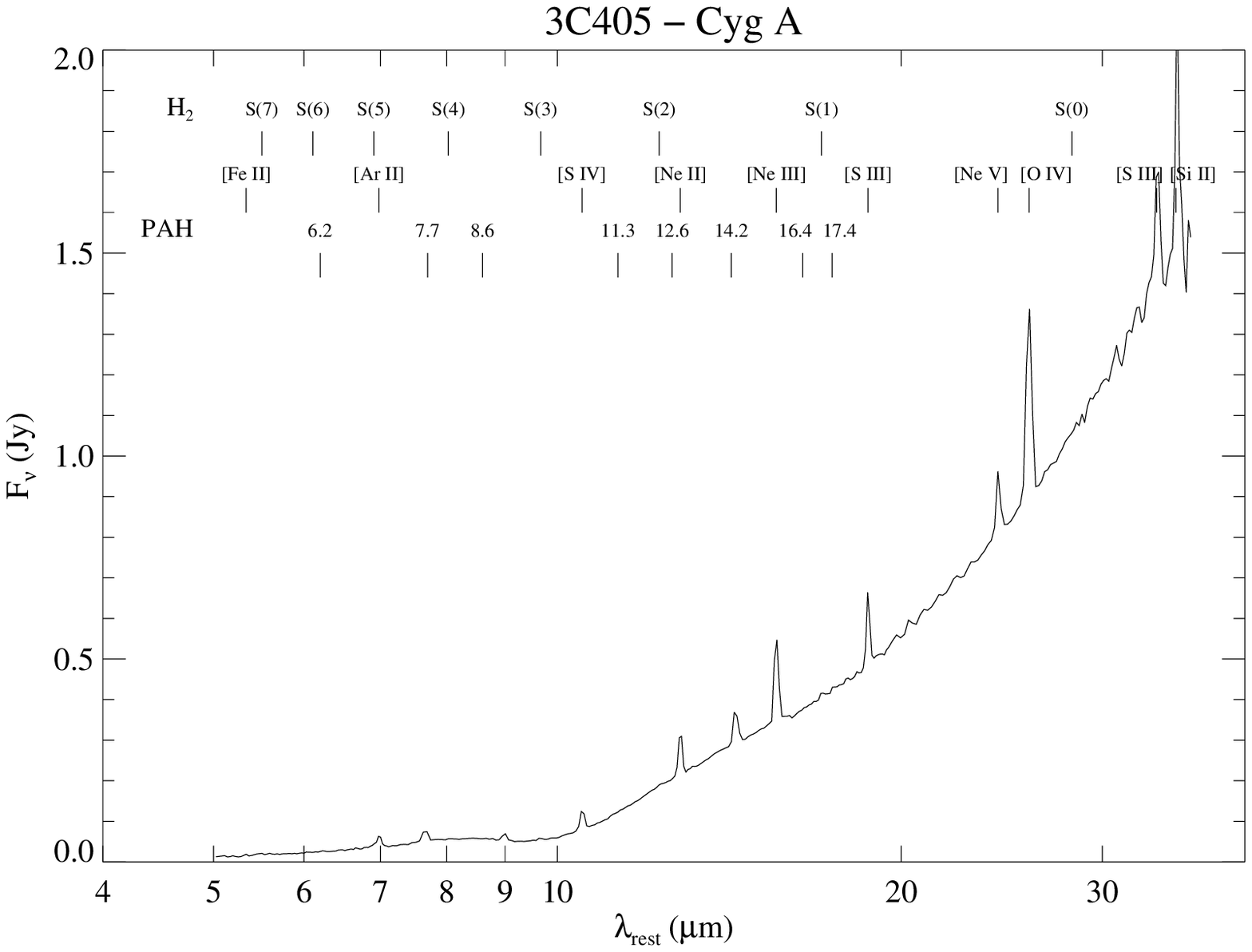}
\caption{Observed MIR spectrum of 3C405, a prototypical hidden
  AGN source\label{cyga_spec}.}
\end{figure}

{\bf 3C84 (Per\,A -- NGC\,1275)\quad}This source is highly core
dominated in the radio \citep{ped90}. In the MIR, however, a strong
thermal spectrum, including weak silicate emission and a steepening
around 12\,$\mu$m is observed. It appears very similar in shape to the
spectrum of the classical hidden AGN source 3C405 (=Cyg\,A;
Fig\,\ref{cyga_spec}) but the continuum appears much stronger in 3C84
relative to the emission lines.

{\bf 3C120\quad} This source is the only 3C FR--I source in our sample
where an AGN is already clearly visible at optical wavelengths as
traced by broad emission lines and a blue optical/UV spectrum
(Tab.\,1). The {\it Spitzer} spectrum reflects this by showing
silicate emission at 9.7\,$\mu$m and 18\,$\mu$m, a rather flat
continuum compared to sources like 3C84 and 3C405 and strong
[\ion{Ne}{5}] and [\ion{O}{4}] emission similar to the MIR spectra of
typical type--1 AGN \citep[e.g.][]{wee05,buc06}.

{\bf 3C218\quad} Hydra\,A is the only source in our sample which is
morphologically clearly a FR--I but with an 178\,MHz luminosity 
that places it among the FR--IIs. Optically a
low ionization galaxy (Tab.\,1) the MIR spectrum shows PAH emission
and [\ion{O}{4}] is also detected. Overall the spectrum looks quite
similar to sources dominated by star formation in the MIR (e.g. 3C31,
3C293). However, here the continuum starts rising already at around 10
to 12\,$\mu$m while for the other star--forming sources such a
steepening in the continuum is observed around $\sim$\,20\,{$\mu$}m
\citep{smi07}. Thus it might have a warmer dust component or stronger 
non--thermal core contributions.

{\bf 3C274 (M\,87 -- Vir\,A -- NGC\,4486)\quad} The MIR spectrum of
M\,87 shows a rather flat slope overlayed with some line emission from
generally low ionization species (except for some weak [\ion{O}{4}]
emission). At short wavelengths some contributions from the host galaxy 
can be identified and there is even some shallow silicate emission 
detected. For this source high--resolution ground based MIR flux measurements 
at $\sim$\,0.5\arcsec~resolution are also available \citep{per01,why04}.

{\bf BL\,Lac\quad} As is common to sources in the class of objects
named after  this prototype, BL\,Lac is dominated by Doppler boosted
synchrotron emission at  least from the radio through the optical. The
MIR spectrum is also dominated  by this synchrotron emission and shows
a virtually featureless power--law continuum ($\alpha$\,$\sim$\,$-0.7$). The
diffuse radio emission suggests a relatively weak FR--I source
\citep{ant86}. We show this object to illustrate a purely non--thermal
spectrum. 

{\bf E1821+643\quad}This source was first classified as a radio--quiet
QSO,  but discovered in a survey at low radio frequencies
\citep{lacy92}. Deep radio imaging, however, revealed a
$\sim$\,300\,kpc FR--I radio structure associated with this source
\citep{blun01}. Being a broad--line object the MIR spectrum looks
quite similar to those of most other quasars or broad--line objects
(e.g. 3C120). We clearly detect silicate emission at 9.7\,{$\mu$}m and
a weak emission feature at 18\,{$\mu$}m. We also detect  [\ion{Ne}{5}]
emission and weak PAH emission (e.g. at 11.3\,{$\mu$}m). The continuum
longward of $\sim$\,10\,{$\mu$}m is redder than for 3C120 but not as
red as for hidden AGN sources like Cyg\,A.
\newline

In Fig.\,\ref{add_corr_spectra} we present the MIR spectra of 3C15 and
3C29. The spectra of these objects suffer from ``spill--over'' due to
saturated peak--up areas. Therefore the flux levels and the slopes in
the SL modules are likely to be compromised. It is
notable, however, that both sources clearly show a red continuum at
wavelengths greater than 15$\,{\mu}$m but no significant PAH
emission. Since the red continuum is detected in the LL orders it is
not an artifact of the spill--over.

\section{MIR emission in FR--I radio galaxies}

In this paper we study the nuclear MIR emission of FR--I radio galaxies
and try to determine its properties and possible origins. Components
considered  are {\sc i}) synchrotron emission from a non--thermal
core, {\sc ii}) warm dust emission resulting from AGN heating, {\sc
iii}) dust heated by star formation as traced by PAH features, and
{\sc iv}) stellar emission (photospheric emission plus dust emission
associated with the circum--stellar envelopes of AGB stars).  We often
summarize contributions {\sc iii} and {\sc iv} as processes from the
host galaxy.  We use the term ``nuclear'' to address central emission
related directly to an AGN\footnote{We note that the definite test for
the nuclear origin of the detected MIR emission would be high spatial
resolution MIR imaging in order to compare the fluxes of any
sub--arcsec nuclear point sources with the {\it Spitzer} fluxes 
(see e.g. \citet{why04} for a discussion of the
contrasting results on M\,87 and Cen\,A).}.

In the framework of models which favor the absence of nuclear dust
with high covering factors in FR--I sources, the MIR spectra should
consist of a combination of non--thermal emission due to a nuclear
synchrotron source and more extended emission related to processes in
the host galaxy. The warm/hot dust component related to an AGN should
be missing or of very low luminosity.  In order to explore any nuclear
MIR emission in FR--I sources, we will here test this scenario by
successively identifying contributions from different mechanisms to
the MIR spectra.

\begin{figure}[t!]
\includegraphics[angle=0,scale=.44]{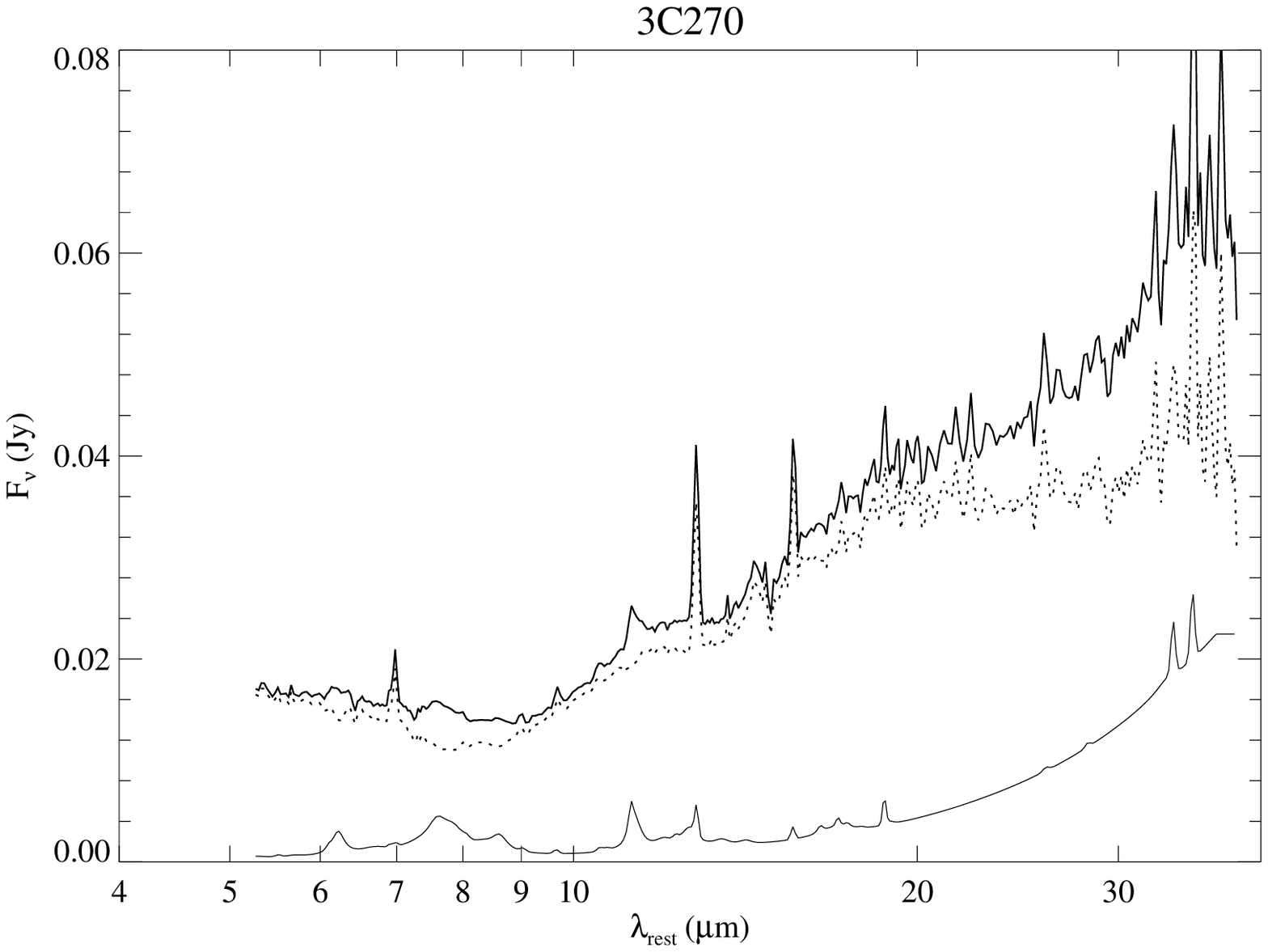}
\caption{3C270 as an example for the subtraction of the star--forming
  template (thin solid line) which is scaled to remove the
  11.3\,$\mu$m PAH feature. The spectrum on the top is the total observed spectrum while 
  the dotted line shows the star--formation corrected spectrum  (note the absence of the
  11.3\,$\mu$m PAH feature).\label{3c270}}
\end{figure}

\begin{figure*}[t!]
\includegraphics[angle=0,scale=.44]{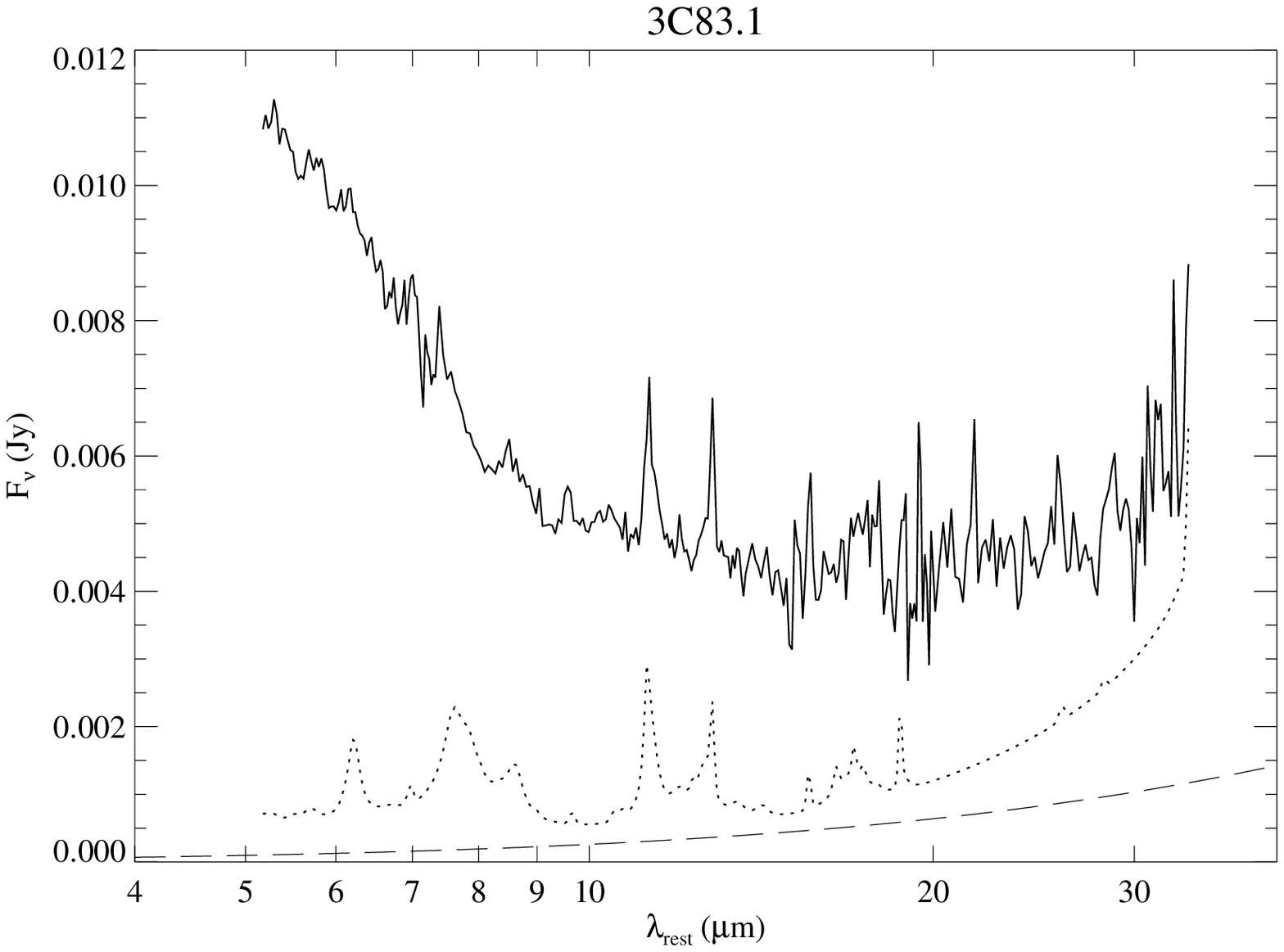}
\includegraphics[angle=0,scale=.44]{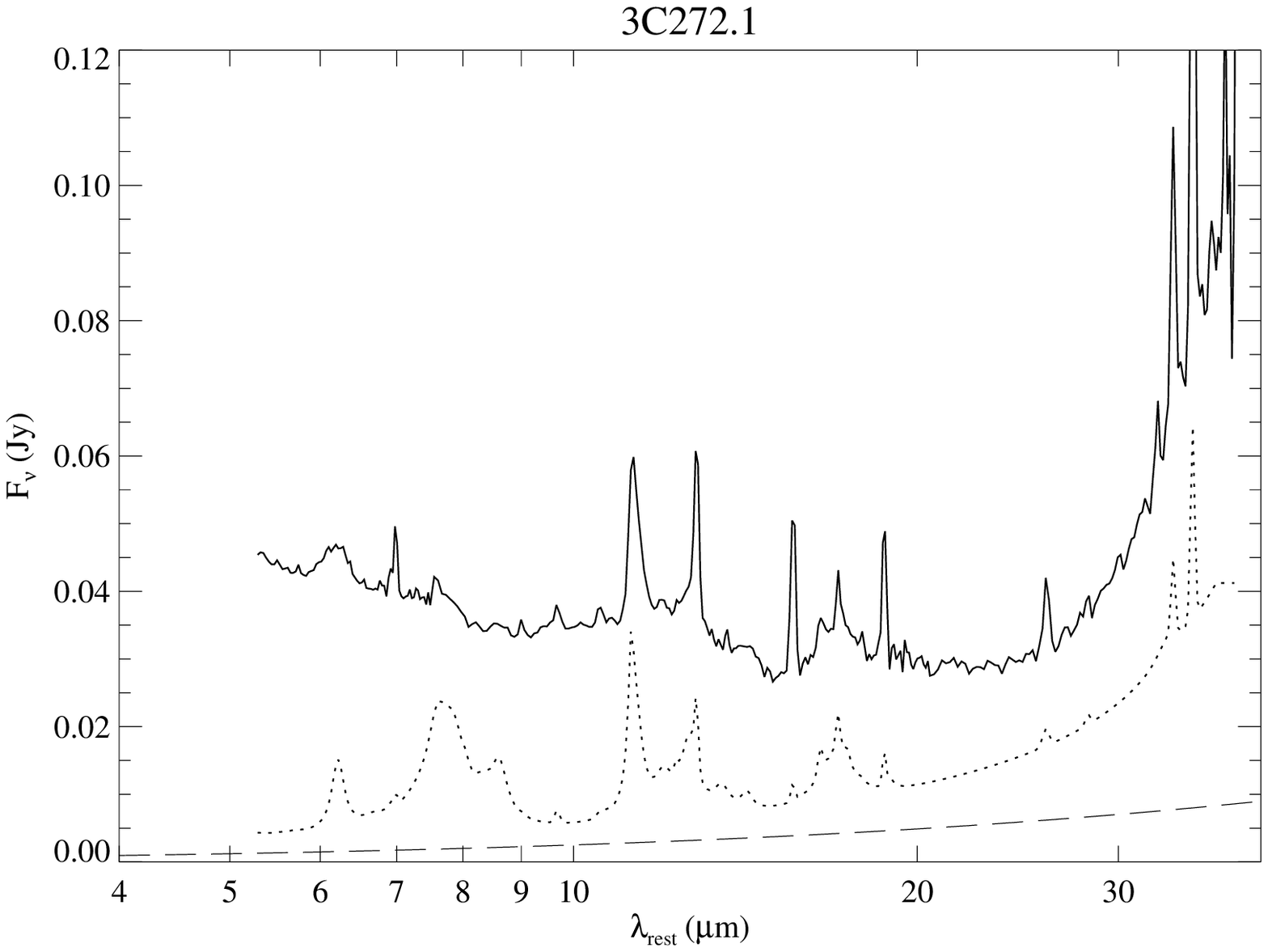}\\
\includegraphics[angle=0,scale=.44]{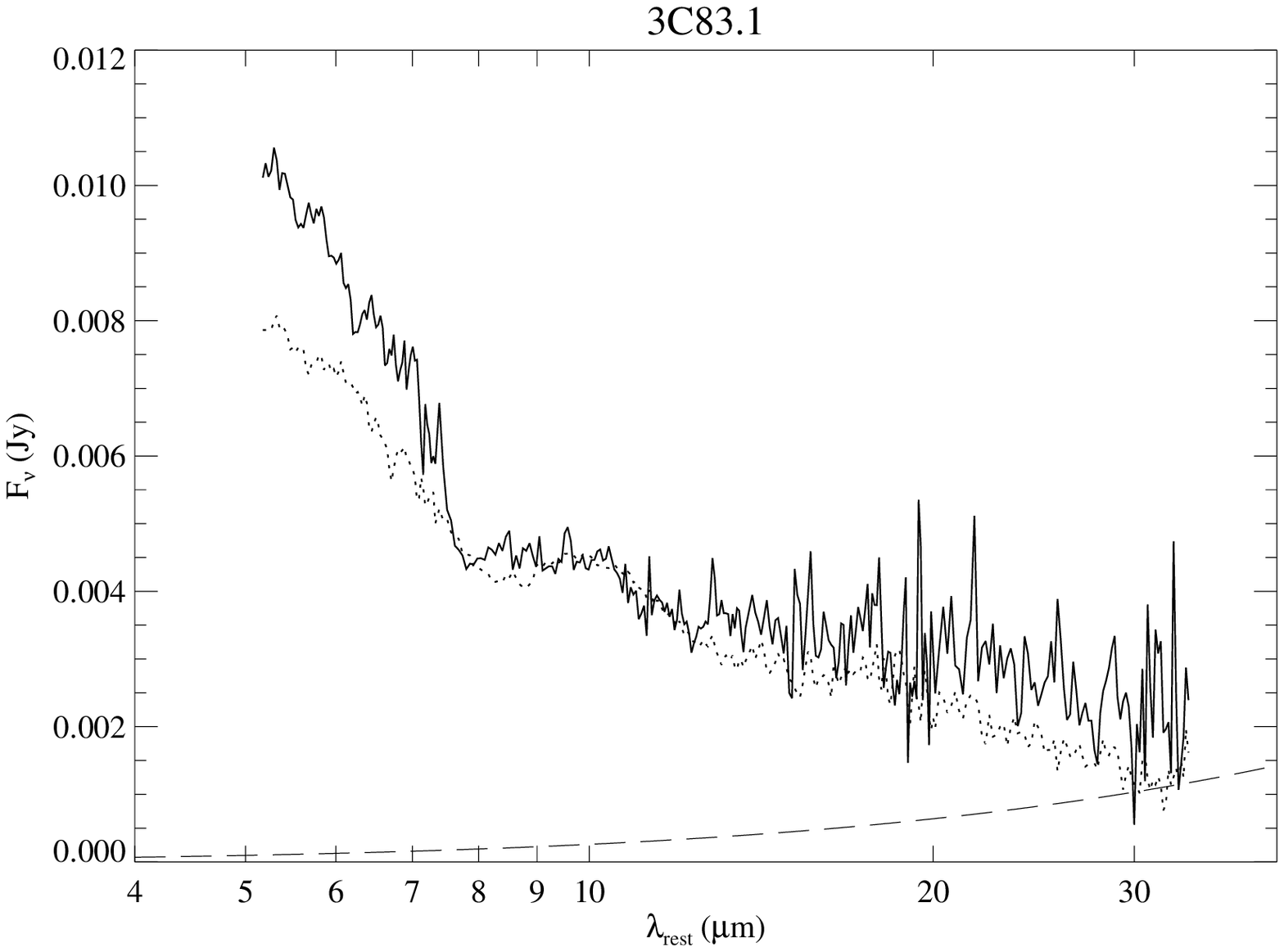}
\includegraphics[angle=0,scale=.44]{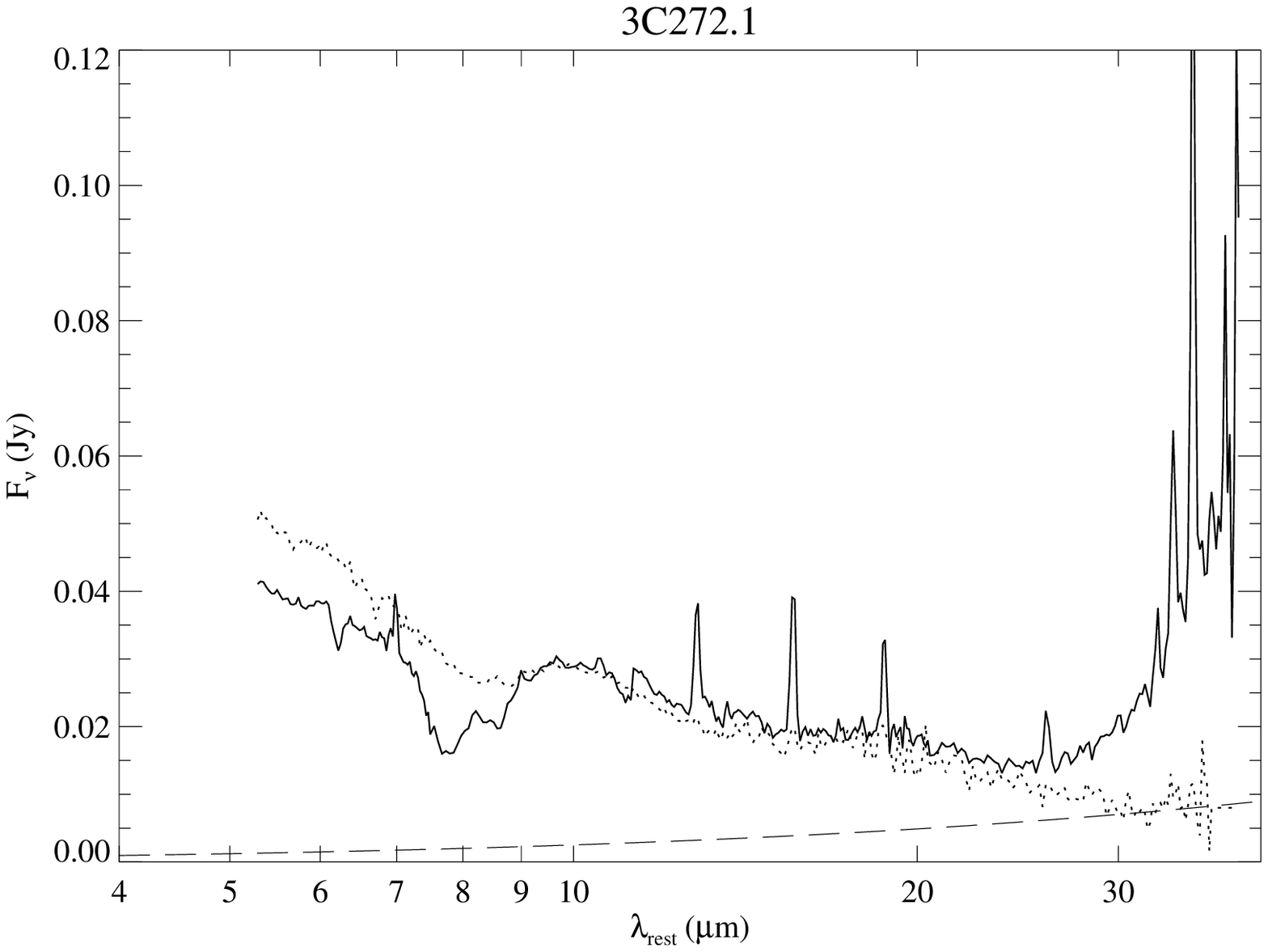}
\caption{The unusual PAH ratios in 3C83.1 ({\it left}) and 3C272.1
({\it right}). The top panel shows the  observed spectrum with the
star--formation template scaled to the 11.3\,$\mu$m PAH feature. The
bottom  panel shows the results of the subtraction and the scaled
spectrum of an early--type galaxy (NGC\,1549).  Approximate
contributions from the synchrotron core (\S4.4) are shown as
long--dashed lines.
\label{pahratio}}
\end{figure*}

\subsection{Contributions from star formation}

Due to the large slit width of IRS (3.7\arcsec~and 10.7\arcsec~for SL
and LL, respectively) contributions from the host galaxy (star
formation and stellar emission) to the MIR continuum can be
significant and even dominant. We need to correct our spectra for 
these host galaxy contributions in order to explore any additional
nuclear MIR emission.

Because emission related to star formation has intrinsically a red
spectral slope it cannot be distinguished easily from the synchrotron
core emission or AGN heated dust. 
Therefore, in order to securely identify any synchrotron or AGN dust
components in the observed MIR spectra, emission from star
formation needs to be corrected for. As an indicator for star--forming contributions
we here use  the PAH features which are commonly
observed in the spectra of star--forming galaxies and are a well known
tracer of star--formation activity. Only if PAHs are detected do we
correct the spectra for emission related to star formation. No
significant PAH emission  usually indicates negligible continuum
contribution from star formation compared  to other processes.

In the case of detected PAH emission we chose to subtract
appropriately scaled average star--formation template spectra taken
from \citet{smi07}.   We used the template with the reddest MIR/FIR
slope which was still in accordance with the data after scaling, thus
being conservative about any residual emission. The template was
scaled to the 11.3\,{$\mu$}m PAH feature which is strong, largely free
of contaminations by atomic emission lines, and close to wavelengths
where AGN related emission usually has a large contribution.  The
star--forming template was scaled in such a way that, after
subtraction, the 11.3\,{$\mu$}m PAH feature is removed from our FR--I
spectrum (see Fig.\,\ref{3c270} for illustration).  In Tables
\ref{tab2}  and \ref{tab4} we give estimates for the percentage
contributions from star formation to the observed MIR flux at
15\,$\mu$m and 30\,$\mu$m.

While this strategy works well for many of our sources, subtraction of
the star--formation template  (scaled to the 11.3\,$\mu$m PAH feature)
leaves a clear depression at the location of the 7.7\,$\mu$m feature
for 3C83.1 and 3C272.1.  This cannot be explained by the shape of
stellar emission and argues for a very low 7.7\,$\mu$m/11.3\,$\mu$m
PAH ratio in these FR--I galaxies (Fig.\,\ref{pahratio}). Apart from
the low PAH ratio both spectra can be explained by a combination of
processes in the host galaxy. We detect no significant nuclear
continuum (but see \S4.3).

Such low PAH ratios have already been noted in local (early--type)
galaxies \citep{kan05,smi07,kan08}. Sources with MIR continua
dominated by star formation usually show no large variation  in the
7.7\,$\mu$m to 11.3\,$\mu$m ratio \citep{bra06,smi07} and similar
star--formation dominated PAH ratios  have been found in luminous
radio--quiet and radio--loud AGN \citep{shi07b}. In absence of strong
contributions from star formation, spectra with unusual
7.7\,$\mu$m/11.3\,$\mu$m values can be observed.  Interestingly, most
sources with such low PAH ratios show low--luminosity AGN activity
\citep{smi07}.  Consistent with this, \citet{stu06} have shown that
IR--faint LINERs (LINERs without much star formation) show these
unusual PAH ratios, while IR--luminous ($L_{\rm IR}/L_{B} \gtrsim 1$
with  $L_{\rm IR}$ being the $8-1000\,\mu$m luminosity) LINERs have
PAH ratios similar to starburst galaxies.  In the former cases it
appears that the nuclear source powers most of the measured PAH
emission. The harder spectrum could potentially lead to the
destruction of the carriers of the 7.7\,$\mu$m band or to the
preferential excitation  of the 11.3\,$\mu$m PAH, thus explaining the
observed relative decrease in the  7.7\,$\mu$m feature \citep[][and
references therein]{smi07,kan08}.  This could in principle indicate that the
PAHs in 3C83.1 and 3C272.1 are  not excited by star formation but by
(low--luminosity) AGN activity instead. Any  corrections made to the
spectrum by the subtraction of a star--forming template would thus
remove MIR continuum emission which is in fact powered by the AGN.

\subsection{Contributions from stellar emission}

Recall that we have already seen in the spectra
(Fig.\,\ref{obs_spectra})  that contributions from stellar 
emission of the host galaxy can be substantial,
particularly at shorter wavelengths. Most or all of our FR--I sources
reside in elliptical galaxies \citep[e.g.][]{gov00,mad06}. The MIR
spectra of early--type galaxies generally show blue colors throughout
the IRS wavelength range \citep[e.g.][]{bre06,kan08} and even if
moderate amounts of residual star formation are observed, the blue
spectral slopes of the stellar emission can remain dominant in the
total spectrum \citep{kan08}. This is supported by the fact that the
SEDs of typical early--type galaxies without signs for considerable
star formation usually show a blue IR slope that, due to emission from
cold dust, turns over into a red slope at wavelengths beyond the IRS
coverage \citep[e.g.][]{dale05,kan07,tem07}. Therefore, the stellar 
contributions in our MIR spectra will have opposite spectral slope compared 
to any underlying synchrotron core or warm dust emission.

As shown by e.g. \citet{bre06} the MIR spectra of non--active
early--type galaxies can be well explained by including the dusty
atmospheres and envelopes of AGB stars into stellar evolution models. The emission
from dust produced in the atmospheres of AGB stars accounts for excess
continuum emission over the long--wavelength quasi Rayleigh--Jeans
extrapolation  of late--type stars as well as for the presence of
silicate emission in the  spectra of early--type galaxies.

We estimate contributions from the stellar population by using
the observed spectrum of a quiescent elliptical galaxy which is scaled
to the blue part of the spectra ($\sim$\,$5-8\,\mu$m). For a template
we chose the quiescent early--type galaxy NGC\,1549 which is fairly
bright at MIR wavelengths, does not show any PAH or atomic line
emission, and has full IRS low--resolution coverage \citep{kan08}.  We
obtained the data for NGC\,1549 from the {\it Spitzer} archive and
extracted a spectrum following the procedures described in
\cite{kan08}. In Tables \ref{tab2} and \ref{tab4} we list estimated
percentage contributions of the stellar component to the observed MIR
flux at two wavelengths.

\subsection{Spectral components in the MIR}

After the subtraction of the star--forming template in the way
described above and after considering the stellar host galaxy emission 
we recognize basically two types of spectra:

(1) sources where the observed MIR emission can be accounted for by
processes in the host galaxy.

The spectra are either dominated by star formation  (3C31, 3C218,
3C264, 3C293) or the observed MIR emission can be explained by a
combination of stellar and star--forming contributions (i.e. after the
subtraction of the star--forming template the residual spectrum is
largely consistent with stellar emission from an early type
galaxy; 3C76.1, 3C386; Fig.\,\ref{pahsub1}).

(2) objects which show significant residual emission at wavelengths 
longer than $\sim$\,$8\,\mu$m with a red spectral slope.

Because even star formation and stellar emission combined cannot
explain these residual continua they must originate from other
processes, for example warm (few hundred K) dust emission or
non--thermal synchrotron emission. In some of the sources where
residual continuum emission at $\lambda>8\,\mu$m could be identified
we also detect PAHs  (3C84, 3C120, 3C129, 3C270, IC4296, NGC\,6251,
E1821+643; Figs.\,\ref{pahsub1},\ref{pahsub2}), but the equivalent
widths of these features are generally small.   For many sources only
red continuum emission at $\lambda>8\,\mu$m (plus possibly stellar
host galaxy emission) was detected  (3C15, 3C29, 3C66B, 3C189, 3C274,
3C317, 3C424, 3C465, BL Lac). No significant PAH emission can be seen
in their spectra.

As mentioned above in the case of 3C83.1 and 3C272.1 we cannot
without doubt attribute the PAH emission to star formation.  We count
these objects as host--dominated sources, an intepretation which is
supported by the shape of their MIR continuum (Fig.\,\ref{pahratio}), but
we emphasize that we cannot be sure at this point if the PAH emission is
excited by star formation or (low--luminosity) AGN activity.

\begin{figure*}[t!]
\centering
\includegraphics[angle=0,scale=.44]{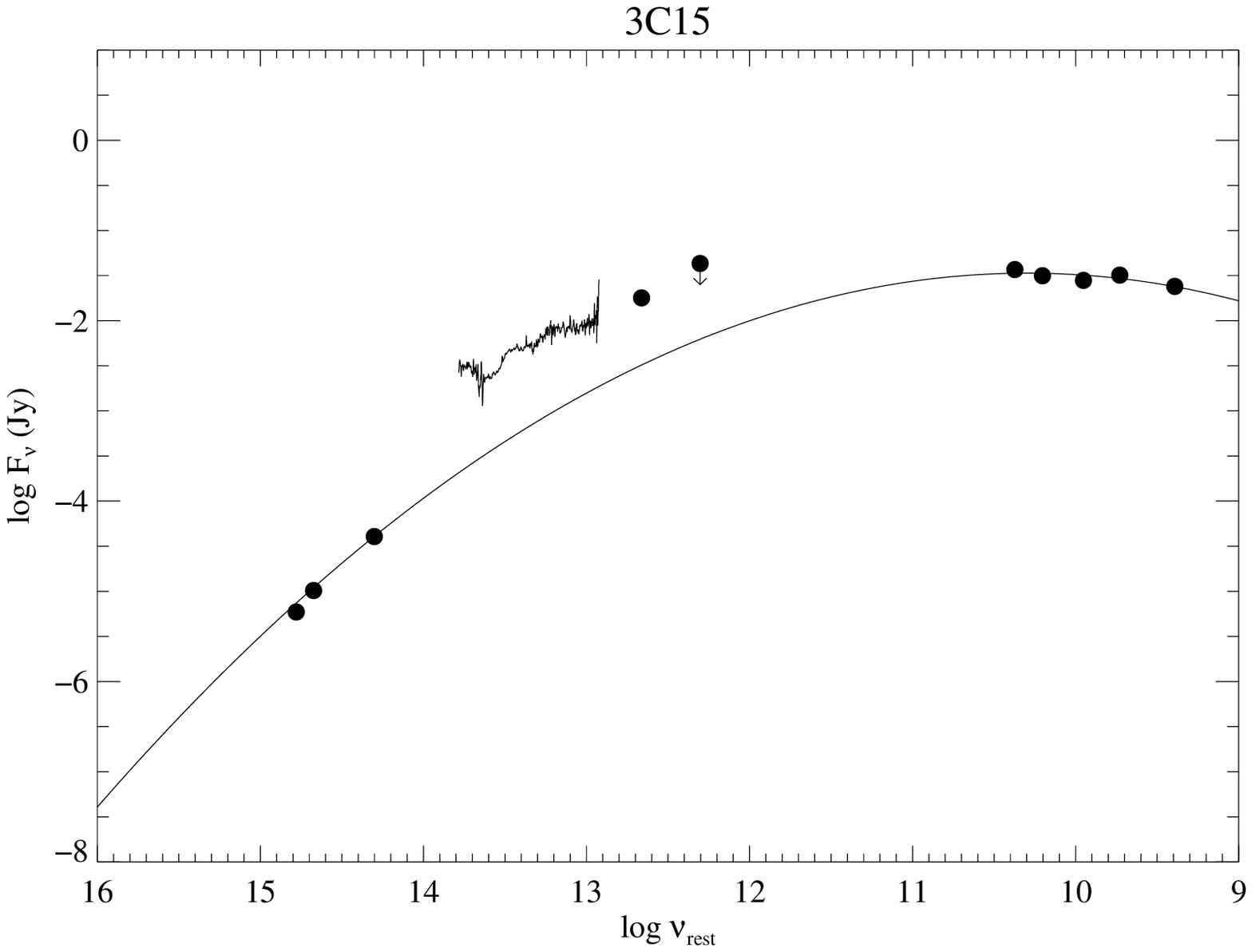}
\includegraphics[angle=0,scale=.44]{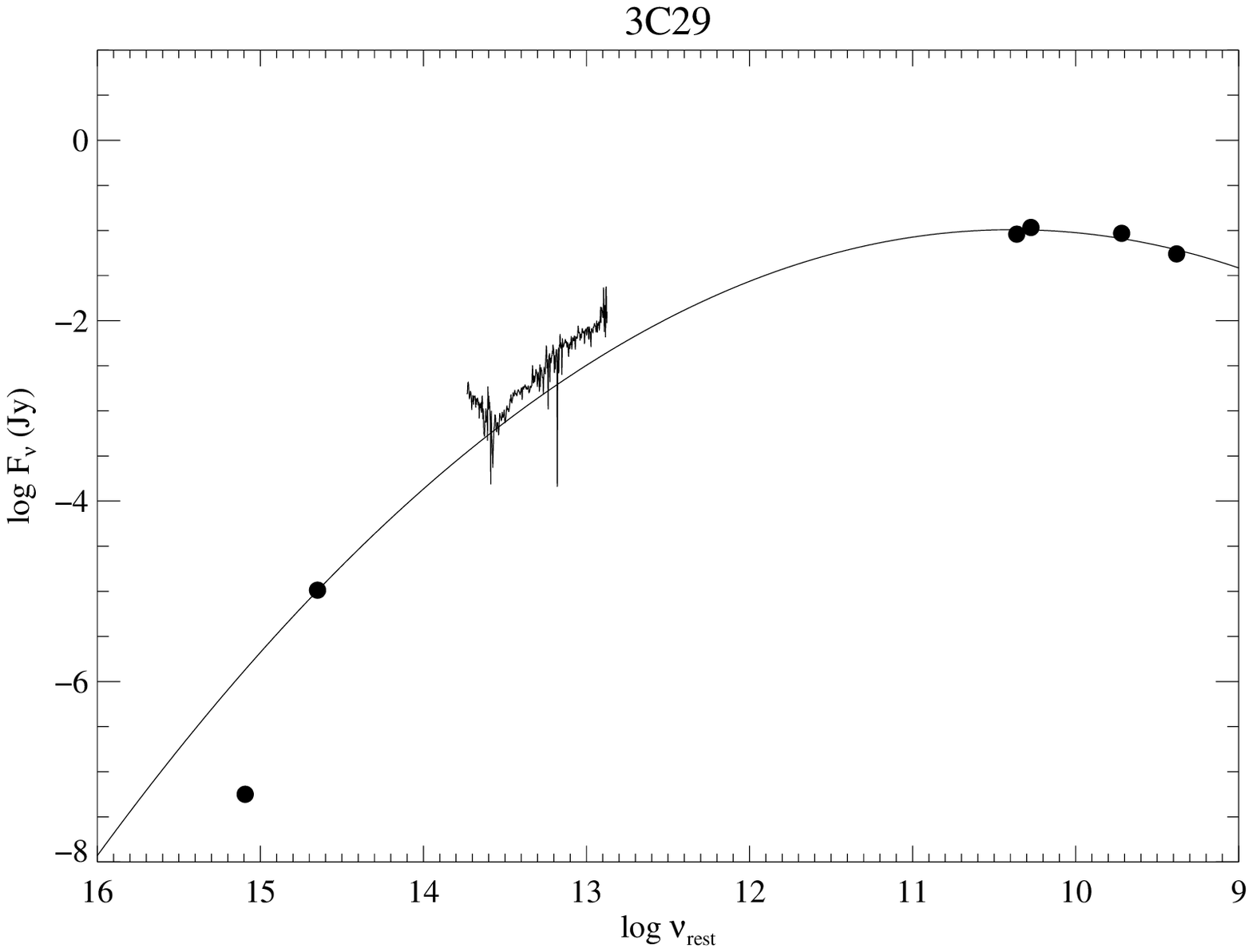}\\
\includegraphics[angle=0,scale=.44]{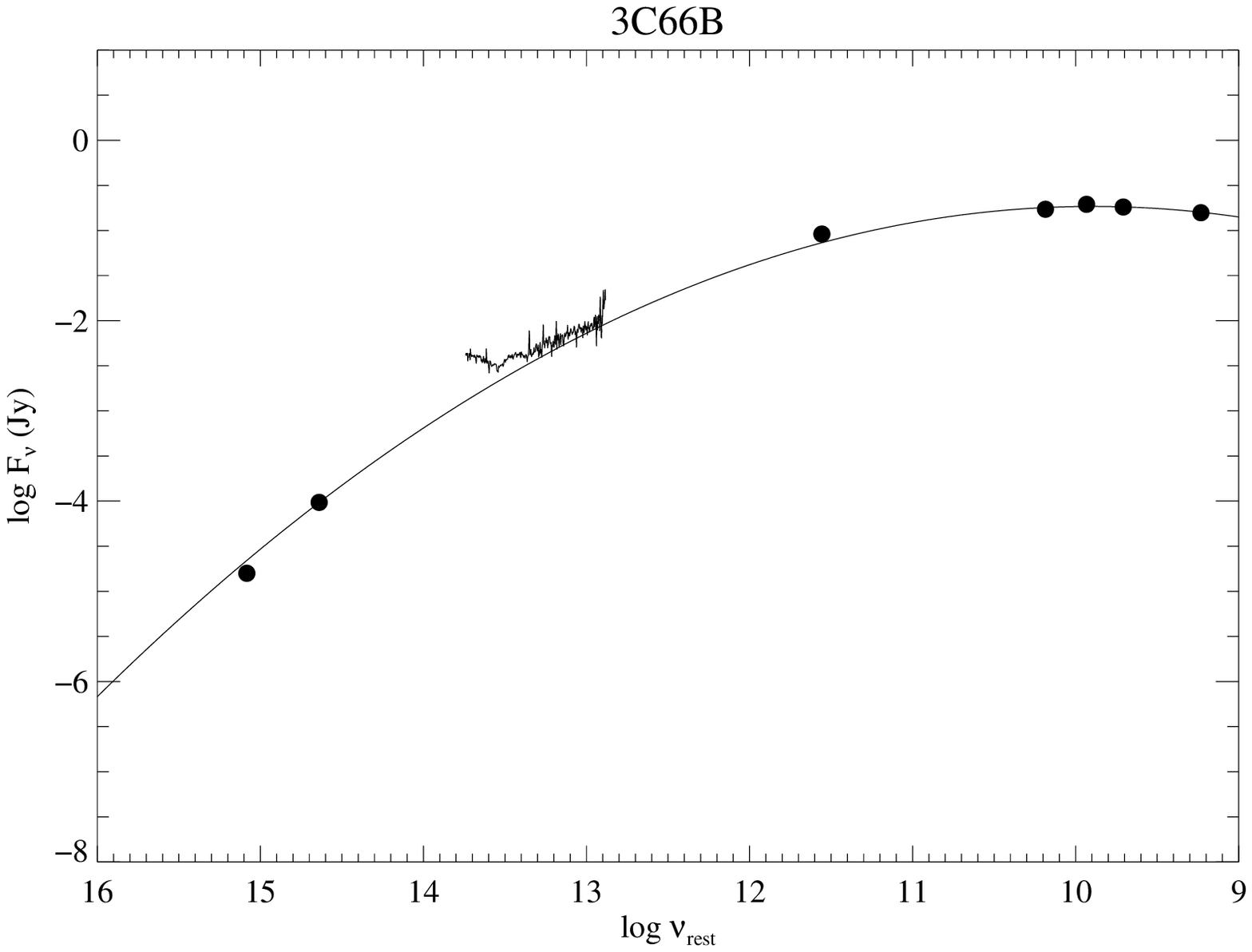}
\includegraphics[angle=0,scale=.44]{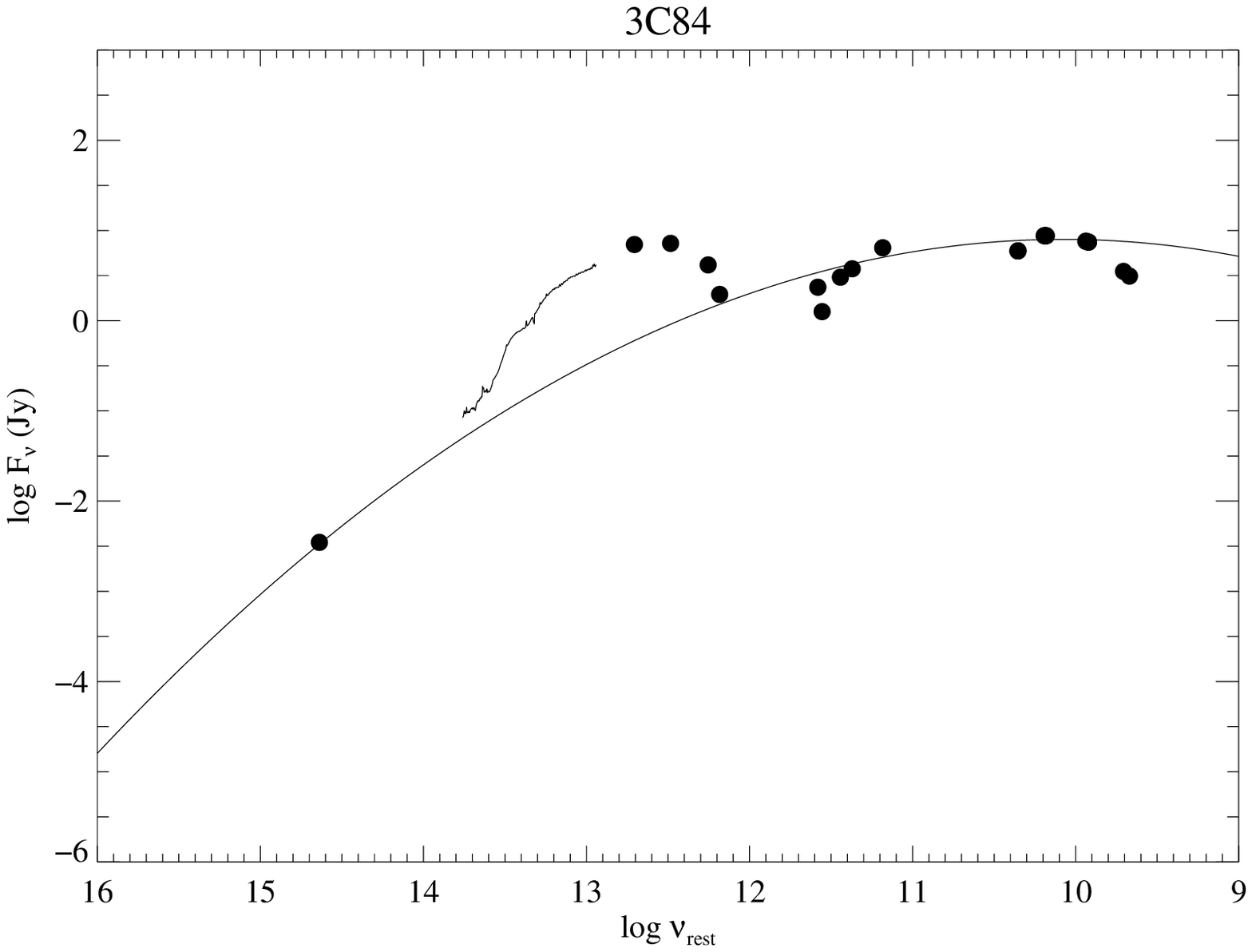}
\caption{Observed nuclear SEDs for sources in which significant residual continuum emission could be identified. 
We used optical fluxes (Tab.\,1), corrected 
for galactic absorption, nuclear NIR fluxes, and 
radio core flux measurements (arcsecond and sub--arcsecond scales). Only optical/NIR and radio core 
measurements were used for the parabolic fits. Nuclear
UV \citep{chia02} fluxes, also corrected for galactic absorption, MIR/FIR, and 
sub--millimeter data (if available) as well as MIR spectra (this work)
are overplotted. For sources with significant PAH features we have subtracted the star formation component 
(3C84, 3C270, IC4296, NGC\,6251; see text). \label{sed1}}
\end{figure*}

\addtocounter{figure}{-1}
\begin{figure*}[ht!]
\centering
\includegraphics[angle=0,scale=.44]{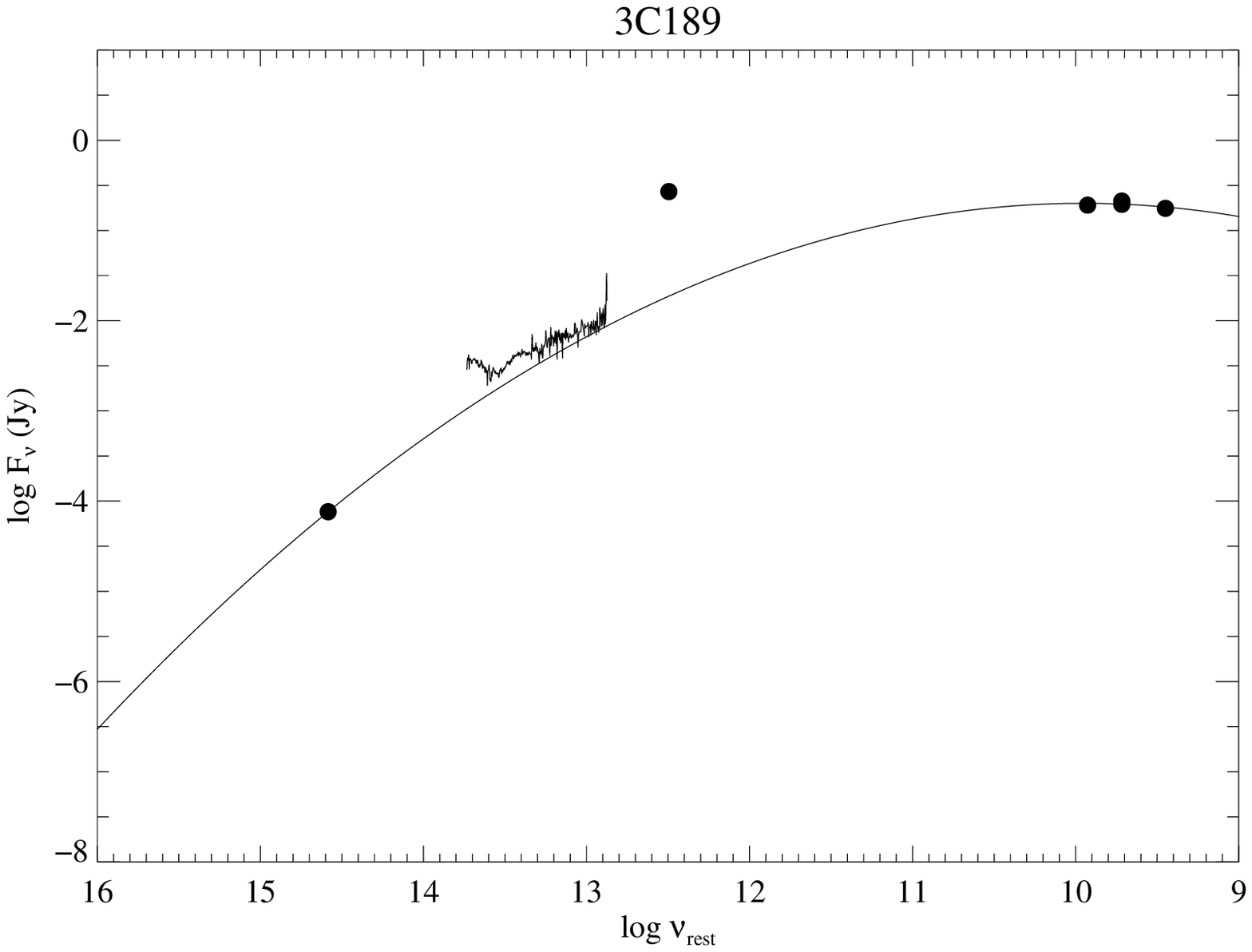}
\includegraphics[angle=0,scale=.44]{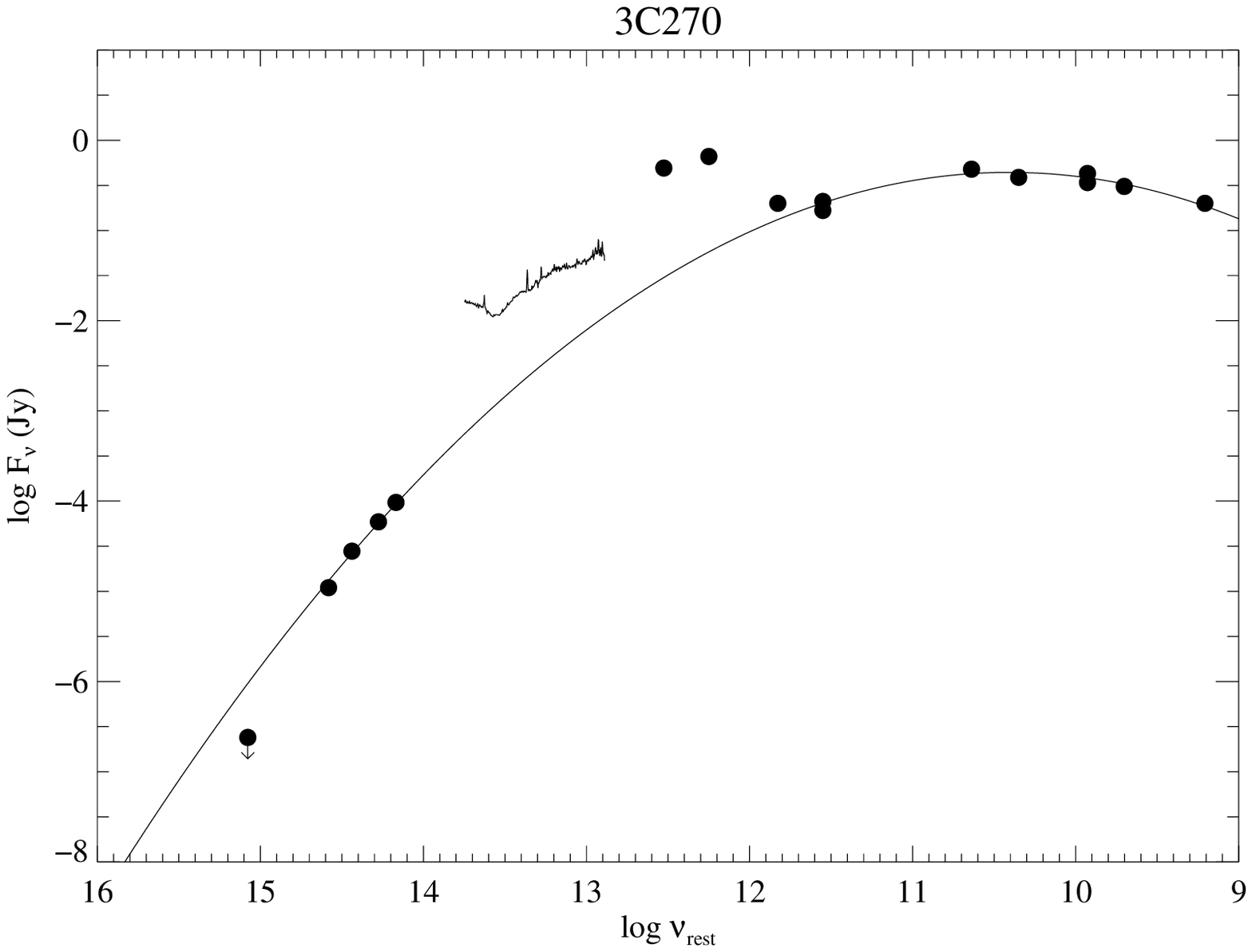}\\
\includegraphics[angle=0,scale=.44]{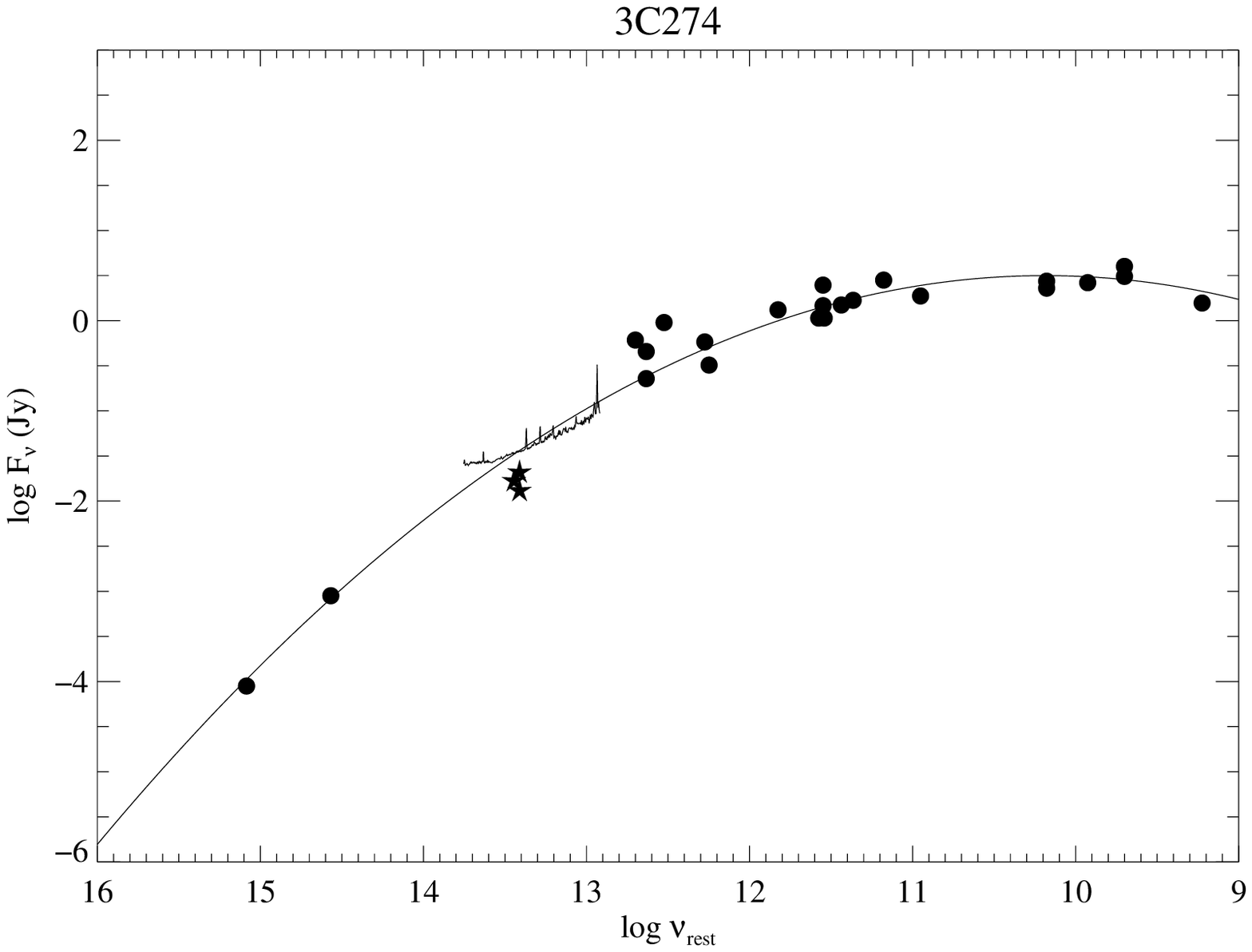}
\includegraphics[angle=0,scale=.44]{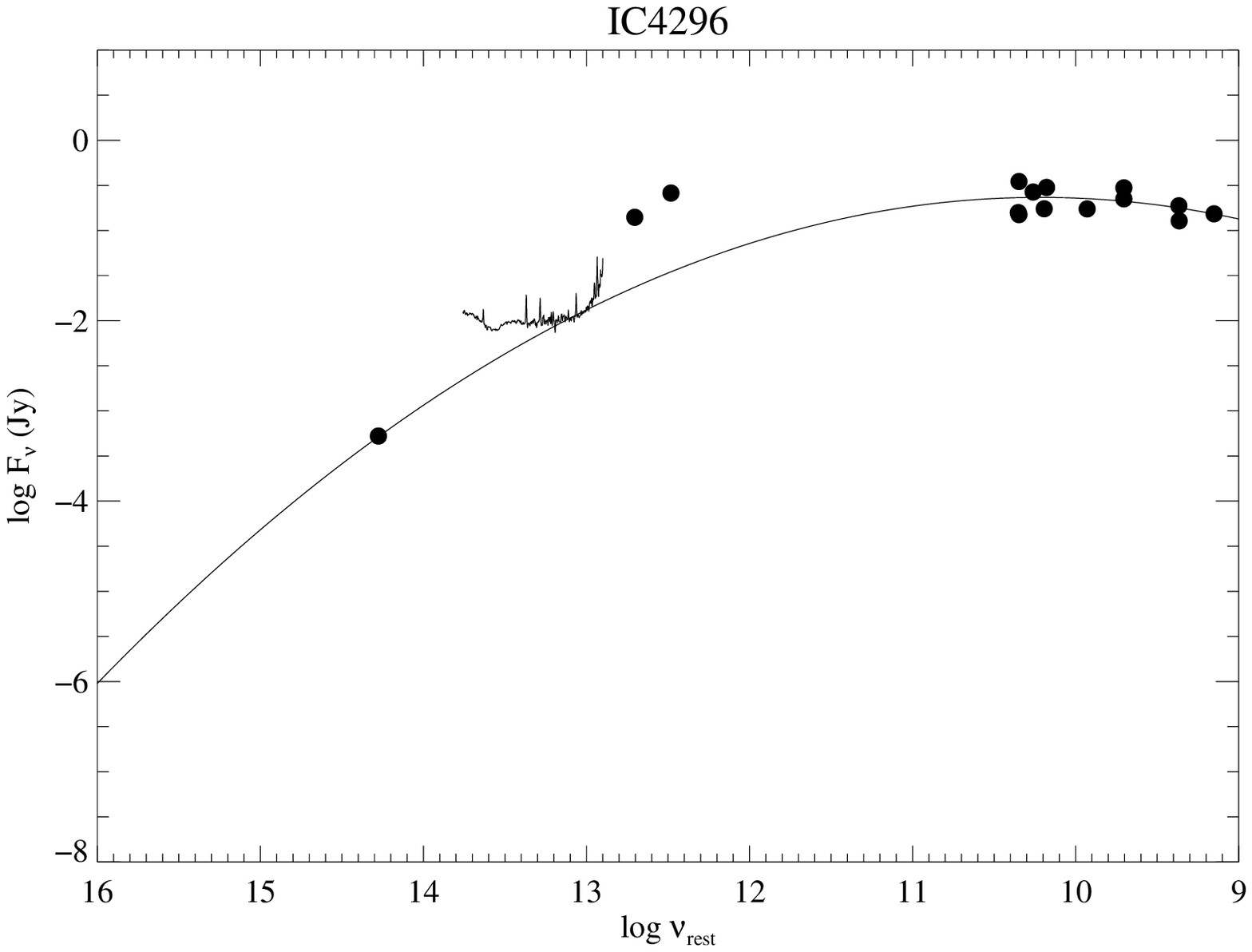}
\caption{{\it continued}.}
\end{figure*}

\addtocounter{figure}{-1}
\begin{figure*}[ht!]
\centering
\includegraphics[angle=0,scale=.44]{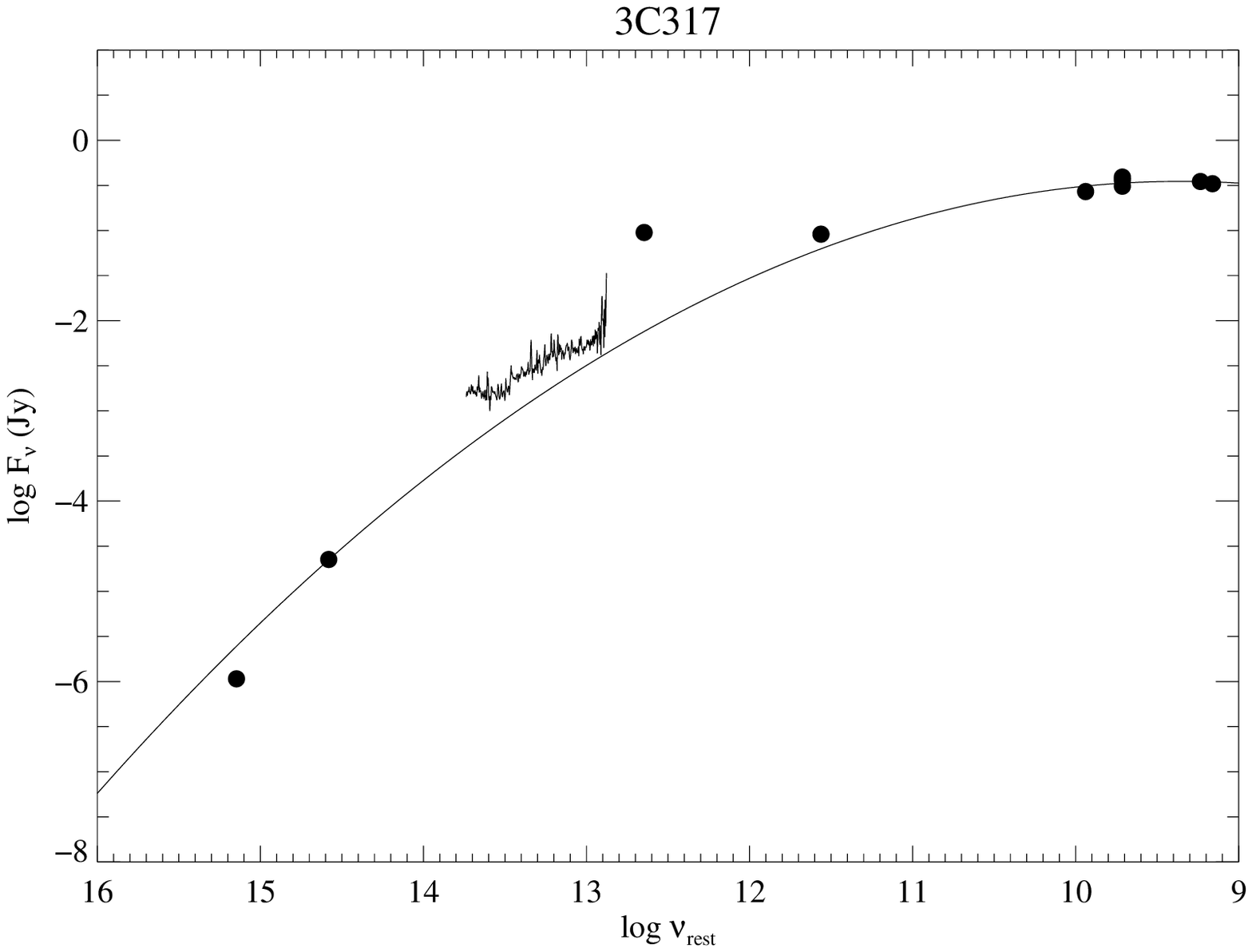}
\includegraphics[angle=0,scale=.44]{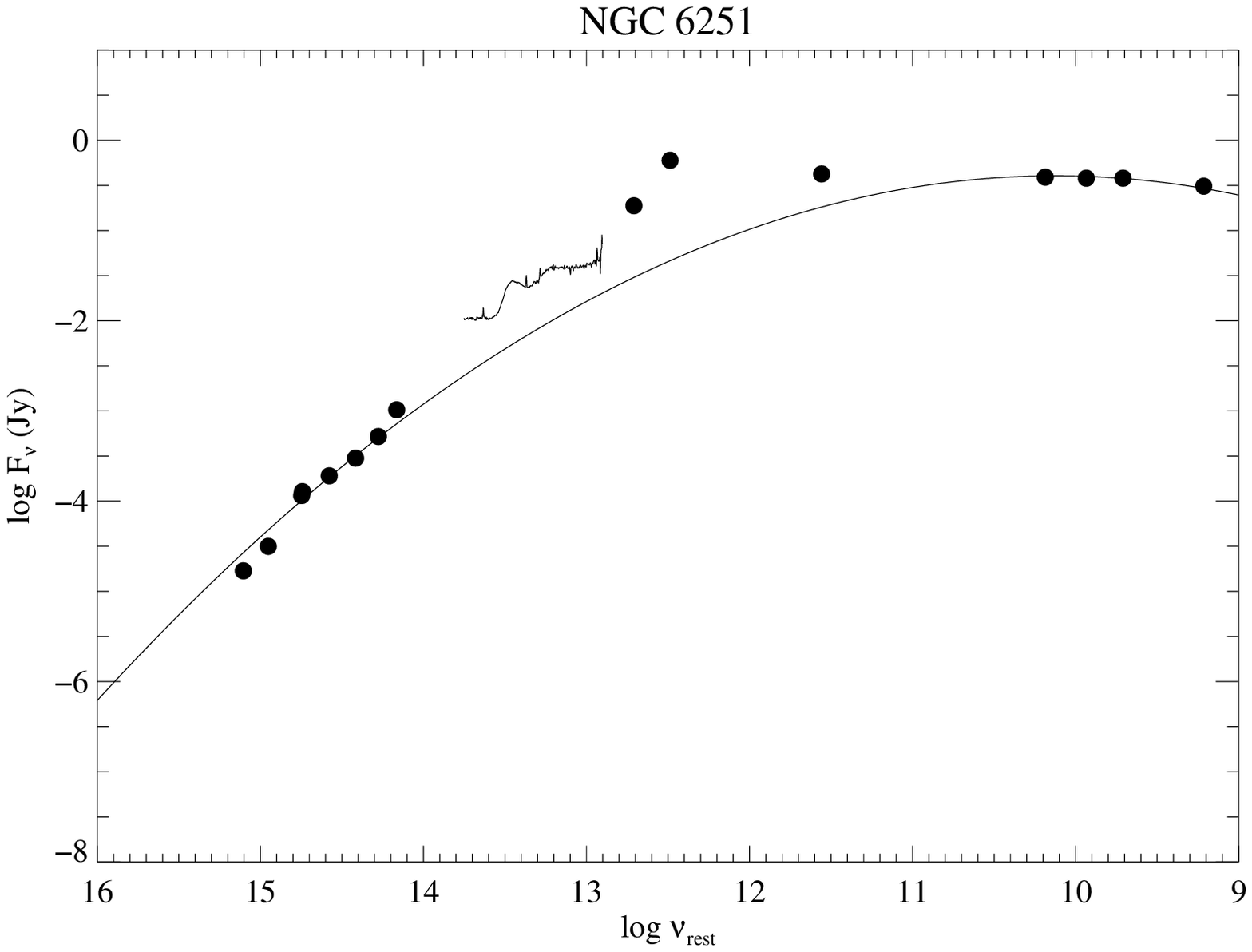}\\
\includegraphics[angle=0,scale=.44]{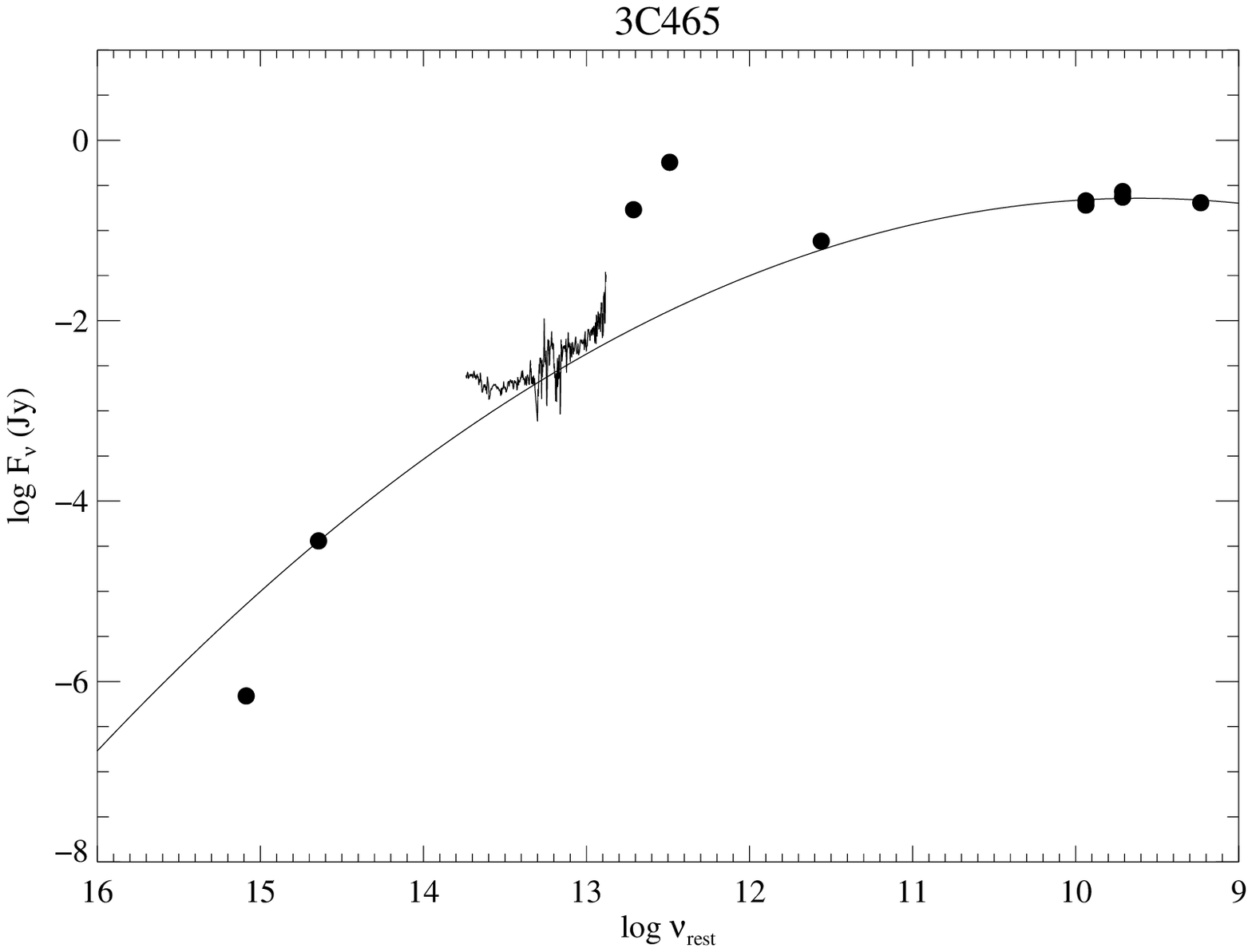}
\includegraphics[angle=0,scale=.44]{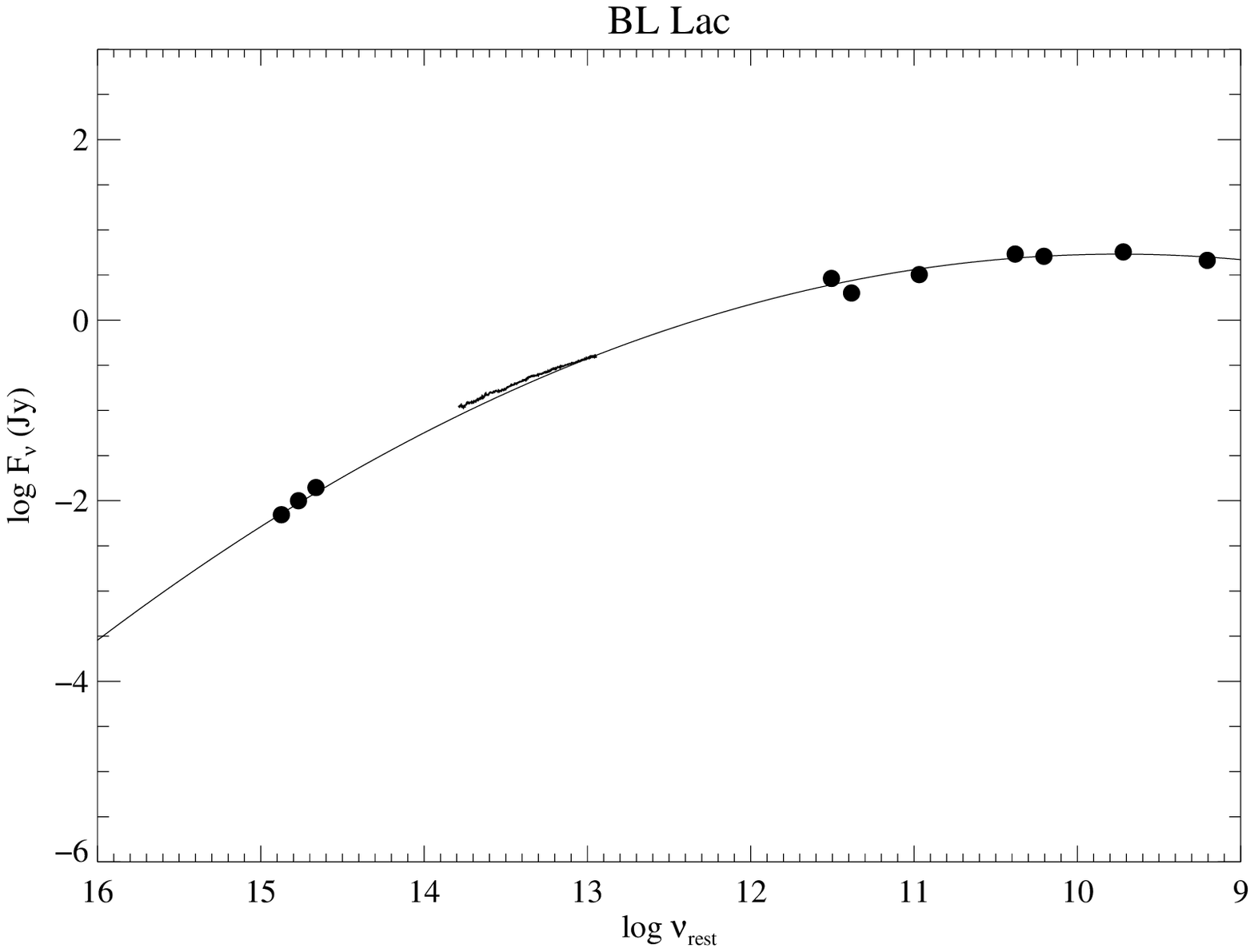}
\caption{{\it continued}.}
\end{figure*}

\begin{figure*}[t!]
\centering
\includegraphics[angle=0,scale=.44]{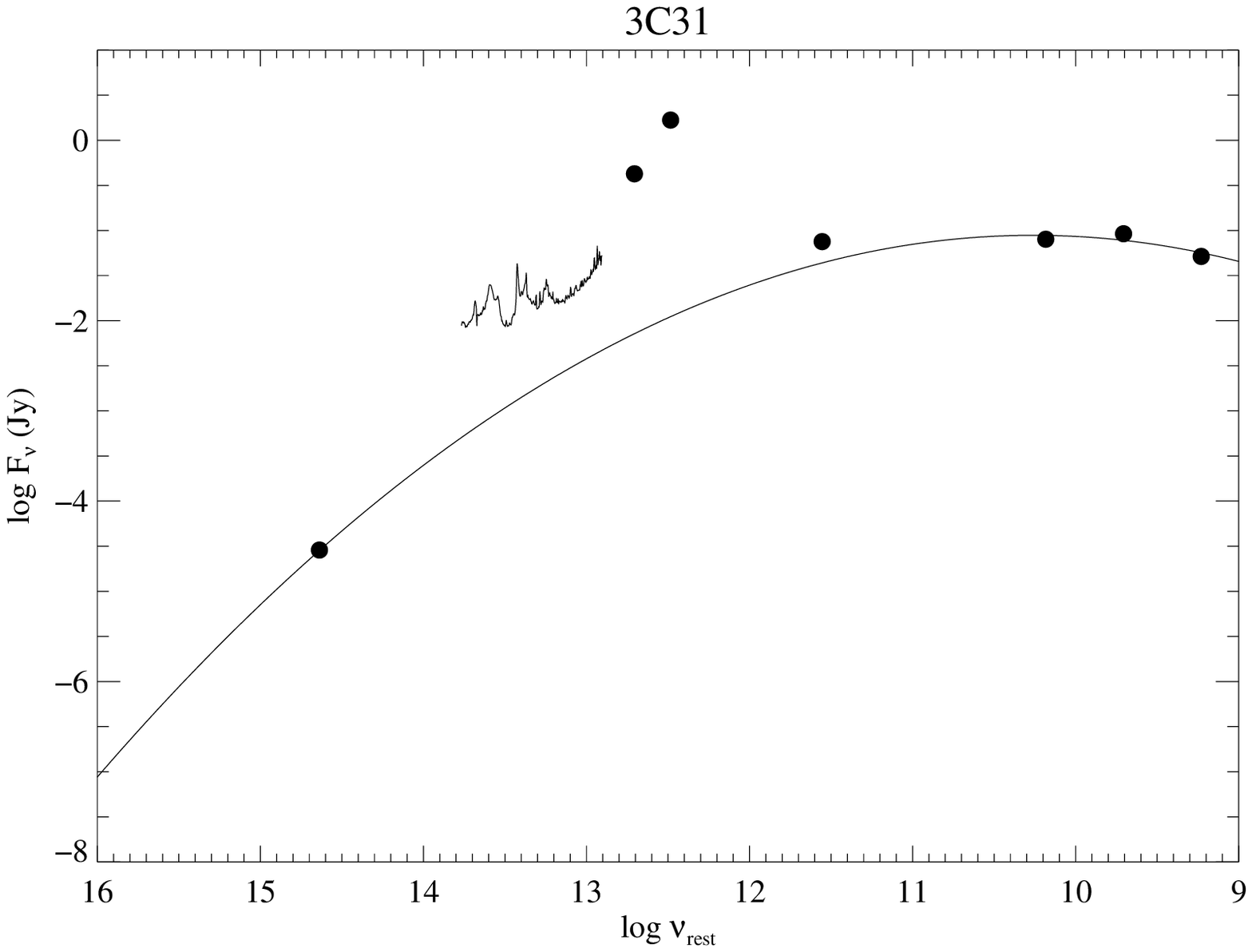}
\includegraphics[angle=0,scale=.44]{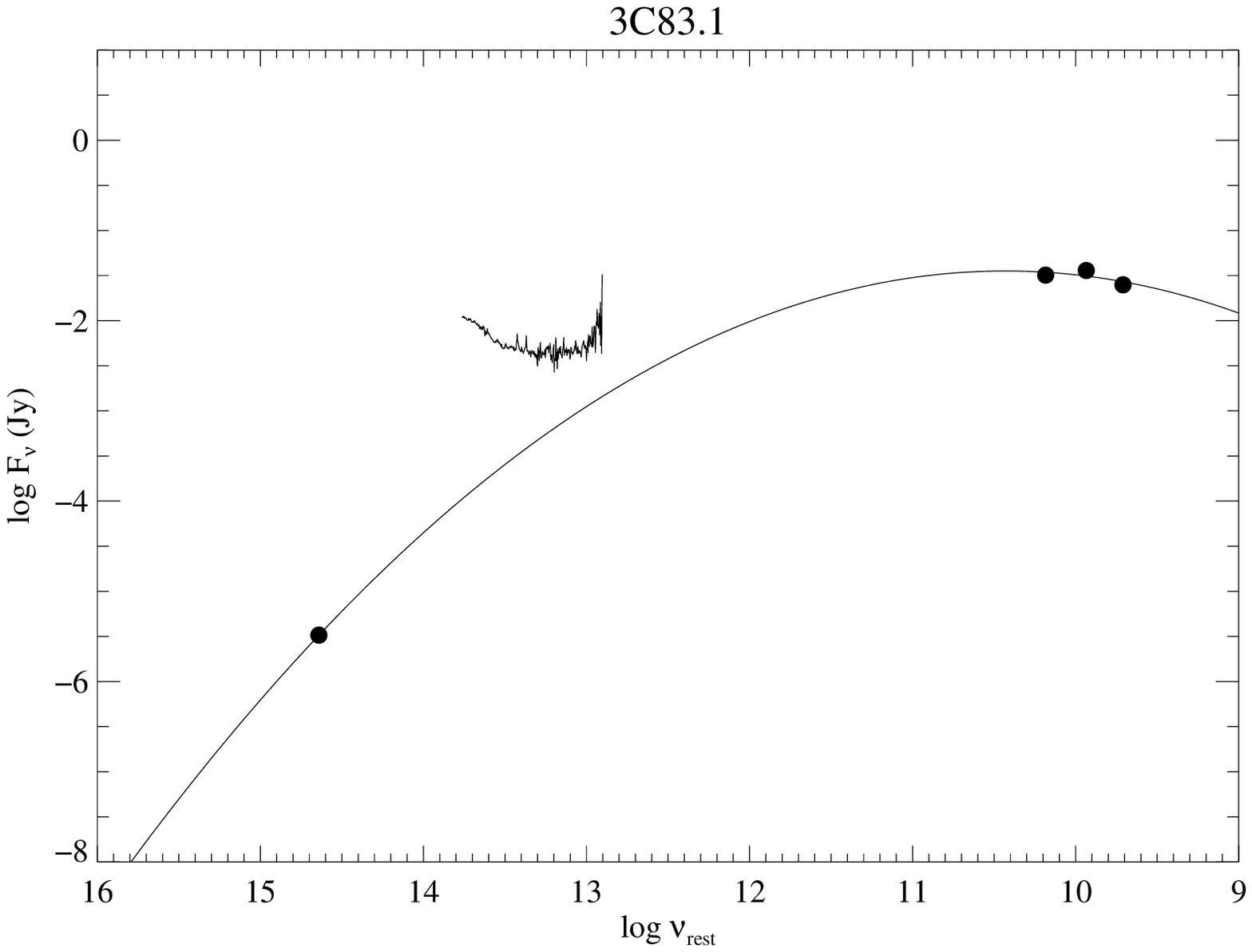}\\
\includegraphics[angle=0,scale=.44]{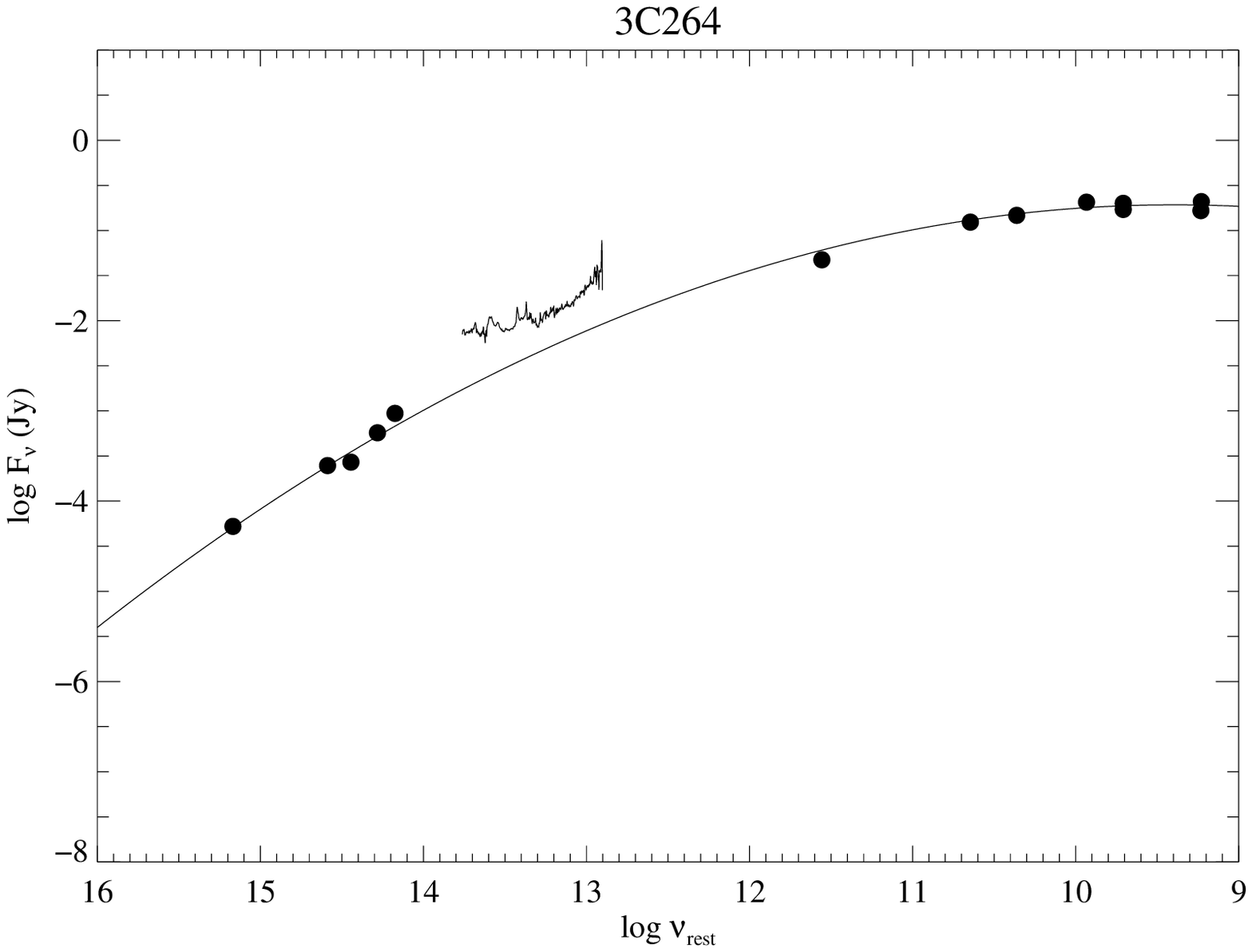}
\includegraphics[angle=0,scale=.44]{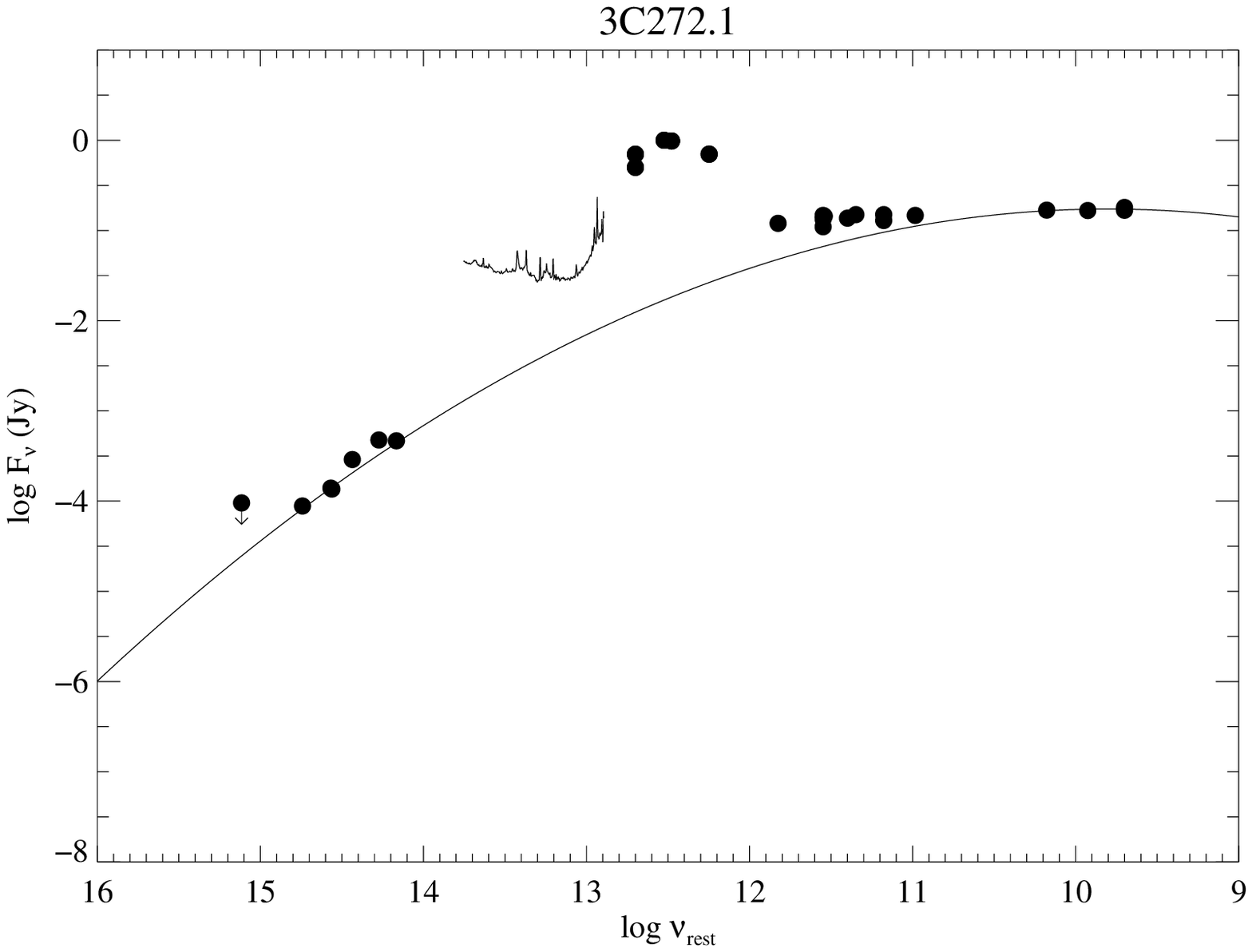}
\caption{Same as in  Fig.\,\ref{sed1} but for sources which are dominated by
  processes in the host galaxy (star formation and/or stellar
  emission). Here, we show the observed MIR spectra. \label{sed2}}
\end{figure*}

\subsection{Comparison with synchrotron prediction}

Since it has been proposed that non--thermal emission might be a
major, if not dominant constituent of the MIR emission in FR--I
sources, we also analyze the MIR spectra in the framework of their
(nuclear) SEDs.  As a comparison for the synchrotron core emission we
chose to use well observed blazars and related objects because they
represent nearly pure synchrotron emission.

It has been shown that, globally over a large frequency interval, the
strongly beamed core emission of blazars can be well approximated by a
parabolic function in log\,$F_{\nu}$ vs. ${\rm log}\,\nu$
\citep[e.g.][]{lan86}. Following these authors we here utilize a
function of the form

\begin{equation}
\label{equ1}
{\rm log}\,{F}_{\nu} = C + ({\rm log}\,\nu - B)^2 / 2A
\end{equation}

to represent the underlying beamed emission from the base of the jet
($C$ is the log of the peak flux, $B$ is the log of the peak frequency
of the parabola, and $A$ represents a ``curvature'' parameter). We
note that a similar parabolic function can also be used to described
the SEDs of BL Lacs in log\,$\nu\,F_{\nu}$ vs. ${\rm log}\,\nu$
\citep[e.g.][]{nie06}.

In order to construct the (nuclear) SEDs, photometric data from the
literature have been compiled from optical through radio
wavelengths. For the radio and optical we only focused on core
emission  and we only compiled these data for sources with detected
optical CCC components. In total our sample has 14 objects in common
with the sample of \citet{chia99}. But for two sources no CCCs could
be identified: for 3C293 \citet{chia99} could not estimate any CCC
flux due to large scale dust structures in the host galaxy, and 3C424
showed a ``radically different'' nuclear behavior (i.e., the FWHM of
the nuclear source is much larger than for other CCCs and it is
spatially resolved). This leaves 12 sources. In following works CCCs
were also detected in 3C15 (R. Baldi as well as M. Chiaberge, private
communication),  in 3C189 \citep{cap02}, in IC4296 \citep{bal06a}, and
in NGC\,6251 \citep{chia03}. Thus, 16 sources here have an optical CCC
detected (Tab.\,1).  Compared to other CCC objects, 3C386 stands out
in the properties of its optical CCC which is exceptionally bright
compared to the radio core. It has been suggested \citep{chia99},
supported by the tentative detection of a broad optical emission line
\citep{simp96} that this source shows additional optical  flux from a
Big Blue Bump component, i.e. showing the central core of a type--1
AGN.  However, it seems more likely that the foreground star which
falls right on top of the nucleus \citep{lyn71,mad06,but09} mimics an
optical CCC because we do not see any signs for such a type--1 AGN in
the MIR. In fact, after subtracting some emission due to star
formation the residual spectrum can be explained well by a quiescent
early--type galaxy (Fig.\,\ref{pahsub1}). In addition, new optical
spectroscopy does not confirm a broad  H$\alpha$ emission--line
component \citep{but09}. Because we cannot tell whether or not 3C386
itself shows a compact optical core we here exclude this object from
further analysis.  This leaves 15 sources where optical CCCs are
securely detected. Eleven of the 15 objects also show residual MIR
emission\footnote{Recall that if our methodology  is correct the
presence of residual emission means they contain AGN heated dust
and/or prominent synchrotron contributions, i.e. a nuclear source of
MIR continuum emission.}, while the spectra of four sources are
dominated by processes from the host galaxy  (stellar emission, star
formation).

We used the parabolic function to fit only the radio core data and the
optical CCCs which are both claimed to be synchrotron emission from
the same source \citep[e.g.][]{chia99,har00}. For these data we know
that they are nuclear and fairly robust while other measurements might
still be affected by  either intrinsic absorption (UV) or aperture
effects (MIR, FIR, sub--mm)\footnote{Because the core flux data were
compiled from the literature they are generally not obtained
simultaneously which can result in variability becoming important for
the quality of the fits. However, in most cases all the multi--epoch
data can be fitted well and any scatter present due to intrinsic
variability seems relatively small.}.

\begin{figure}[t!]
\centering
\includegraphics[angle=0,scale=.44]{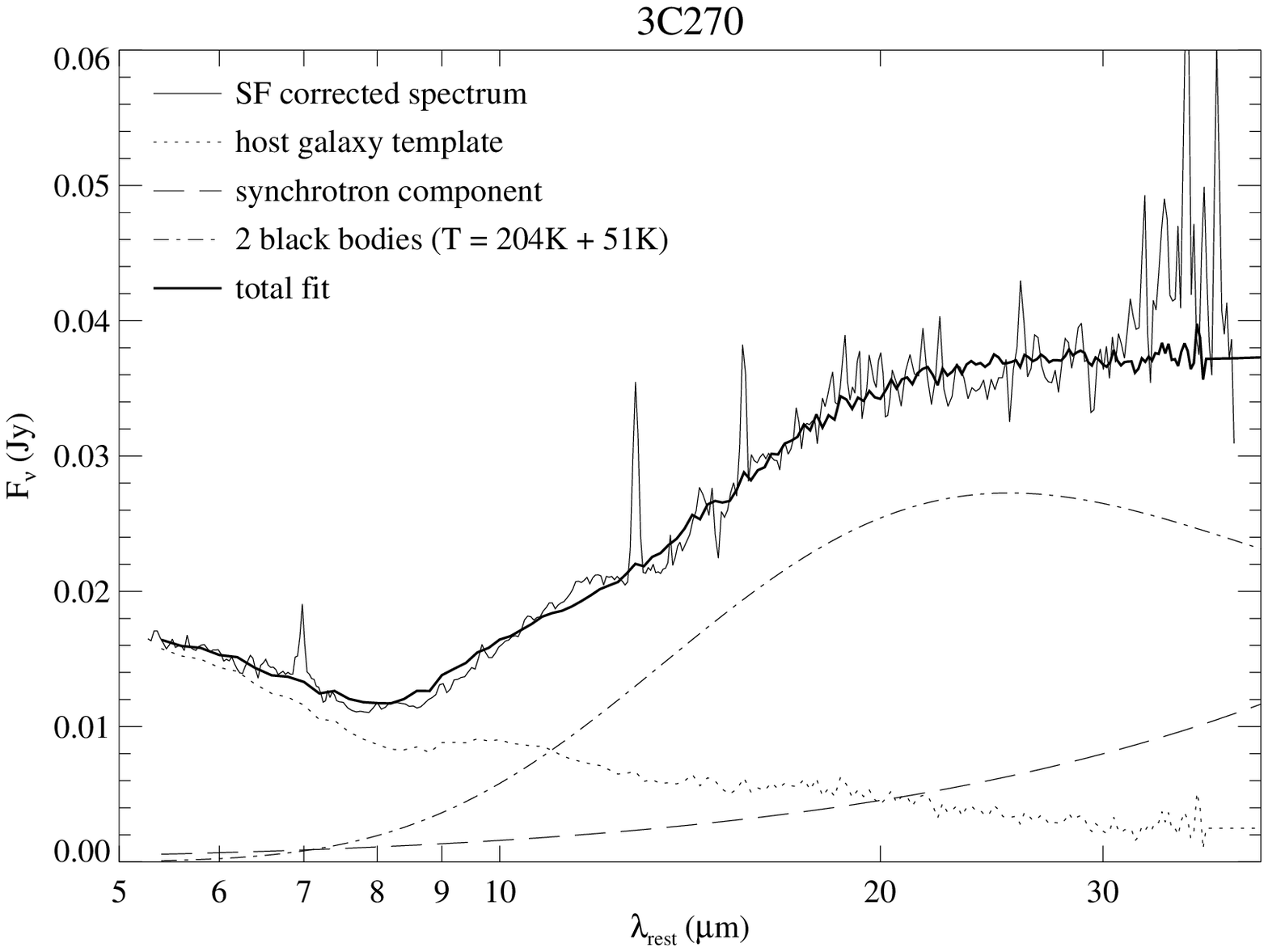}
\caption{The star--formation corrected spectrum of 3C270 is fitted with the spectrum of the 
  elliptical galaxy NGC\,1549 (dotted line), the synchrotron component from the core (dashed line), and two 
  black bodies (dot--dashed line; see text for details).\label{3c270_fit}}
\end{figure}

\subsubsection{Sources with residual MIR emission}

In the  following we will discuss the SEDs on an object to object
basis. References  to flux measurements used for the SED plots
(Fig.\,\ref{sed1}) in addition to those already given in
Tab.\,\ref{tab1} are also noted. In Tables \ref{tab2} and 
\ref{tab4} we give  the percentage contributions of the estimated
non--thermal component to the  observed MIR flux.

{\bf 3C15\quad} This source shows MIR continuum emission which is not
due to either star formation or stellar emission
(Fig.\,\ref{add_corr_spectra}).  Using optical and NIR CCC
measurements (provided by R. Baldi as well as M. Chiaberge, private
communication) and radio core data the non--thermal core component
clearly underestimates the observed MIR flux.

However, we point out that 3C15 has a strong arcsecond--scale jet
\citep{lea97} which is also detected in the optical \citep{mar98} and
in X--rays \citep{kat03}. Since the extended jet is included in both IRS slits
(SL slit oriented along the jet axis, LL slit oriented perpendicular to
the jet axis) it can potentially make additional non--thermal
contributions to the MIR spectrum. The inspection of the 2--D spectral
images from IRS does not reveal any secondary component in the SL
profile but the spectrum appears dominated by a single source (the
size of the jet is almost comparable to the resolution of {\it
Spitzer} at IRS  wavelengths).

At 8.4\,GHz the total radio flux of the arcsecond jet is greater than the core
flux by a factor of $\sim$\,10 \citep{lea97,har98}. Assuming a
typical jet spectral index of $\alpha$\,$\sim$\,$-0.6$ \citep{bri84}  this
would lead to a non--thermal jet contribution of 2\,mJy and 3\,mJy at
15\,$\mu$m and 30\,$\mu$m, respectively, in the case of a single power
law.  For this conservative estimate we used a spectral index which is
quite flat for a jet. In fact, the radio--optical spectral index
of the X--ray brightest jet knot (``Knot C'',~\citealt{kat03}) is
argued to be steeper than our chosen value \citep{kat03,dul07}.

Thus, even under the assumption of a conservatively flat spectral index
for the  jet the total non--thermal contribution (core + jet) to the
MIR emission within the IRS apertures is still a factor of $\geq$\,2
smaller than the total observed emission. In addition, the spectrum
itself shows a change in slope around $\sim$\,20\,$\mu$m (which could
be identified with a 18\,$\mu$m silicate feature) which argues for
thermal contributions to the total flux.

Considering the appearance of the spectrum (including the potential
18\,$\mu$m silicate feature), the absence of notable PAH emission, and
the difference between total flux and (conservatively estimated)
non--thermal contributions we therefore argue that 3C15 shows thermal
MIR emission from warm nuclear dust, possibly heated by an AGN.  

We also note that \citet{rin05} observe the X--ray core of 3C15 to be 
fairly weak ($L_{2-10\,{\rm keV}}$\,$\sim$\,$5\times10^{40}$\,erg/s intrinsic) but 
absorbed by a significant column density of $N_{\rm H}$\,$\sim$\,$9\times10^{22}$\,cm$^{-2}$.
{\it References:} \citet{dic08,har98,mor97}

{\bf 3C29\quad}The parabolic fit matches the radio data quite well but
greatly overpredicts the nuclear UV measurement.  Assuming that this is due to
extinction, dereddening of the UV flux in order to match it with the
synchrotron fit would also result in increasing the optical flux
somewhat.  Such a synchrotron SED could in principle account for the
small excess in the MIR spectrum.  In fact, \citet{chia02}
note that the UV point in  3C29 might be absorbed by a thin, extended
dust lane. However, the mismatch of the UV measurement could also
arise from  variability. While the optical and UV data 
were obtained more than 5 years apart, the difference between the
synchrotron estimate and the actual UV measurement corresponds to a
factor of $\sim$\,30 in flux. This seems high to be solely due to
variability. Considering that some absorption in the  UV/optical
bands is present, that the slope of the MIR spectrum agrees reasonably well
with the synchrotron fit, and that no host corrections were applied
the MIR emission in this source is likely to be mostly non--thermal.
{\it References:} \citet{mor97,ric06}


{\bf 3C66B\quad}Here all data points (including the UV and MIR) agree
very well with the parabolic fit.  Thus, there does not seem to be
significant extinction present in this object and the residual MIR
emission is very likely to be non--thermal. At the shortest
wavelengths of the MIR spectrum the stellar
population of the host galaxy can be seen.  3C66B has an
arcsecond--scale jet \citep{har96} which is also seen in the
optical \citep{but80,mac91,per06}, in  X--rays \citep{har01}, and even
at MIR wavelengths a detection with {\it ISO} is claimed \citep{tan00}. However, unlike  the
case of  3C15, the jet in 3C66B is weaker than the core. \citet{har01}
fit the radio through X--ray  jet spectrum with a broken power--law
model where the steepening from $\alpha$\,$\sim$\,$-0.5$ to 
$\alpha$\,$\sim$\,$-1.35$ occurs  in the infrared. The radio--IR spectral slope agrees
reasonably with the 14.5\,$\mu$m MIR jet flux of $\sim$\,1.7\,mJy
estimated by \citet{tan00}. Considering these values the jet
contribution in the MIR is only $<$\,50\% of that of the core. Thus,
the inclusion of the jet in the IRS spectral aperture does not alter
our conclusion for the dominance of non--thermal MIR emission in
nucleus of 3C66B. {\it References:} \citet{jack93,xu00,quil03,kharb05}


{\bf 3C84\quad}While this source is strongly core dominated in the
radio, a clear dust emission bump is observed which reaches from the
FIR well into the MIR. The SED is very complex and suffers from
resolution effects  (sub--mm vs. VLA vs. VLBA), which reduce the
quality of the  synchrotron core fit. In addition the VLBA data
suggest self absorption in the  core making the SED asymmetric which
cannot be fitted well with a parabolic function.  Therefore, the fit
was limited to the optical core and the radio core measurements
excluding  the 5\,GHz data where the downturn in the SED is already
quite prominent. Besides the resolution  effects and the self absorbed
core, variability  might also be important and the fit cannot be more
than a rough approximation  of the synchrotron core component in this
case. However, the thermal MIR and FIR bump is  very pronounced in
3C84 and variations in the underlying synchrotron component are almost
negligible at these wavelengths.

The MIR spectrum itself does not show significant contribution from
stellar emission, only minor indications for star formation, and
overall has a spectral appearance quite similar to ``thermal'' spectra
of AGN powered sources. The fact that the spectrum is on average
almost an order of magnitude brighter than the underlying non--thermal
core emission strengthens the argument for thermal dust emission in
3C84, powered by a central engine. In addition, \cite{ho97b} find
broad emission--line components in H$\alpha$ and other permitted
lines. {\it References:} \citet{kna91,quil03,haas04,sti04,tay06}


{\bf 3C189\quad}In the observed MIR spectrum this source is very similar to 3C66B
(Fig.\,\ref{obs_spectra}). Also  in the context of the SED we
see that the fit to the synchrotron core matches with the MIR
emission in shape as well as in flux. An {\it IRAS} 100\,$\mu$m flux
measurement indicates  the presence of cool (possibly extended) dust. 
{\it References:} \citet{bri78,kna90,gio94}

{\bf 3C270\quad}Including nuclear NIR measurements as well as (large
aperture) FIR and (sub)--mm data, the SED has very good coverage. We
see that the radio core data, the NIR, and the  (sub)--mm data fit
well with the synchrotron prediction, arguing for their common non--thermal
origin.  The MIR spectrum, however, shows a very clear excess over the
underlying non--thermal emission. Not only does the excess in MIR flux indicate
additional thermal emission from warm dust in this source, but also
the shape of the {\it Spitzer} spectrum makes its nature as
non--thermal emission very unlikely. The pronounced change of slope
around $\lambda$\,$\sim$\,$20\,\mu$m rest wavelength (Fig.\,\ref{3c270})
strongly suggest thermal emission  from warm dust to be present in
this source. 

In fact, after subtracting the star--forming template from the
spectrum the MIR continuum can be well fitted as a
combination of the underlying synchrotron contribution and the
elliptical template for the host galaxy, plus a black body with a
temperature of $\sim$\,200\,K (Fig.\,\ref{3c270_fit})\footnote{The fit
to the star--formation corrected spectrum has only 3 variable
parameters: the scaling of the elliptical template as well as the
temperature of the two black bodies. Because the synchrotron component
was determined independently from radio/optical core measurements its
flux was held fixed. We note that the colder black body is not well
constrained but including it improves the fit for
$\lambda$\,$>$\,20\,$\mu$m. However, the presence or absence of this second
black body leaves the temperature of the warmer component virtually
unchanged as it is strongly constrained by the continuum between 10
and 20\,$\mu$m.}.  This emission from warm dust is not powered by star
formation because this would result in accompanying PAHs. At
15\,$\mu$m the continuum emission of this residual thermal MIR component has a
luminosity of $\nu L_{\nu, 15\,\mu{\rm m}}$\,$\sim$\,$4 \times10^{41}$
erg\,s$^{-1}$ (see also Tab.\,\ref{tab4}).

In the optical, \citet{bar99} report a broad H$\alpha$ line in polarized
light. This (tentative) detection further supports the presence of a
classical hidden AGN. Moreover, a geometrically thick nuclear dust
disc has been observed in silhouette \citep{jaf93}. We also note that
the optical emission--lines in the nuclear region have flux ratios
expected for Seyferts (\citealt{fer96}; see \S4.8).

Several studies are in broad agreement on the
basic X--ray properties like the moderate intrinsic luminosity of
$L_{2-10\,{\rm keV}}$\,$\sim$\,$1\times10^{41}$\,erg/s and the moderate
absorption  on the order of $N_{\rm H}$\,$\sim$\,$5\times10^{22}$\,cm$^{-2}$
\citep{chia03,gli03,sam03,rin05,zez05}. The origin of the observed
X--ray emission is, however, controversially discussed: While
\citet{chia03} interpret the X--ray power--law component as due to the
jet, \citet{gli03} argue in favor of an accretion flow as the most
likely source for the  bulk of the X--ray emission. On the other hand,
\citet{zez05} state that both processes are able to explain  the
observed X--ray properties. A Fe K$\alpha$ line is (marginally) detected 
with an equivalent width of $\sim$\,230\,eV
\citep{gli03,sam03,rin05} which is consistent with the upper limits
from other studies \citep[e.g.][]{chia03,zez05}.

While the warm dust might represent the same nuclear absorbing
material detected in X--rays we note that due to the large {\it Spitzer} slit
width our spectra include the prominent  $\sim$\,300\,pc nuclear dust
disc. In principle the 200\,K dust observed in the MIR could
correspond to this dust disc. Because in situ star formation as a
heating mechanism can be largely excluded (no PAHs) another source of sufficient
energy needs to be present in the nucleus of 3C270. Radiation from a
(possibly hidden) active nucleus is an obvious possibility. 
However, in the case of heating by the low--to--medium luminosity 
AGN in 3C270 simple energy budget 
arguments strongly suggest the $\sim$\,200\,K dust to reside on much smaller 
scales than $\sim$\,300\,pc.
{\it References:}
\citet{jon97,cap00,jon00,quil03,haas04,kharb05}


{\bf 3C274\quad} Interestingly, the slope of the {\it Spitzer}
spectrum appears flatter than what one would expect  according to the
synchrotron core fit. After accounting for host galaxy light
using the elliptical template the MIR continuum  alone can be well
fitted with a single power law of $\alpha$\,$\sim$\,$-0.80$ for
wavelengths shorter  than $\sim$\,$25\,\mu$m. At longer wavelengths
the spectral slope steepens with respect to the  power law which might
be due to an additional dust component \citep{per07}.  As discovered
by \citet{why01} and later discussed by \citet{per01} and
\citet{why04}, an unresolved   $\sim$\,$0.5\,\arcsec$ MIR core is
present in 3C274. The MIR spectrum is discussed in  detail by
\citet{per07}.  {\it References:}
\citet{kna91,xu00,nag01,quil03,haas04,lis05,per01,per07,shi07a,why04}


\begin{table*}[t!]
\setlength{\tabcolsep}{1.7mm}
\begin{center}
\caption{Estimated contributions to the MIR spectra for sources without detected AGN dust components.\label{tab2}}
\begin{tabular}{l|r@{.}lccc|r@{.}lccc|l}
\tableline\tableline
Object    & \multicolumn{5}{c|}{F$_{15\,\mu{\rm m}}$ (rest frame)} & \multicolumn{5}{c|}{F$_{30\,\mu{\rm m}}$ (rest frame)} & \\
          & \multicolumn{2}{c}{total\tablenotemark{a}}         & SF   & stars & synch\tablenotemark{b} & \multicolumn{2}{c}{total\tablenotemark{a}}         & SF   & stars & synch\tablenotemark{b} & comment              \\
          & \multicolumn{2}{c}{mJy} & \% & \% & \% &   \multicolumn{2}{c}{mJy} & \% & \% & \% & \\               
\tableline
\multicolumn{12}{c}{sources without optical core measurements}\\
\tableline
3C76.1    &      1&72 & \hspace*{0pt}  21 & \hspace*{2pt} 0 & \nodata           &         3&53 & \hspace*{0pt}  30 & \hspace*{0pt}  0 & \nodata & low S/N spectrum  \\              
3C129     &      3&26 & \hspace*{0pt}  17 & \hspace*{0pt}46 & \nodata           &        11&44 & \hspace*{0pt}  14 & \hspace*{0pt}  6 & \nodata &   \\              
3C218     &      4&49 & \hspace*{-1pt}100 & \hspace*{2pt} 0 & \nodata           &        15&91 & \hspace*{-2pt}100 & \hspace*{0pt}  0 & \nodata & SF dominated spectrum  \\              
3C293     &     20&10 & \hspace*{-1pt}100 & \hspace*{2pt} 0 & \nodata           &        57&98 & \hspace*{-2pt}100 & \hspace*{0pt}  0 & \nodata & SF dominated spectrum \\             
3C386     &      1&97 & \hspace*{0pt}  16 & \hspace*{0pt}74 & \nodata           &         1&78 & \hspace*{0pt}  52 & \hspace*{-2pt}37 & \nodata & foreground star mimics CCC \\              
3C403.1   & $<$\,1&34 & \nodata           & \nodata         & \nodata           &    $<$\,1&53 & \nodata           & \nodata          & \nodata & not detected \\             
3C424     &      1&52 & \nodata           & \hspace*{0pt}12 & \nodata           &         5&06 & \nodata           & \hspace*{0pt}  2 & \nodata &  \\             
\tableline
\multicolumn{12}{c}{sources with optical core measurements}\\
\tableline
3C29      &      2&48 & \nodata           & \hspace*{0pt}15 & \hspace*{0pt}  58 &         8&16 & \nodata           & \hspace*{0pt}  2 & \hspace*{0pt}  42 & peak--up spill--over\tablenotemark{c}, UV abs.\\
3C31      &     13&99 & \hspace*{0pt}  87 & \hspace*{2pt} 0 & \hspace*{0pt}  13 &        29&30 & \hspace*{0pt}86   & \hspace*{0pt}  0 & \hspace*{0pt}  14 & SF dominated spectrum \\               
3C66B     &      4&76 & \nodata           & \hspace*{0pt}21 & \hspace*{0pt}  79 &         8&72 & \nodata           & \hspace*{0pt}  5 & \hspace*{0pt}  85 & \\              
3C83.1    &      4&08 & \hspace*{0pt}  18 & \hspace*{0pt}64 & \hspace*{0pt}  12 &         5&37 & \hspace*{0pt}40   & \hspace*{-2pt}22 & \hspace*{0pt}  22 & low PAH ratio \\              
3C189     &      4&80 & \nodata           & \hspace*{0pt}20 & \hspace*{0pt}  71 &         8&73 & \nodata           & \hspace*{0pt}  5 & \hspace*{0pt}  79 & cold dust\\             
3C264     &      9&11 & \hspace*{0pt}  52 & \hspace*{2pt} 0 & \hspace*{0pt}  48 &        23&88 & \hspace*{0pt}67   & \hspace*{0pt}  0 & \hspace*{0pt}  33 & SF dominated spectrum \\              
3C272.1   &     28&08 & \hspace*{0pt}  30 & \hspace*{0pt}57 & \hspace*{0pt}  13 &        46&70 & \hspace*{0pt}53   & \hspace*{-2pt}15 & \hspace*{0pt}  15 & cold dust, low PAH ratio \\              
3C274     &     43&25 & \nodata           & \hspace*{2pt} 7 & \hspace*{0pt}  93 &        81&38 & \nodata           & \hspace*{0pt}  2 & \hspace*{0pt}  98 & \\              
IC\,4296  &     10&02 & \hspace*{2pt}   6 & \hspace*{0pt}30 & \hspace*{0pt}  69 &        19&88 & \hspace*{2pt} 9   & \hspace*{0pt}  7 & \hspace*{0pt}  68 & cold dust \\             
3C317     &      2&90 & \nodata           & \hspace*{0pt}16 & \hspace*{0pt}  50 &         5&86 & \nodata           & \hspace*{0pt}  3 & \hspace*{0pt}  57 & \\             
3C465     &      3&08 & \nodata           & \hspace*{0pt}19 & \hspace*{0pt}  68 &         7&47 & \nodata           & \hspace*{0pt}  3 & \hspace*{0pt}  59 & cold dust\\
BL Lac    &    246&45 & \nodata           & \hspace*{2pt} 0 & \hspace*{-2pt}100 &       381&92 & \nodata           & \hspace*{0pt}  0 & \hspace*{-2pt}100 & \\   
\tableline
\end{tabular}
\tablenotetext{a}{Flux measured from the observed MIR spectrum.}
\tablenotetext{b}{Flux of the synchrotron component shown in
Figs.\,\ref{sed1} and \ref{sed2}.}   
\tablenotetext{c}{Due to the peak--up spill-over the spectral slope and flux at short wavelengths might be compromised. This 
adds some uncertainty to the scaling of the stellar template.  }
\tablecomments{The contributions
of the individual components do not necessarily add up to
100\,\%. This is due to the nature of our appraoch which uses  fixed
templates in a step--by--step manner in order to identify and correct
successively for their contributions to the observed spectra. This
naturally leaves room for variations between individual objects which
can manifest as residuals or slight overcorrections. Also, components
not considered might have an effect (e.g. cold dust, particularly at
$30\mu{\rm m}$). In addition, for sources with significant
contributions from non--thermal emission variations in the nuclear
components and obscuration for UV/optical data might result in some
mismatch and residual emission. See text for comments on individual
sources where this might be the case.}
\end{center}
\end{table*}

{\bf IC4296\quad}For IC4296 the synchrotron fit falls very close to
the MIR spectrum, but only for a very narrow wavelength interval
($\lambda$\,$\sim$\,$20-30\mu$m).  In this source a star--forming
template has been subtracted because PAHs are clearly detected and the
spectrum seems to be dominated by stellar emission shortwards of
$\lambda$\,$\sim$\,$10\,\mu$m (Fig.\,\ref{pahsub2}).  Above 25\,$\mu$m
a considerable steepening is observed in the spectrum suggesting the
presence of a thermal bump in the SED due to cooler dust which is
supported by FIR measurements.  Despite the influence of processes in
the host galaxy at some IRS wavelengths, this source has a very strong
non--thermal contribution of greater than $>$\,60\% at 15\,$\mu$m and
30\,$\mu$m (Tab.\,\ref{tab2}). Although the  synchrotron signature is
weaker than for sources like 3C66B or 3C189 it manifests as clear
residual  MIR emission.  Interestingly, \citet{pel03} argue that the
X--ray core in IC4296  ($L_{2-10\,{\rm
keV}}$\,$\sim$\,$1.6\times10^{41}$\,erg/s intrinsic) originates  from
the jet and the X--rays are observed to be only mildly absorbed
($N_{\rm H}$\,$\sim$\,$1\times10^{22}$\,cm$^{-2}$).  {\it References:}
\citet{kil86,kna89,mor97,ven00,pel03,ric06}


{\bf 3C317\quad}For this source the MIR spectrum falls very
close to the  synchrotron fit in flux. The synchrotron curve
overestimates  the nuclear UV measurement somewhat and a similar
situation of absorption intrinsic to the source or variability as in 3C29 could be
imagined. In fact, \citet{chia02}  reported a factor of $\sim$\,$10$ in
variability for the nuclear UV point source.  Thus, the discrepancy of
the CCC measurements (and the offset in MIR flux) is likely to be caused 
by variability. The similarity in spectral shape and the
reasonable match in flux (combined with the absence of star--formation tracers) strongly
suggest the dominance of non--thermal emission in 
the MIR. {\it Spitzer} FIR measurements at 70\,$\mu$m suggest the presence of cold dust 
in this object.
{\it References:} \citet{zhao93,ven00,quil03,quil08}

{\bf NGC\,6251\quad}Using UV, optical, and NIR measurements
\citep{chia03} and multi--frequency  radio core measurements
\citep[VLBI,][]{eva05}, a well sampled nuclear SED can be constructed
for NGC\,6251. As for other sources, variability might be an issue for
the non--thermal core measurements and in fact variability with a
factor of $\sim$\,2 over a few years has been reported for the nuclear
(VLBI) radio component \citep{eva05}. However, the radio measurements
shown here were obtained  simultaneously and the fit also agrees well
with most of the multi--epoch {\it HST}  measurements.  For this
source the MIR spectrum shows a clear excess over the synchrotron core
fit, even after  the correction for some minor star formation as
traced by PAH emission of small equivalent width
(Fig.\,\ref{obs_spectra}). The small contribution from star formation,
the absence of signs for significant contributions from stellar
emission, and the overall shape of the spectrum (including the strong
silicate emission features) strongly suggests a considerable component
of thermal MIR emission in the nuclear regions of NGC\,6251.  The
spectrum also looks quite different from other sources which are
dominated by non--thermal emission in the MIR (e.g. 3C66B, 3C189 in
Fig.\,\ref{sed1}).

At 30\,$\mu$m, where the strong silicate emission features will have
only a minor impact on the continuum luminosity
\citep[e.g.][]{spo07}, we measure $\nu L_{\nu} = 5.6\times10^{42}$
erg\,s$^{-1}$ in the spectrum.  On the other hand, the luminosity of
the synchrotron component at this wavelengths  is about  $\nu
L_{\nu}$\,$\sim$\,$2.2\times10^{42}$ erg\,s$^{-1}$, which is
consistent with  the  $\nu L_{\nu}$\,$\sim$\,$2.5\times10^{42}$
erg\,s$^{-1}$ given by the synchrotron self--Compton SED fit of
\citet{chia03}. We note however, that the factor of $\sim$\,2.5
luminosity difference between the synchrotron core and the MIR
spectrum is of the order of the measured radio core variability.  In
principle, the excess of the spectrum could be accounted for by
variability of the underlying synchrotron core component. Although
this would change the thermal to non--thermal ratio in the MIR (with
the non--thermal part potentially even dominating at times) the
appearance of the MIR spectrum still argues for the existence  of warm
nuclear dust in NGC\,6251. 

\citet{fer99} report the tentative
detection of  a broad component in H$\alpha$ (this result seems to be
strongly dependent  on the assumed modelling of the narrow lines
though). In X--rays the power--law component in NGC\,6251 shows only very
little absorption ($N_{\rm H}$\,$\sim$\,$5\times10^{20}$\,cm$^{-2}$)
and  has an intrinsic luminosity of $L_{2-10\,{\rm
keV}}$\,$\sim$\,$4.8\times10^{42}$\,erg/s  \citep{gli04,eva05}.  The
additional detection of a broad Fe K$\alpha$ line (EW\,$\sim$\,220\,eV
and FWHM\,$\sim$\,0.6\,keV) led \citet{gli04} to argue in favor of the
presence of a standard accretion disk. In their study the base of the
jet  is ruled out as the sole origin of the nuclear X--ray
emission. \citet{eva05} on the other hand question the significance of
the K$\alpha$ detection and  favor the jet as the source for the bulk
of the observed X--ray core emission. However,  they do not exclude
some contributions from an accretion flow.  {\it References:}
\citet{gol88,chia03,quil03,eva05}

{\bf 3C465\quad} The observed spectrum falls very close to the
synchrotron core fit. Assuming that the observed nuclear UV
component is also non--thermal, significant absorption seems
to be present in this source. Alternatively, variability of a factor of 
$\sim$\,10 could cause this UV flux difference. As noted before,
both processes can in principle account for the remaining offset between the
MIR spectrum and the synchrotron fit. A steepening of the spectrum is
observed for wavelengths larger than $\sim$\,30\,$\mu$m indicating a
thermal bump at FIR wavelengths. This is supported by {\it IRAS} 60 and 
100\,$\mu$m fluxes. 
{\it References:} \citet{gol88,ven95,quil03,jet06}

{\bf BL Lac\quad}Using quasi--simultaneous flux measurements from
\citet{lan83} we construct an SED of this object, which is here used 
as an example for a strongly aligned, highly beamed FR--I source. Note that 
although no {\it HST} data were used the (ground--based) optical fluxes 
will be dominated by the core emission. The measurements can be well fit by the 
parabolic function and even though the MIR spectrum was obtained at 
a different epoch, it fits extremely well with the estimate for the synchrotron 
core emission. The spectrum can be fitted with a power law with $\alpha$\,$\sim$\,$-0.7$.

{\it References:} \citet{lan83}

\begin{table*}
\setlength{\tabcolsep}{1.3mm}
\begin{center}
\caption{Estimated contributions to the MIR spectra for sources with detected AGN dust components (see notes to Tab.\,\ref{tab2}).\label{tab4}}
\begin{tabular}{l|r@{.}lcccc|r@{.}lcccc|l}
\tableline\tableline
Object    & \multicolumn{6}{c|}{F$_{15\,\mu{\rm m}}$ (rest frame)} & \multicolumn{6}{c|}{F$_{30\,\mu{\rm m}}$ (rest frame)} & \\
          & \multicolumn{2}{c}{total}  & SF  & stars & synch & AGN\tablenotemark{a} & \multicolumn{2}{c}{total}       & SF & stars & synch & AGN\tablenotemark{a} & comment              \\
          & \multicolumn{2}{c}{mJy} & \% & \% & \% & \% &   \multicolumn{2}{c}{mJy} & \% & \% & \% & \% & \\               
\tableline
3C15      &      5&52 & \nodata & \hspace*{0pt}19 & 14      & 67 &         9&15 & \nodata         & 5 & \hspace*{0pt}18 & 77 &  peak--up spill--over\tablenotemark{b}\\
3C84      &   1307&88 &       1 & \hspace*{2pt} 0 & 12      & 86 &      3859&70 & \hspace*{2pt} 1 & 0 & \hspace*{2pt} 9 & 90 &  \\              
3C120     &    345&79 &       2 & \hspace*{2pt} 0 & \nodata & 98 &       653&08 & \hspace*{0pt}11 & 0 & \nodata         & 89 & type--1 AGN  \\              
3C270     &     27&45 &       7 & \hspace*{0pt}21 & 11      & 61 &        52&14 & \hspace*{0pt}27 & 5 & \hspace*{0pt}16 & 52 &  \\              
NGC\,6251 &     27&88 &       3 & \hspace*{2pt} 0 & 29      & 68 &        49&68 & \hspace*{2pt} 5 & 0 & \hspace*{0pt}33 & 62 &  \\              
E1821+643 &    333&78 &       2 & \hspace*{2pt} 0 & \nodata & 98 &       540&74 & \hspace*{2pt} 9 & 0 & \nodata         & 91 & type--1 AGN  \\              
\tableline
\end{tabular}
\tablenotetext{a}{The contribution of the AGN component (which dominates in these objects) was 
determined as the total flux minus the other three contributions considered here. }
\tablenotetext{b}{See comment on 3C29 in Tab.\,\ref{tab2}.}
\end{center}
\end{table*}

\subsubsection{Sources without residual MIR emission}

In the following we will present the sources with detected CCCs but
where no significant residual MIR continuum emission can be
identified. The observed spectra can be explained by star formation
and/or stellar emission. The SED plots are shown in
Fig.\,\ref{sed2}.

{\bf 3C31\quad}This source is strongly dominated by star formation in
the host galaxy (Fig.\,\ref{obs_spectra}). Its observed spectrum is
almost an order of magnitude brighter than then underlying
non--thermal component. Interestingly, the SL aperture for the  MIR
spectrum is roughly consistent with the size of the molecular  disk
observed in the center of this source \citep{oku05}. {\it IRAS} 
60\,$\mu$m and 100\,$\mu$m measurements  indicate substantial amounts
of colder FIR dust.  {\it References:} \citet{bur77,xu00,eva05a}


{\bf 3C83.1\quad}The few existing data points for the core emission
are  well approximated by the parabolic fit. The total light spectrum
is dominated by  processes in the host galaxy and lies well above the
synchrotron fit. The subtraction of the star--forming template leaves
a negative residual at the location of the 7.7\,$\mu$m PAH feature
(Fig.\,\ref{pahratio}). As discussed above, this  cannot be
accounted for by the stellar spectrum of the host galaxy and argues
for a low 7.7\,$\mu$m to 11.3\,$\mu$m PAH ratio which might in
fact  be powered by low--luminosity AGN activity instead of star
formation (see \S4.1).  {\it References:} \citet{odea86,xu99}


{\bf 3C264\quad}While the observed spectrum of this source is
dominated by star formation, it falls much closer to the synchrotron
component than in 3C31.  From Fig.\,\ref{sed2} we see that the
non--thermal contribution might not be negligible
(see also Tab.\,\ref{tab2}). However, without prior knowledge of
this fact the source spectrum can be fitted by star formation only due
to the wide variety in the MIR spectral properties of star--forming
galaxies \citep[e.g.][]{dale05,bra06,smi07}.

We note that this source has an  arcsecond--scale radio jet
\citep[e.g.][]{lara04} which is also detected at optical  wavelengths
\citep{cra93,baum97}. However, in the radio the core is stronger than
the total jet emission and has a significantly flatter spectrum
\citep{lara04}.  While the jet emission steepens even more at optical
wavelengths the contributions  in the MIR are negligible compared to
the core.  {\it References:} \citet{xu00,lara04,kharb05}


\begin{figure*}[t]
\includegraphics[angle=0,scale=.44]{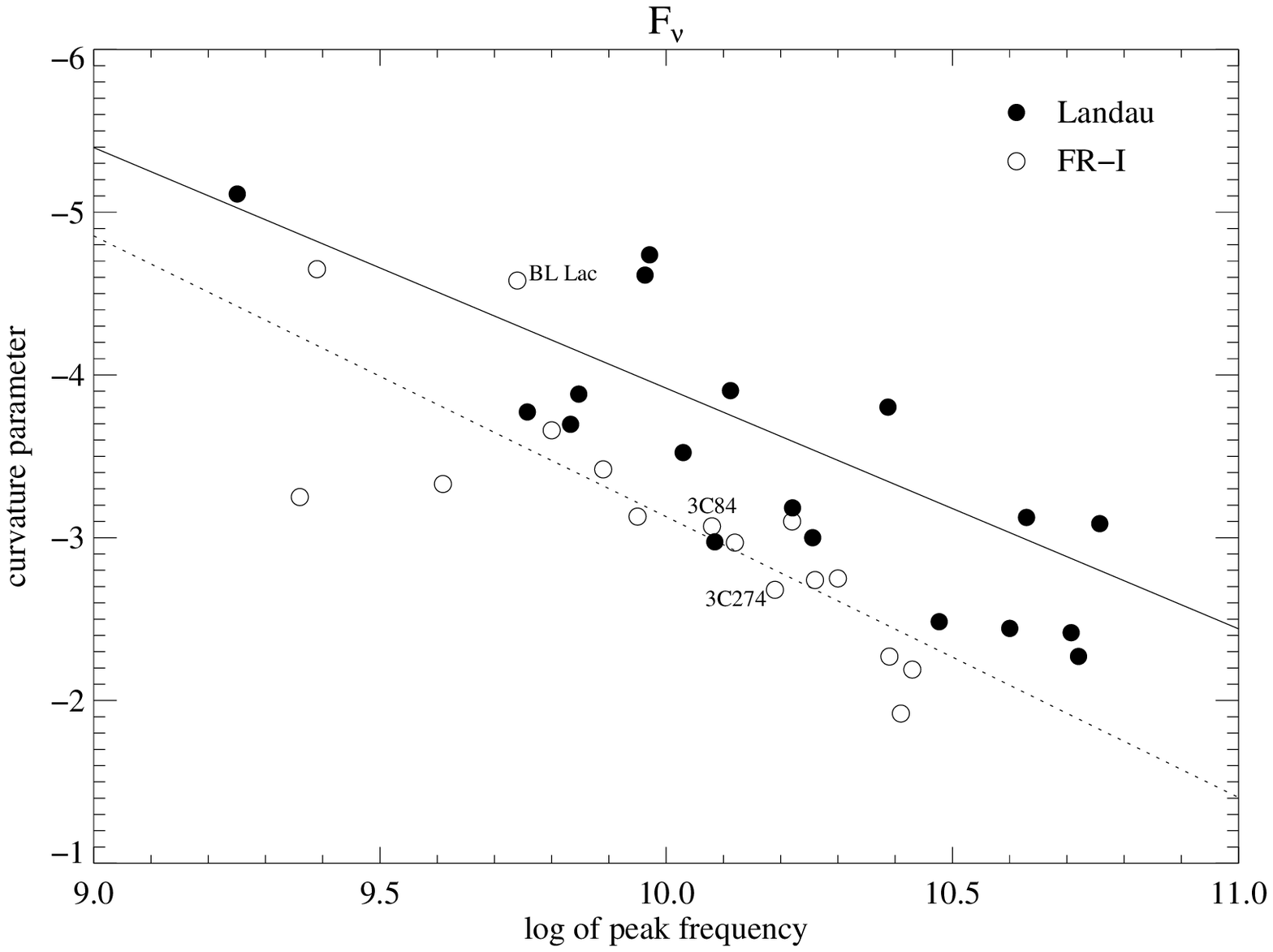}
\includegraphics[angle=0,scale=.44]{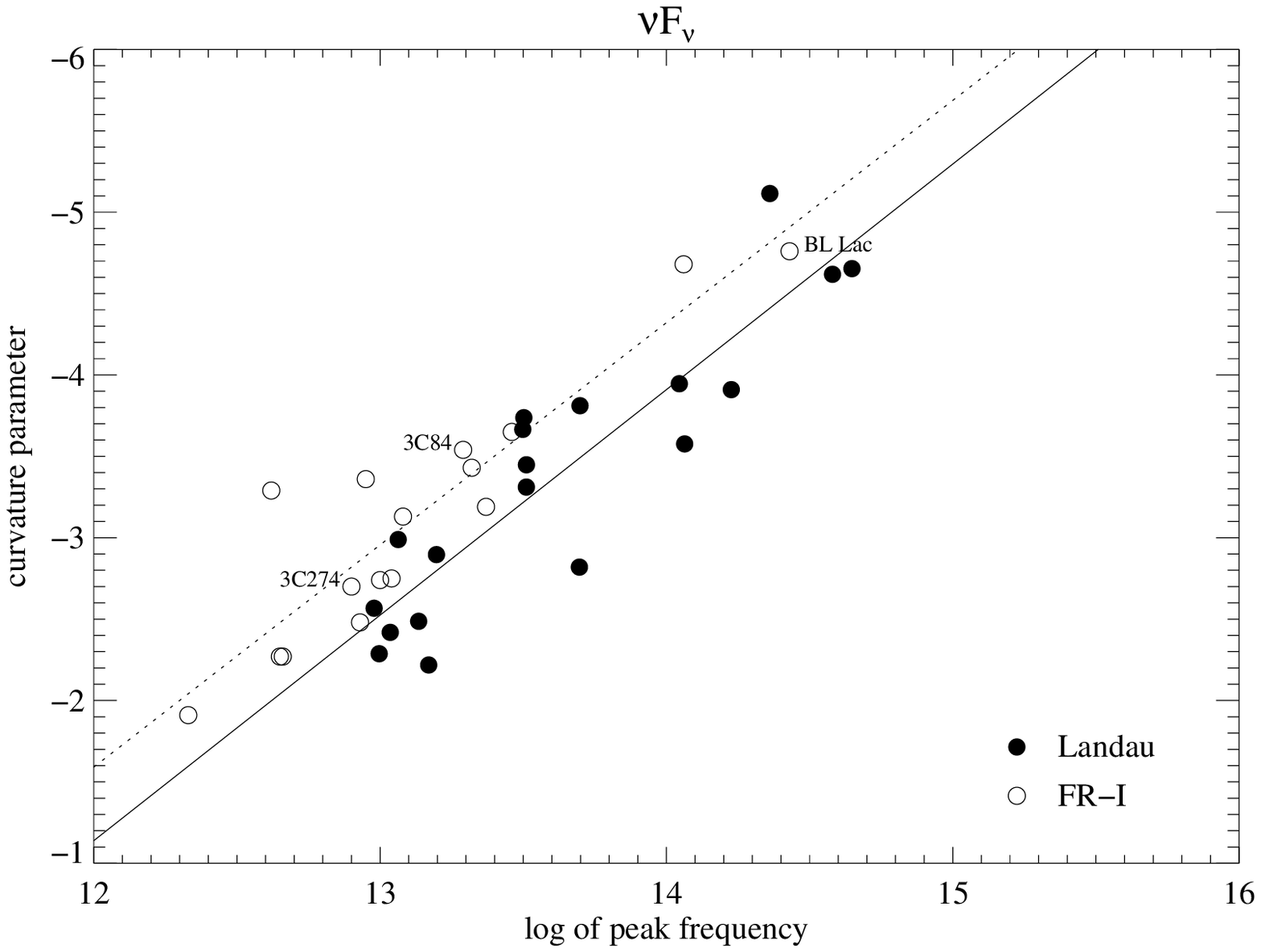}
\caption{Curvature parameter plotted against the logarithm of the peak
frequency for the core SED fits in F$_{\nu}$ ({\it left}) and
$\nu$F$_{\nu}$ ({\it right}). Open symbols represent  FR--I sources,
filled symbols the blazars from \citet{lan86}. The solid and dashed
lines show linear least--square fits to the blazar and FR--I data
points, respectively. Note that a smaller absolute number for the
curvature parameter corresponds to a more strongly curved SED. We
marked the sources BL Lac, 3C84, and 3C274 which are quite different from the
lobe--dominated objects of our core sample but show
extended radio emission of FR--I type.\label{peak_vs_curv}}
\end{figure*}

{\bf 3C272.1\quad}The unusual sharp turn from a slightly blue slope
into a steep, red slope observed in the MIR spectrum of this source 
is supported in the framework of the SED
where several FIR measurements indicate the presence of a strong
thermal bump from cool dust.  The spectrum lies well above the core
synchrotron fit at almost all wavelengths and comes closest near
$\lambda$\,$\sim$\,25\,$\mu$m where the slope change takes place 
(Fig.\,\ref{pahratio}).  The combination of strong PAH emission and a
very low PAH 7.7\,$\mu$m to 11.3\,$\mu$m  ratio results in a prominent
depression  at the location of the 7.7\,$\mu$m feature after
subtraction of the star  forming template. As discussed above
(\S4.1) the unusual PAH ratio might point toward  AGN
excitation instead of star formation but we count this source here as
host dominated.  {\it References:}
\citet{bow00,lee00,nag01,haas04,lee04,doi05}


\subsection{Sources without CCC measurements }
In \S4.4 we study the contributions of the synchrotron cores  to the
MIR emission in our sources. However, this test is limited to  objects
where optical CCC measurements are available. When using the CCC
fluxes in combination with radio core data, a nuclear SED can be
constructed which bridges the MIR wavelengths. This way a fairly
robust estimate of the non--thermal component can be achieved which
helps to distinguish between thermal and non--thermal contributions to
any residual MIR emission (i.e. emission not resulting from star formation 
or stellar processes).

Where no CCC measurements are available, a reliable estimate of the
synchrotron core contribution to the MIR emission is not
possible. However, the  overall shape of the spectrum might still
provide indications on which process -- thermal or non--thermal --
dominates the MIR emission. Out of the 25 sources considered in this
paper, 7 objects do not have CCC measurements available. We here
briefly comment on these seven sources:

{\bf 3C76.1\quad}The MIR spectrum of this source can be well explained
by a combination of star formation and stellar emission. No
significant residual continuum can be identified. However, this
spectrum  suffers from low S/N which limits the
reliability of any conclusions regarding the  MIR continuum
constituents.

{\bf 3C129\quad}After correcting for some contributions from star
formation and accounting for  stellar emission from the host galaxy,
3C129 shows strong, red continuum emission longwards of
$\sim$\,15\,$\mu$m (Fig.\,\ref{pahsub1}). Inspecting
Fig.\,\ref{pahsub1} reveals that we could in principle have used a
star--formation template with a steeper MIR/FIR slope which would
reduce the level of residual continuum. However, given the lack of any
strong optical emission lines, the general appearance of the optical
spectrum, and the clear dominance of an old stellar population 
\citep[as determined by][]{but09}, it seems unlikely to us that this source
harbors the stronger star formation which could power the redder
MIR/FIR colors of a different template. However, while residual nuclear 
continuum emission is observed in 3C129 we cannot discriminate between a
thermal or non--thermal origin because we lack optical CCC
measurements. But we note that the residual MIR spectrum shows some
concave upward curvature (Fig.\,\ref{pahsub1}) and is significantly different in 
appearance compared to objects like 3C66B or 3C189 in which the MIR continuum is 
dominated by non--thermal emission (Figs.\,\ref{obs_spectra},\ref{sed1}). 
This argues in support of the thermal origin of the residual emission.

{\bf 3C218\quad}The MIR spectrum of Hydra A can be well explained by
star formation only, perhaps with some minor contributions from
stellar emission at the shortest MIR wavelengths 
(Fig.\,\ref{add_spec}). Considering the equivalent width of the  PAH
features the continuum at $\lambda > 20\,\mu$m is in agreement with
what can be expected for normal star forming galaxies. The presence of 
a very young stellar population is also seen in optical spectra 
\citep{mel97,are01,wil04}.

{\bf 3C293\quad}The MIR spectrum of 3C293 is strongly dominated by
star formation and no residual nuclear continuum can be identified
(Fig.\,\ref{obs_spectra}). In these properties it is quite
similar to 3C31.

{\bf 3C424\quad}For 3C424 no significant contributions from star
formation can be identified in the MIR spectrum and a weak and noisy,
but securely detected, continuum can be seen.

{\bf 3C120} and {\bf E1821+640}  are broad-line AGN, which have MIR spectra
     typical of their class, displaying strong silicate emission
     features and high ionization emission lines (Fig.\,\ref{add_spec}).

\subsection{Comparison with blazars and related objects}

It has been suggested that most (but not all) FR--I radio sources are
the misaligned counterparts of BL Lac objects (\citealt{kol92};
\citealt{urr95}; \citealt{cas99}; \citealt{ant02a}). As outlined in
\S4.4, the SEDs of these strongly beamed, highly core dominated
sources are well described by the same parabolic function we utilized
to fit the cores of the FR--I objects (which is here used just as a
fitting function for descriptive purposes).

In order to allow the comparison between the blazars and the FR--I
cores  we used the quasi--simultaneous data from \cite{lan86} to fit
the blazars in the same way as the FR--Is.  However, we limited the
parabolic fit for the blazars to the radio (2\,cm through 20\,cm) plus
optical $R$--band measurements, thus maximizing the comparability with
the  FR--I fits (the wavelengths of the nuclear FR--I data used to
constrain their  fits were very similar). Overplotting all the blazar
data onto the  fits obtained using only the limited data points shows
that this approach yields a  reasonable representation of the curved
SEDs.

We will now compare the resulting fit parameters for peak frequency
and SED curvature of the (putatively) misaligned sources with the
respective data of their highly aligned (supposed) siblings. Moreover,
we performed  the SED fits not only in F$_{\nu}$ but also in
$\nu$F$_{\nu}$ using the same parabolic function as given in equation
(1) and using the same set of nuclear data.

As already shown in e.g. \citet{lan86} a correlation of peak frequency
and curvature in F$_{\nu}$ exists for blazars.  For FR--I sources a
very similar behavior is detected (Fig.\,\ref{peak_vs_curv}). However,
while both types of objects cover about the same range in F$_{\nu}$
peak frequencies there is a systematic  offset for the FR--Is towards
more strongly curved SEDs. A comparable result has been obtained using
broad band spectral indices of FR--Is and low--energy peaked BL
Lacs. These two types of objects  overlap in their radio/optical
power--law slopes but there is a clear tendency for FR--Is to have
steeper $\alpha_{\rm ro}$ than the BL Lacs \citep{har00,tru03,bal06a}.
Because for a parabola peaking at radio wavelengths (which is
typically the case for the sources considered here) a steeper
radio/optical single power--law slope corresponds to a higher
curvature, our results shown in Fig.\,\ref{peak_vs_curv} are
consistent with these previous findings.

In $\nu$F$_{\nu}$ the blazars also show a correlation between peak
frequency and curvature, but now the FR--Is seem to continue this
correlation at  lower peak frequencies. However, we see no trend in
the blazar distribution with FR morphology\footnote{Using
measurements for the extended radio luminosity  associated with the
blazars \citep{ant85} we introduce a luminosity  threshold (and
varying this threshold between log\,$L_{\rm 1.4\,GHz, ext}$  of 31.5
and 32.5 erg/s/Hz) for identifying ``FR--I blazars''. Taking those
blazars with extened radio luminosities below the threshold as the
possible aligned counterparts for our misaligned FR--I sources we see
no  trend of blazar FR type in Fig.\,\ref{peak_vs_curv}.}.  In
particular, blazars with extended radio luminosities typical for FR--I
objects do not necessarily concentrate towards the ``classical'' FR--I
objects in Fig.\,\ref{peak_vs_curv}, but seem to be randomly
distributed within the blazar sample.

The lower peak frequency of FR--Is in $\nu$F$_{\nu}$ could be
understood as a result of different  beaming factors for the emission
dominating in blazars and FR--I cores.  This seems to be
in agreement with the model of e.g. \citet{chia00b} who  assume that
the jet has a velocity structure with a fast moving central spine
(which is strongly beamed and dominates in aligned sources) and a
slower -- but still relativistic -- outer layer of more isotropic,
less strongly beamed emission which dominates in the misaligned cases.
Due to a shift in peak frequency of the emission, different amounts of
relativistic beaming for aligned (i.e BL Lac) and non--aligned (i.e. FR--I) soures 
would also affect the observed broad--band spectral indices. This can serve as one possible
explanation for the difference in $\alpha_{\rm ro}$ (or SED curvature)
between FR--Is and BL Lacs \citep[e.g.][]{chia00b,tru03,bal06a}.

\begin{table}[t!]
\setlength{\tabcolsep}{1.5mm}
\begin{center}
\caption{24\,$\mu$m and [\ion{O}{3}] emission--line fluxes.\label{taboiii}}
\begin{tabular}{l|r@{.}lcc}
\tableline\tableline
Object & \multicolumn{2}{c}{F$_{\rm rest}$({24\,$\mu$m})} & F([\ion{O}{3}]) & ref\\
       & \multicolumn{2}{c}{in mJy}                  & in erg/s/cm$^2$ & \\       
\tableline
     3C15 &    8&63 &                    3.15$\times10^{-15}$ & 1 \\ 
     3C29 &    6&80 &                    2.69$\times10^{-15}$ & 1 \\
     3C31 &   19&16 &                    4.71$\times10^{-15}$ & 1 \\
    3C66B &    7&42 &                    1.10$\times10^{-14}$ & 1 \\
   3C76.1 &    2&92 & \hspace*{-9pt}$<$\,2.91$\times10^{-15}$ & 1 \\
   3C83.1 &    4&46 & \hspace*{-9pt}$<$\,1.85$\times10^{-15}$ & 1 \\
     3C84 & 3130&13 &                    5.89$\times10^{-13}$ & 1 \\
    3C120 &  618&30 &                    3.47$\times10^{-12}$ & 7 \\
    3C129 &    6&95 & \hspace*{-9pt}$<$\,6.94$\times10^{-15}$ & 1 \\
    3C189 &    7&57 &                    2.05$\times10^{-15}$ & 5 \\
    3C218 &    8&77 &                    3.80$\times10^{-15}$ & 8 \\
    3C264 &   15&05 &                    1.47$\times10^{-15}$ & 1 \\
    3C270 &   42&59 &                    7.57$\times10^{-15}$ & 2 \\
  3C272.1 &   29&52 &                    5.11$\times10^{-15}$ & 1 \\
    3C274 &   63&16 &                    2.41$\times10^{-14}$ & 1 \\
   IC4296 &   13&83 & \hspace*{-9pt}$<$\,1.74$\times10^{-15}$ & 7 \\
    3C293 &   34&60 &                    1.36$\times10^{-15}$ & 1 \\
    3C317 &    4&74 &                    8.32$\times10^{-15}$ & 1 \\
NGC\,6251 &   41&98 &                    5.20$\times10^{-15}$ & 6 \\
    3C386 &    1&45 & \hspace*{-9pt}$<$\,2.53$\times10^{-14}$ & 1 \\
    3C424 &    2&60 &                    1.52$\times10^{-15}$ & 1 \\
    3C465 &    5&12 &                    3.11$\times10^{-15}$ & 1 \\
E1821+643 &  527&33 &                    2.45$\times10^{-13}$ & 3 \\
   BL Lac &  332&69 &                    3.82$\times10^{-15}$ & 4 \\
\tableline
\end{tabular}
\end{center}
\tablerefs{
  (1) \citealt{but09};\\
  (2) \citealt{ho97};
  (3) \citealt{kol91};\\
  (4) \citealt{law96};
  (5) SDSS;\\
  (6) \citealt{shu81};
  (7) \citealt{tad93};\\
  (8) \citealt{wil04}}
\end{table}

\begin{figure}[t]
\includegraphics[angle=0,scale=0.55]{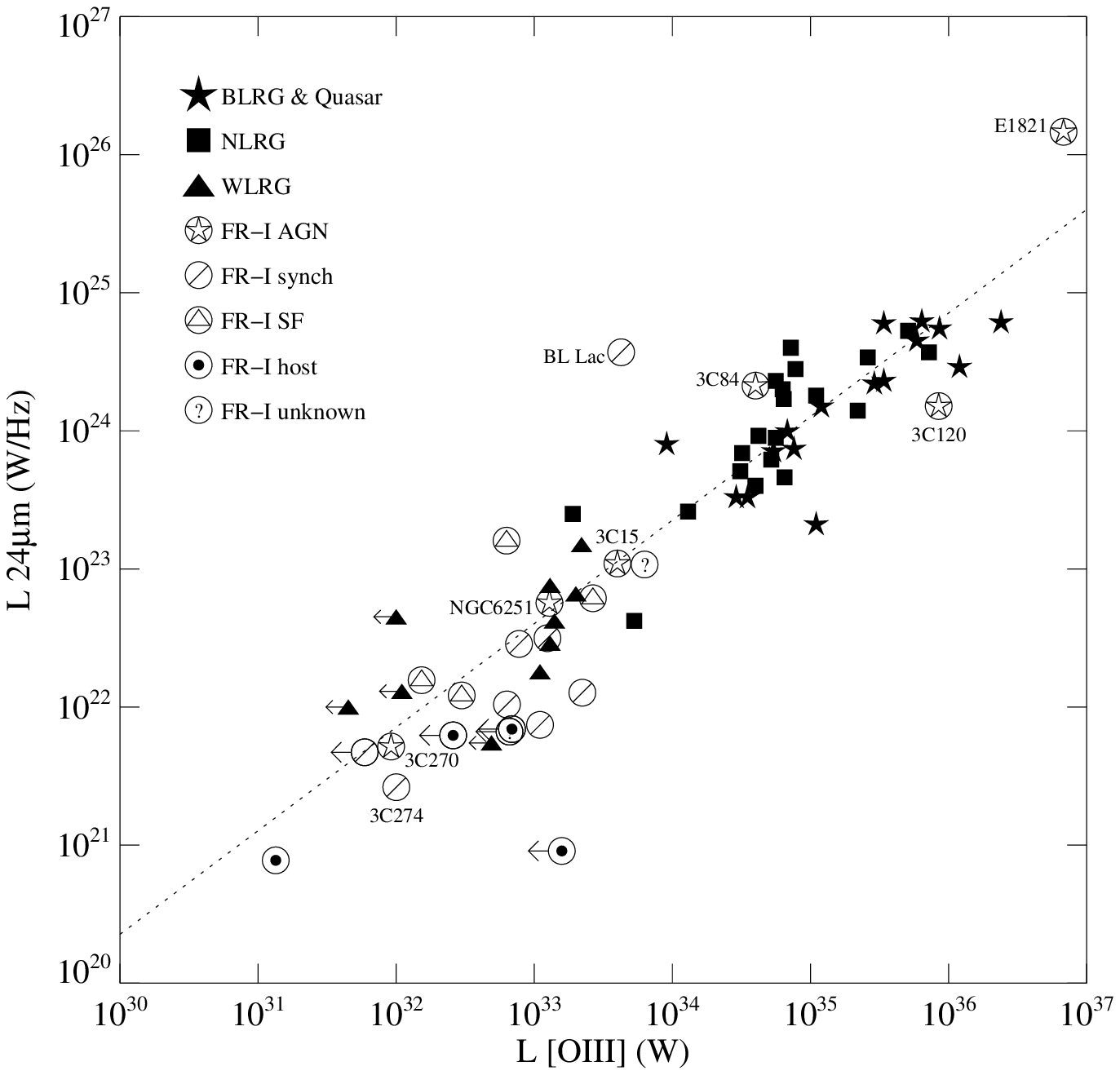}
\caption{MIR luminosity at 24\,$\mu$m plotted versus the [\ion{O}{3}]
emission--line luminosity. We show the objects presented in \citet{dic09} as
filled symbols and the FR--Is from this paper as open symbols. The
FR--I points also indicate the dominating factor of the MIR
emission: AGN -- AGN--heated dust; synch -- synchrotron emission; SF -- star formation; 
host -- stellar emission (plus possibly star formation). 
Sources labeled ``unknown'' (3C129 and 3C424) have MIR
continuum emission in excess of host galaxy contributions but due to
missing CCC measurements we cannot securely distinguish between AGN or
synchrotron emission. The dashed line is the bisector of a linear least square fit 
to the filled symbols (see text for details).
\label{mir_vs_oiii}}
\end{figure}

\subsection{MIR and emission--line luminosity}

Under the assumption that the narrow--line region of AGN is extended
and not affected by nuclear obscuration it has been argued that the
luminosity of the [\ion{O}{3}]\,$\lambda5007$\,\AA~emission line can
serve as an isotropic tracer of the intrinsic radiative AGN luminosity
\citep{raw91,tad98}. It has also been shown that for lumninous AGN the
radiative power, as traced by [\ion{O}{3}], is correlated with the MIR
luminosity at 24\,$\mu$m \citep{tad07,dic09}\footnote{For distant
FR--II radio galaxies and quasars an interesting thread in the
literature offers good evidence for the anisotropy of [\ion{O}{3}],
which we do not pursue here since we concentrate on FR--I objects.
But these interesting references are e.g. \citet{jack90},
\citet{hes93}, \citet{diser97}, \citet{simp98}. See also
\citet{haas05} for an infrared/optical comparison.}.
 
In Fig.\,\ref{mir_vs_oiii} we plot the luminosity at 24\,$\mu$m versus
the luminosity in the [\ion{O}{3}] emission--line for the the objects
presented in \citet[filled symbols]{dic09} and for our FR--I sources
(open symbols). We also show as a dashed line the bisector of a linear
least square fit to the filled symbols\footnote{The bisector was
determined using only the filled symbols and following the same
strategy as \citet{dic09}: Sources with upper limits on their
[\ion{O}{3}] emission were excluded as well as sources with clear
starburst signatures in their optical spectra. For the rest of this
section we refer to this sample of 33 objects as the ``D09
sample''.}. While the AGN FR--Is on average fall close to the
bisector the visual inspection of Fig.\,\ref{mir_vs_oiii} also 
reveals that the bulk of the remaining FR--I sources tend to have lower
24\,$\mu$m luminosities. In fact, a Kolmogorov--Smirnoff test of the
displacement in $L_{24\,\mu{\rm m}}$ from the fitted line gives only
a $\sim$\,1\,\% chance that two samples drawn from the same parent
population would show a difference as strong as we see for the D09
sample and the non--AGN FR--Is\footnote{In order to match the
selection criteria  of those two samples we excluded FR--I sources
with [\ion{O}{3}]  upper limits from the K--S statistics as well as
sources which are dominated  by star formation (plus BL Lac).}.  
The FR--I AGN sources on the other hand  are statistically consistent 
with being drawn from the same parent population as the D09 sample. We
note that this comparison assumes that the 24\,$\mu$m flux is
dominated by an AGN component. While this is a reasonable assumption
for the majority of the sources in the D09 sample (which mainly
consists of NLRGs and quasars) we have shown here that it is certainly
not the case for the non--AGN FR--Is. For the latter the bulk of the
MIR emission can be accounted for by other processes
(Tab.\,\ref{tab2}).

\begin{figure*}[t]
\centering
\includegraphics[angle=0,scale=0.43]{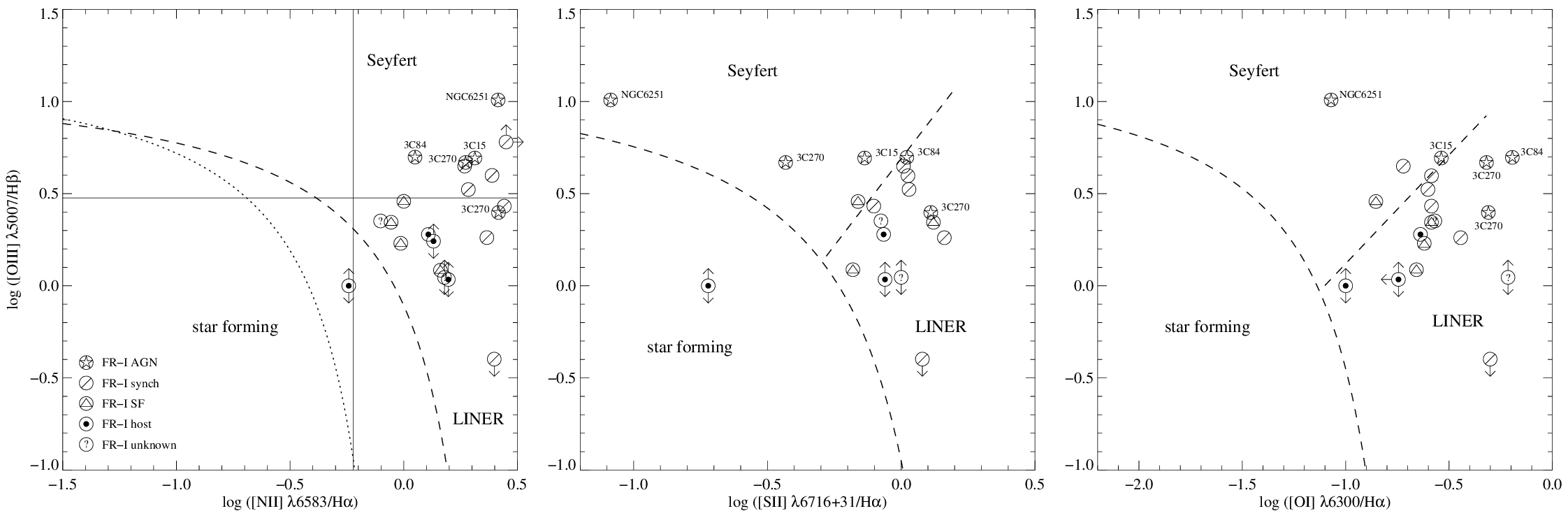}
\caption{Diagnostic optical emission--line diagrams. On the left the 
dotted line is the dividing line used by \citet{kau03} and the dashed line corresponds to the 
dividing line of \citet{kew01}. Sources with [\ion{N}{2}]/H$\alpha$\,$>$\,0.6 are often defined 
as Seyferts for [\ion{O}{3}]/H$\beta$\,$>$\,3 and as LINERs for [\ion{O}{3}]/H$\beta$\,$<$\,3. For 
the diagrams in the middle and on the right the dividing lines from \citet{kew06} are used. The 
symbols are the same as in Fig.\,\ref{mir_vs_oiii}. For 3C270 the data point with the higher 
[\ion{O}{3}]/H$\beta$ ratio corresponds to the small aperture {\it HST} data while the lower 
flux ratio represents the ground--based results (see text).
\label{fri_bpt}}
\end{figure*}

The differences between FR--Is with and without warm dust can
also be established by comparing their respective
MIR--to--[\ion{O}{3}] luminosity ratios: We find  log\,$[\nu
L_{\nu}$(24\,$\mu$m)/$L$([\ion{O}{3}])] =  $2.74 \pm 0.06$ for the
four AGN FR--Is but only  a ratio of $2.06 \pm 0.20$ for FR--Is
without clear signs for warm dust emission\footnote{We here exclude
the luminous broad--line AGN 3C120 and E1821+643, the synchrotron
dominated sources BL Lac and 3C274 as well as the four sources which
are dominated by emission from star formation (Tab.\,\ref{tab2}). Note
also that the slope of the fitted line shown in
Fig.\,\ref{mir_vs_oiii} is not equal one meaning that the
24\,$\mu$m/[\ion{O}{3}] ratio changes with luminosity.}.  The results
of this exercise are just as we would expect from the spectral
analysis and provide a strong consistency check for our method.

\subsection{Diagnostic diagrams}

Optical emission--line diagnostics are commonly used to identify the
dominant activity type in line--emitting sources
\citep[e.g.][]{bal81,vei87,kew01,kau03}.  Although we here largely
focus on the MIR continuum properties of our FR--I sources, a brief
study of their optical emission--lines ratios with respect to their
MIR spectral ``type'' seems promising. We take data for the
emission--line ratios either from the references given in
Tab.\,\ref{taboiii} or from \citet[3C218]{han95} and from
\citet[IC4296]{lew03}.

In Fig.\,\ref{fri_bpt} we show three emission--line ratio
diagrams  for the FR--I objects (minus the broad--line sources 3C120
and E1821+643 and excluding BL Lac). Virtually all FR--Is, regardless
of their MIR properties, fall into the  LINER or Seyfert region. Even
the star--forming sources occupy the  LINER region rather than showing
HII--type emission--line ratios (this  is discussed in the context of
ULIRGs in e.g. \citealt{ant02a,ant02b}).  Only one source (3C386)
occupies the transition region (Fig.\,\ref{fri_bpt}, {\it left}) or
the HII region (Fig.\,\ref{fri_bpt}, {\it middle}).

For the FR--Is with warm dust emission, 3C15 and NGC\,6251 fall into
the Seyfert region for all three diagrams, which historically requires
the  presence of an active nucleus in these objects. 3C84 and 3C270 on
the other hand can be classified as Seyferts or as LINERs depending on
which diagram is used (Fig.\,\ref{fri_bpt}). This, however, also shows
that sources which are LINERs optically can have significant emission
from warm dust in the MIR. Interestingly, both of these LINERs with
warm  dust have broad emission--lines detected: 3C84 shows a broad
H$\alpha$ component \citep{ho97b} while \citet {bar99} report  a
(tentative) broad line in polarized light for 3C270. This is
consistent  with the findings of \citet{stu06} who show that type--1
LINERs on average  have an additional warm dust component compared to
type--2 LINERs. Therefore  we can speculate that LINERs with warm dust
emission will also show broad emission  lines.

Because nuclear emission--line ratios can be diluted by
more extended emission from e.g. star formation, especially for the
weak and elusive sources considered here, we demonstrate the importance
of aperture effects in the case of 3C270 which appears twice on each diagram: For
every line--ratio combination shown (Fig.\,\ref{fri_bpt}) 3C270 is always
classified as a LINER when using the the emission--line fluxes
measured in a 2\arcsec$\times$4\arcsec~aperture \citep{ho97}.
However, considering only the line ratios for an {\it HST} spectrum of
the very nuclear 0.09\arcsec$\times$0.09\arcsec~\citep{fer96} places
3C270 in the Seyfert region for two out of the three diagrams.

Fig.\,\ref{fri_bpt} also reveals that some of the FR--Is which are
dominated by synchrotron emission in the MIR (and which lack clear
signs for substantial warm dust) exhibit Seyfert--like emission--line
ratios and many fall very close to the LINER/Seyfert transition line.

\section{Summary and conclusions}

The MIR spectra of 25 FR--I radio galaxies are presented. Many of
these sources show contributions from stellar emission at the shortest
MIR wavelengths observed. The spectra generally turn over from a blue
to a red continuum slope (in F$_{\nu}$) around $\sim$\,10\,$\mu$m.
Signs for star formation as traced by PAHs is detected in several
objects and the relative contribution to the total spectra range from
not detected (e.g. 3C66B, 3C317) to minor (e.g. 3C129, 3C270) to
dominant (e.g. 3C31, 3C293).

Focusing on sources with detected optical compact cores, the nuclear
SEDs for 15 objects are fitted with a parabolic function in order to
estimate the synchrotron contribution to the MIR emission.  For 7/15
objects (47\,\%) the nuclear MIR emission turned out to be dominated by
non--thermal processes: Apart from the stellar contributions at short
wavelengths, the MIR spectra of these sources are consistent with the
optical and radio core measurements.   Mid--infrared continuum
emission in excess over the synchrotron cores  is detected in the 8/15
remaining sources. For four out of those eight objects the spectra are
dominated by processes related to the host galaxy (star formation,
stellar emission). The remaining four sources (27\,\%) -- 3C15, 3C84, 3C270, and
NGC\,6251 -- show clear signatures of thermal MIR emission in excess
over the non--thermal core contributions. This excess cannot be
explained by either stellar emission or star formation, which argues
for the presence of thermal dust emission powered by an AGN. At
least one of these objects has been (tentatively) confirmed to have 
a hidden type-1 AGN with spectropolarimetry.

Thus, for most of our sources in which the nuclear MIR component  can
be identified (and is not masked by host galaxy emission) this
emission is likely to be of non--thermal origin. But some FR--I
sources do exist which show evidence for a hidden AGN, consistent with
findings at other wavelengths.

Parabolic functions were fitted to the radio and
optical core data in order to estimate the nuclear synchrotron
component. These fits were performed in F$_{\nu}$ as well as in
${\nu}$F$_{\nu}$. Comparing the fit parameters  for the FR--Is with
those for a set of blazars and related objects, it appears that in
F$_{\nu}$ both types of objects show a similar  relation between the
peak frequency and the curvature of the SED. However,  while FR--Is
cover about the same peak frequency range as the blazars used here
for comparison, they show systematically stronger curvature in their
SEDs.  This reflects in the fit parameters for ${\nu}$F$_{\nu}$ where
the FR--Is  show on average smaller peak frequencies and again
stronger SED curvature. Different amounts of relativistic beaming 
in BL Lacs and FR--Is could serve as a possible explanation for 
these results.

The comparison with a correlation between MIR and
optical emission--line luminosity found for luminous AGN reveals that
the AGN dominated FR--Is are consistent with that relation while the
non--AGN FR--Is deviate substantially. In addition we show that 
FR--Is with and without warm dust differ significantly in their 
average 24\,$\mu$m/[\ion{O}{3}] luminosity ratio.
The conclusion about the presence of an AGN in some of the FR--Is is 
further strengthened by their optical emission--line ratios.  
These tests provide a strong consistency check for our method.

One important conclusion of this paper, as well as of other results in
the literature, is that while most FR--Is lack powerful type--1 AGN,
it is not tenable to generalize on associations between FR--I galaxies
and ``non--thermal only'' AGN. A significant fraction of FR--Is do
have warm dust emission which could by attributed to hidden 
type--1 nuclei, a conclusion also anticipated on other
grounds \citep[e.g.][]{ant02a}.

\acknowledgments  This work is based on observations made with the
{\it Spitzer Space Telescope}, which is operated by the Jet Propulsion
Laboratory, California Institute of Technology under a contract with
NASA. Support for this work was provided by NASA through an award
issued by JPL/Caltech. We have  also made use of the NASA/IPAC
Extragalactic Database (NED), which is operated by the Jet Propulsion
Laboratory, California Institute of Technology, under contract with
NASA. We thank Ann Wehrle for a critical reading of a previous version
of this paper, and also Alan Marscher and Luis Ho for advice on
specific issues. We are greatful to Marco Chiaberge for comments
on a previous version of the paper and for newly measuring the optical
compact  core flux of 3C15. We are thankful for the critical comments
of the referee which helped to improve the paper.  This reaserch has
made use of SDSS data products. Funding for the SDSS and SDSS-II has
been provided by the Alfred P. Sloan Foundation, the Participating
Institutions, the National Science Foundation, the U.S. Department of
Energy, the National Aeronautics and Space Administration, the
Japanese Monbukagakusho, the Max Planck Society, and the Higher
Education Funding Council for England. The SDSS Web Site is
http://www.sdss.org/. 

{\it Facilities:} \facility{Spitzer (IRS)}

\appendix

\section{Appendix: Examples for the subtraction of the star-forming template}
Here we present additional plots on the subtraction of the star--formation component.  
The object 3C270 is already shown in
Figs.\,\ref{3c270} and \ref{3c270_fit} while 3C83.1 as well as 3C272.1 can be found in
Fig.\,\ref{pahratio}. This leaves 3C129, 3C386, and IC4296  as
objects in which a significant correction due to emission from star
formation was applied.  We do not show objects which are clearly
dominated by star formation in their MIR spectra  (see
Tab.\,\ref{tab2}). We also do not show objects where the contributions
from star formation -- as traced  by the PAH features -- is very minor
and thus only a very small correction was made (3C84, 3C120, NGC\,6251,
E1821+643; Tabs.\,\ref{tab2},\ref{tab4}).

\begin{figure*}[h!]
\centering
\includegraphics[angle=0,scale=.44]{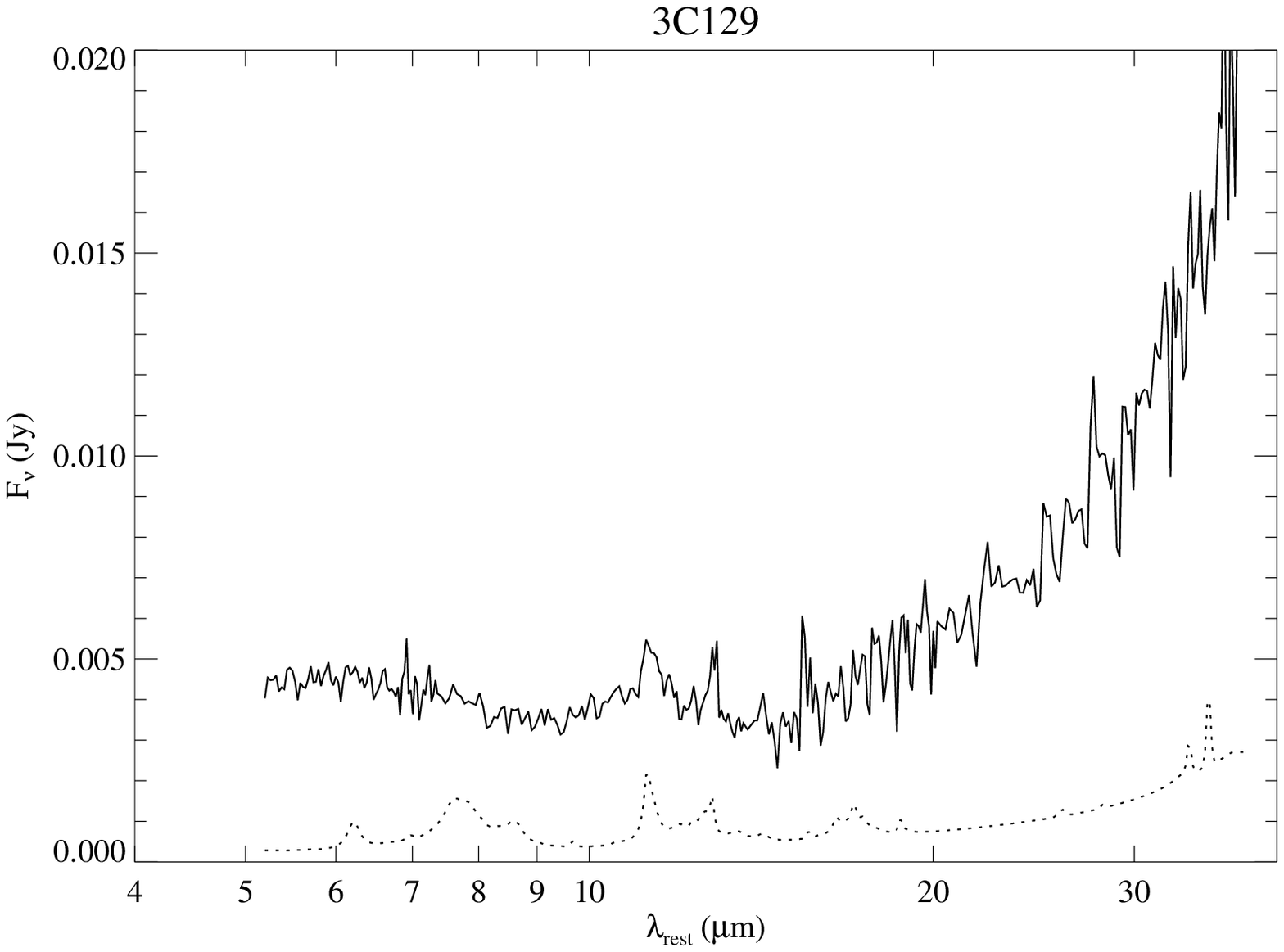}
\includegraphics[angle=0,scale=.44]{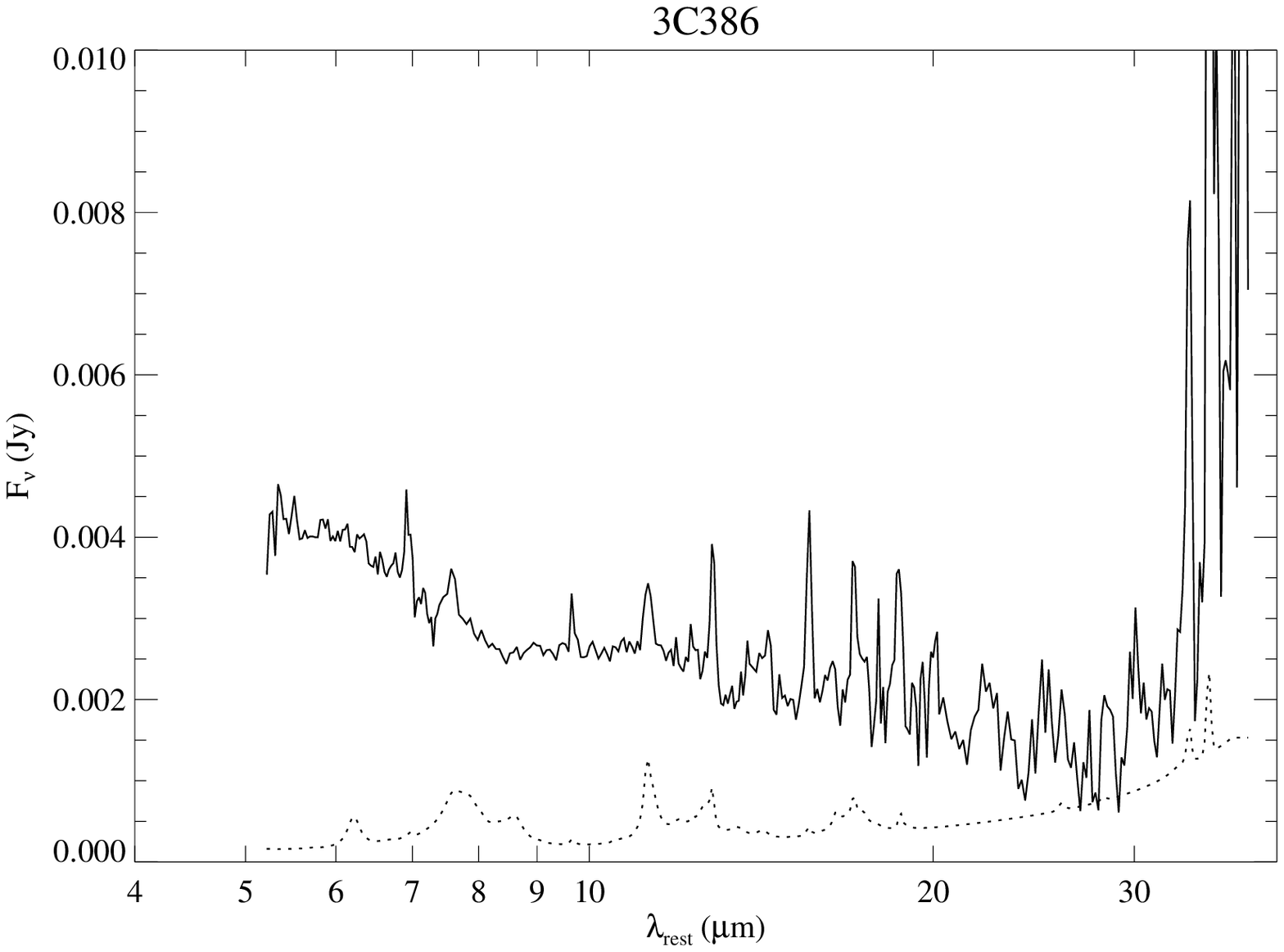}\\
\includegraphics[angle=0,scale=.44]{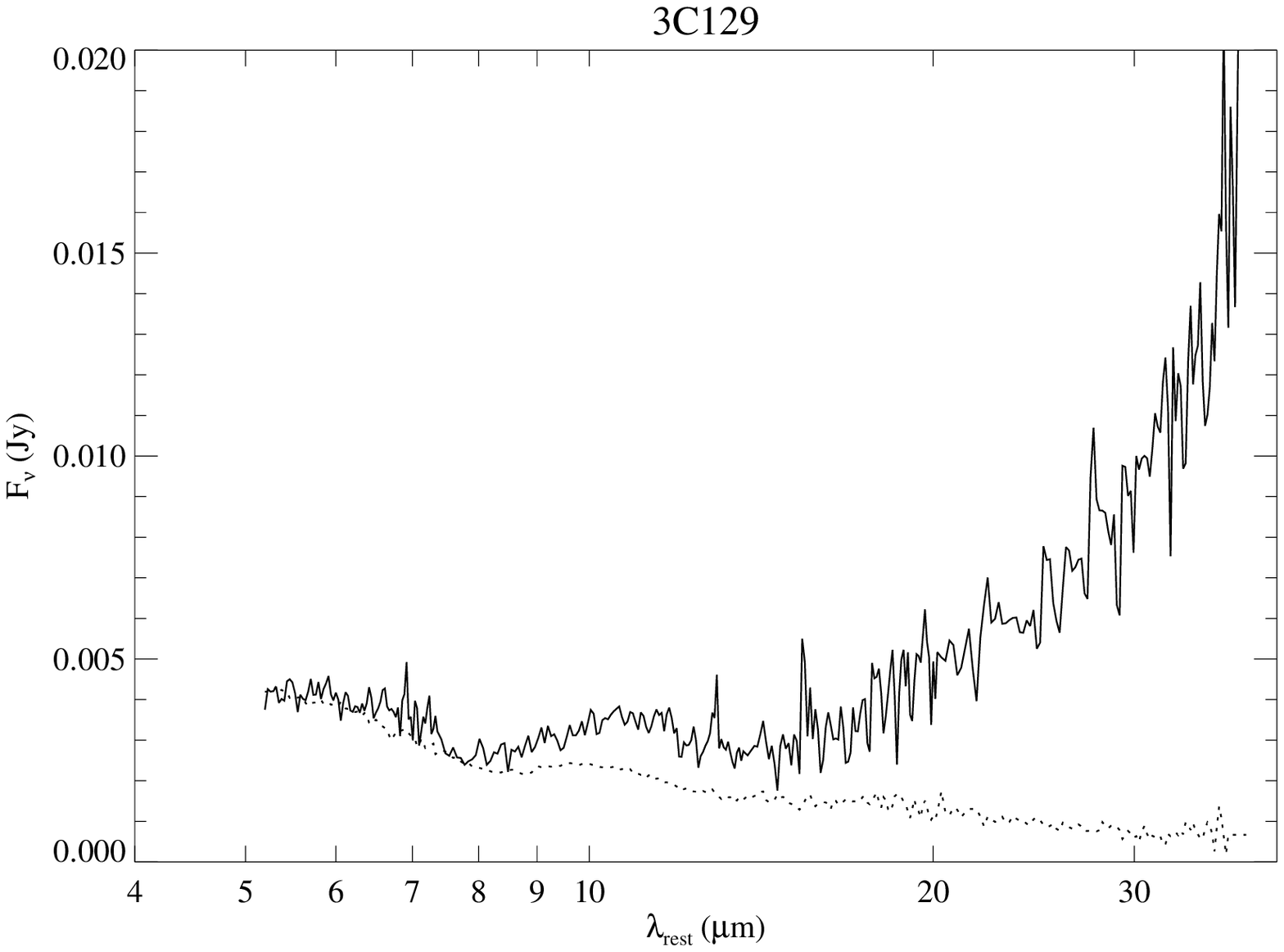}
\includegraphics[angle=0,scale=.44]{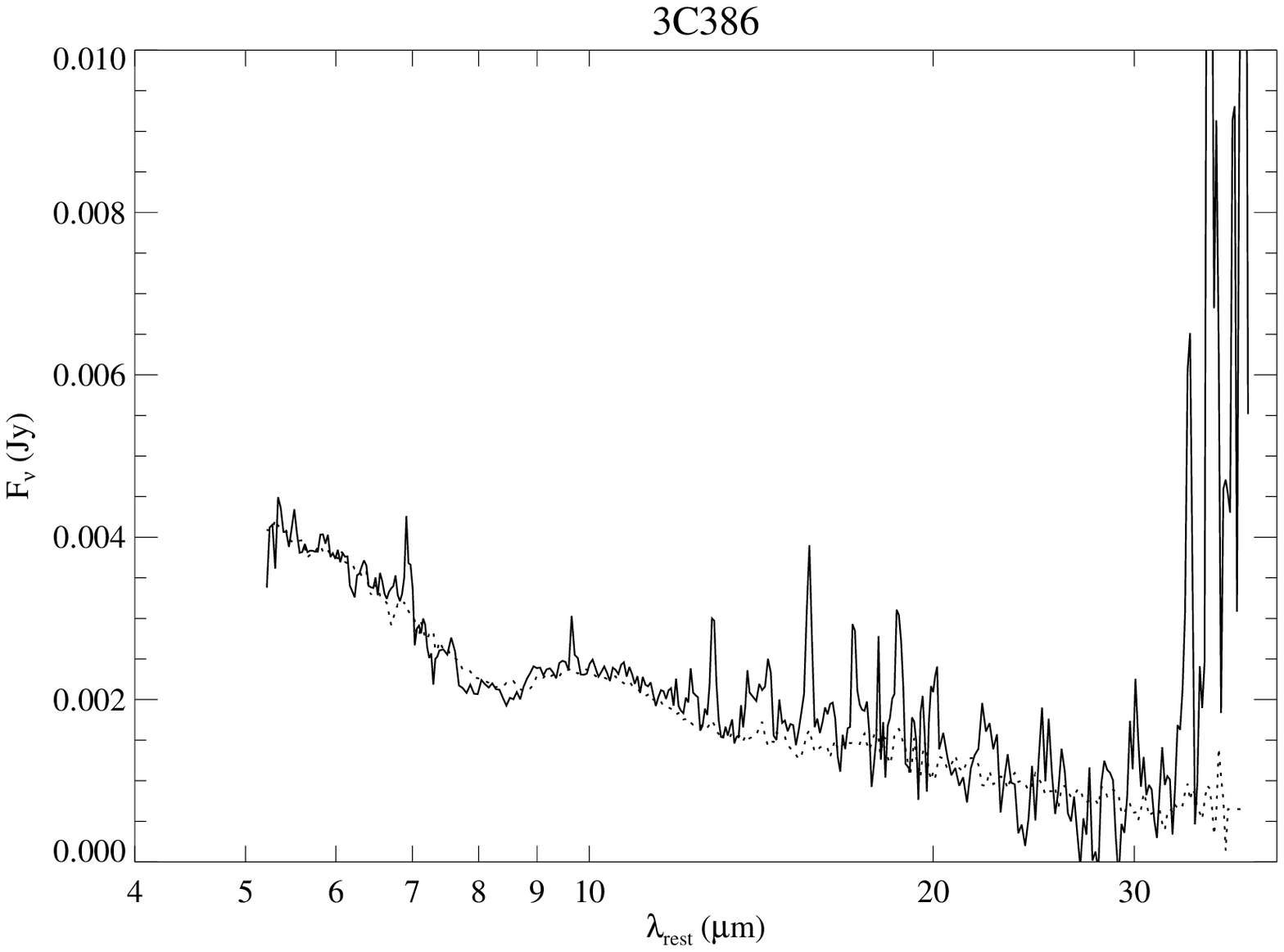}\\
\caption{Subtraction of the star-forming template for 3C129 ({\it
 left}) and 3C386 ({\it right}).  The top panel shows the observed
 spectrum with the scaled star--forming template. The bottom  panel
 shows the corrected spectrum with a scaled spectrum of an early--type
 galaxy (NGC\,1549).
\label{pahsub1}}
\end{figure*}

\begin{figure}[ht!]
\centering \includegraphics[angle=0,scale=.44]{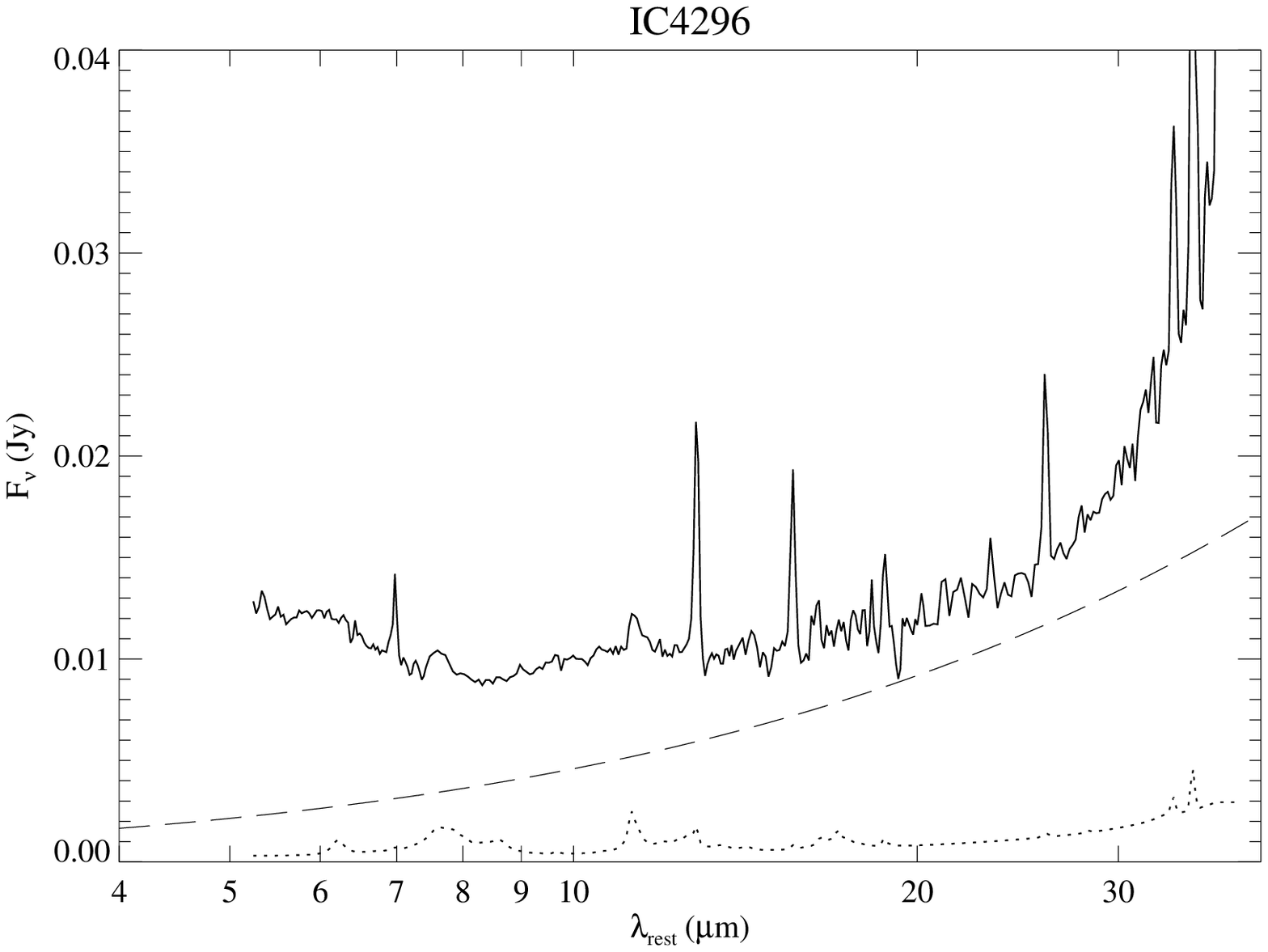}
\includegraphics[angle=0,scale=.44]{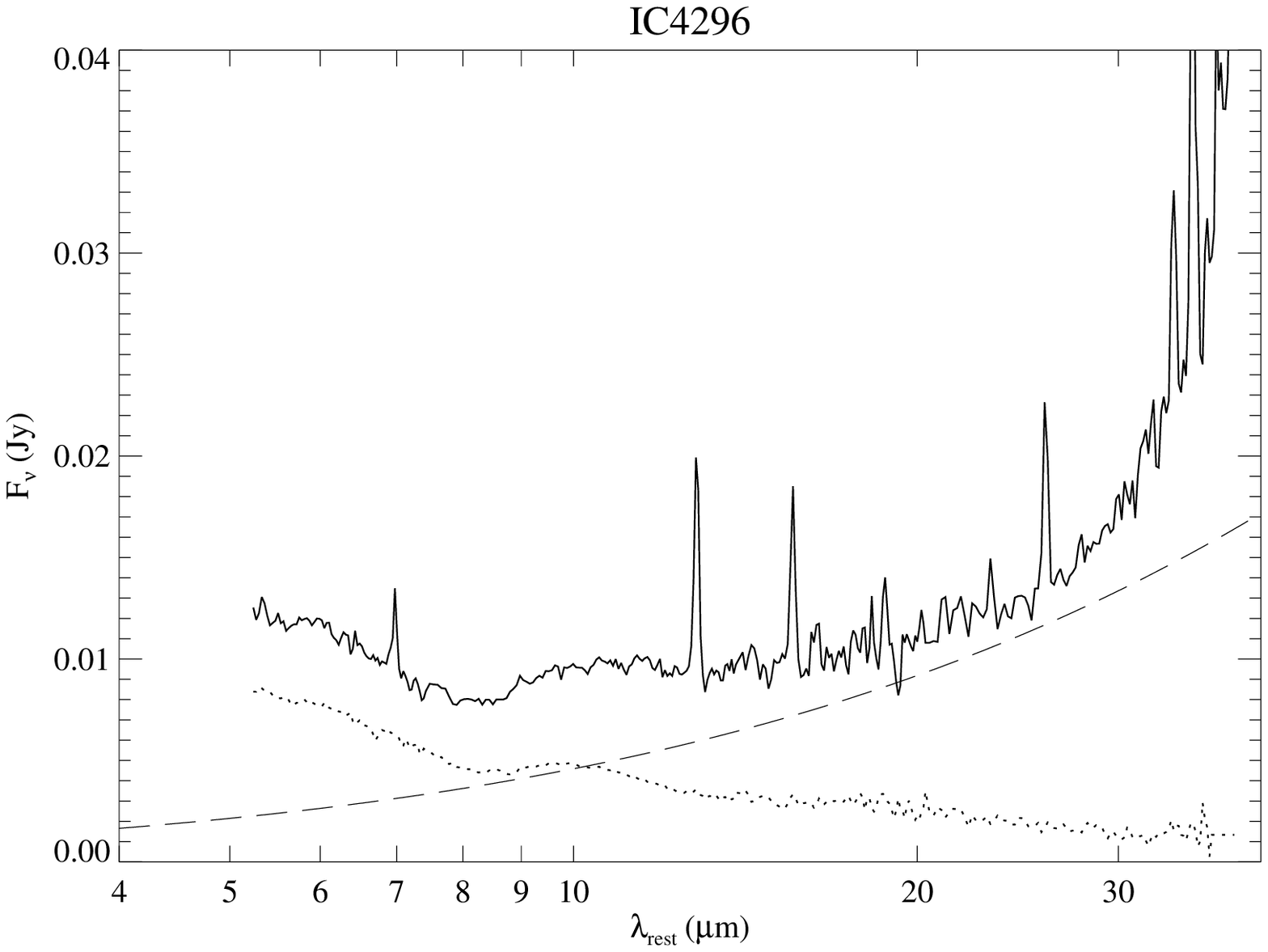}
\caption{Subtraction of the star--forming template for IC4296. The
panels are organized the same way as in Fig.\,\ref{pahsub1}. Here it
becomes obvious what is mentioned in the note to Table \ref{tab2}: Due
to the strong relative contribution from the synchrotron core (long--dashed line; Fig.\,\ref{sed2}) the
scaling of the templates to {\it features} in the spectrum can result
in  overcorrections.  Also, the steep slope at $\lambda>30\,\mu{\rm
m}$ hints at substantial emission from cold dust (which is supported by FIR measurements; Fig.\,\ref{sed1}) might
influence the measurement at 30\,$\mu$m.\label{pahsub2}}
\end{figure}

\end{document}